\documentclass[prd,aps,twocolumn,nofootinbib,showpacs,amsmath,superscriptaddress]{revtex4-1}
\usepackage{graphicx}
\usepackage{epstopdf}
\usepackage{bm}
\usepackage{epsfig}
\usepackage{graphics}
\usepackage{xspace}
\usepackage{amssymb}
\usepackage{amsmath}
\usepackage{xcolor}
\usepackage[utf8]{inputenc}
\usepackage[normalem]{ulem}
\usepackage{stmaryrd}
\def\OMIT#1{}

\newcommand{\nn}{\nonumber}

\newcommand{\bea}{\begin{eqnarray}}
\newcommand{\eea}{\end{eqnarray}}

\newcommand{\gsim}{\mathrel{\rlap{\lower4pt\hbox{\hskip1pt$\sim$}}\raise1pt\hbox{$>$}}}

\newcommand{\be}{\begin{equation}}
\newcommand{\ee}{\end{equation}}

\begin{document}
\title{Dynamic Jet Charge}
\author{Zhong-Bo Kang}
%\email{zkang@g.ucla.edu}
\affiliation{Department of Physics and Astronomy, University of California, Los Angeles, CA 90095, USA}
\affiliation{Mani L. Bhaumik Institute for Theoretical Physics,
University of California, Los Angeles, CA 90095, USA}
\affiliation{Center for Frontiers in Nuclear Science, Stony Brook University, Stony Brook, NY 11794, USA}

\author{Xiaohui Liu}
%\email{xiliu@bnu.edu.cn}
\affiliation{Center of Advanced Quantum Studies, Department of Physics, Beijing Normal University, Beijing 100875, China}
\affiliation{Center for High Energy Physics, Peking University, Beijing 100871, China}
\author{Sonny Mantry}
%\email{sonny.mantry@ung.edu}
\affiliation{Department of Physics and Astronomy,
                   University of North Georgia,
                   Dahlonega, GA 30597, USA}
\author{M.C. Spraker}
%\email{mark.spraker@ung.edu}
\affiliation{Department of Physics and Astronomy,
                   University of North Georgia,
                   Dahlonega, GA 30597, USA}
\author{Tyler Wilson}
%\email{trwils5018@ung.edu}
\affiliation{Department of Physics and Astronomy,
                   University of North Georgia,
                   Dahlonega, GA 30597, USA}

\begin{abstract}

We propose a modified definition of the jet charge, the ``dynamic jet charge", where the constant jet momentum fraction weighting parameter, $\kappa$, in the standard jet charge definition is  generalized to be a function of a dynamical property of the jet or the individual jet constituents. The dynamic jet charge can complement analyses based on the standard definition and give improved discrimination between quark and gluon initiated jets and between jets initiated by different quark flavors. We focus on the specific scenario where
each hadron in the jet contributes to the dynamic jet charge with a  $\kappa$-value dynamically determined by its jet momentum fraction. The corresponding dynamic jet charge distributions have qualitatively distinct features and are typically characterized by a multiple peak structure. For proton-proton collisions, compared to the standard jet charge, the dynamic jet charge gives significantly improved discrimination between quark and gluon initiated jets and comparable discrimination between $u$- and $d$-quark initiated jets. In Pythia simulations of heavy ion collisions, the dynamic jet charge is found to have higher jet discrimination power compared to the standard jet charge, remaining robust against the increased contamination from underlying event. We also present phenomenological applications of the dynamic jet charge to probe nuclear flavor structure.
\end{abstract}

\maketitle

\section{Introduction}

Identifying the electric charge of partons that emerge from the hard scattering process to initiate the formation of jets can be useful in the search for new physics and testing various aspects of the Standard Model. Earlier experimental application of jet charge~\cite{Field:1977fa} was in deep inelastic scattering studies~\cite{Berge:1979qg,Berge:1980dx,Allen:1982ze,Albanese:1984nv,Barlag:1981wu,Erickson:1979wa}, finding evidence for quarks in the nucleon. The jet charge observable has also been applied in measurements of the charge asymmetry~\cite{Stuart:1989db,Decamp:1991se}, in tagging the charge of bottom quark jets~\cite{Braunschweig:1990cv,Abazov:2006vd,Aaltonen:2013sgl,Aad:2013uza} and hadronically decaying $W$ bosons~\cite{Khachatryan:2014vla,Chen:2019uar}, determination of electroweak parameters~\cite{Buskulic:1996kx}, and in testing aspects of perturbative Quantum Chromodynamics (QCD)~\cite{Krohn:2012fg,Waalewijn:2012sv, Aad:2015cua,Sirunyan:2017tyr,Fraser:2018ieu}. Jet charge has also been used to probe nuclear medium induced jet quenching  on quark and gluon initiated jets in heavy ion collisions~\cite{Gyulassy:1993hr,Wang:1994fx,Chien:2016led,Connors:2017ptx,Li:2019dre,Sirunyan:2020qvi,Du:2020pmp}. Most recently, a new theoretical framework~\cite{Kang:2020fka,Kang:2020xez} was introduced to use jet charge as a probe of the quark flavor structure of the nucleon.

The jet charge is one of a variety of  jet substructure tools used for jet discrimination~\cite{Gallicchio:2011xq,Sirunyan:2020qvi,Brewer:2020och}. A comprehensive review of other jet substructure techniques can be found in Ref.~\cite{Larkoski:2017jix}. The utility of the jet charge observable for jet discrimination has also prompted the development of machine learning techniques~\cite{Fraser:2018ieu,Chen:2019uar} for extracting the jet charge.  Recently, a color tagger was introduced~\cite{Buckley:2020kdp} as another jet discriminant in order to distinguish between color singlet states  decaying into two jets from dijet backgrounds.   In a recent analysis~\cite{Chien:2018dfn}, it was shown that the quark-gluon jet discrimination power of the various jet substructure techniques  typically worsens in heavy ion collisions, compared to proton-proton ($pp$) collisions, but suggested a systematically improvable framework for studying medium modification for quark and gluon initiated jets. 

In the context of these various tools developed for jet discrimination, we introduce a modified definition of the standard jet charge, that we refer to as the \textit{dynamic jet charge}. This new definition can complement analyses based on the standard jet charge and allows for improved discrimination between quark and gluon initiated jets and between $u$-quark and $d$-quark initiated jets. We present simulation results to illustrate the characteristic properties of the dynamic jet charge and its discrimination power at the LHC.  We also present corresponding results for Pythia simulations of heavy ion PbPb-collisions and find that, unlike for the standard jet charge, the discrimination power of the dynamic jet charge remains similar to that found in $pp$-collisions,  being largely unaffected by the significantly greater underlying activity. While Pythia simulations account for the increased underlying event activity in heavy ion collisions, they do not include nuclear medium effects and the related jet quenching effects. Further studies  of the dynamic jet charge for heavy ion collisions could be done using the JEWEL~\cite{Zapp:2013vla} or JETSCAPE~\cite{Putschke:2019yrg} simulation which includes hot nuclear medium (i.e. quark-gluon plasma) effects, and is left for future work.
%%%%%%%%%%%%%%%%%%%%%%%%%
\begin{figure}
    %\centering
        \includegraphics[scale=0.8]{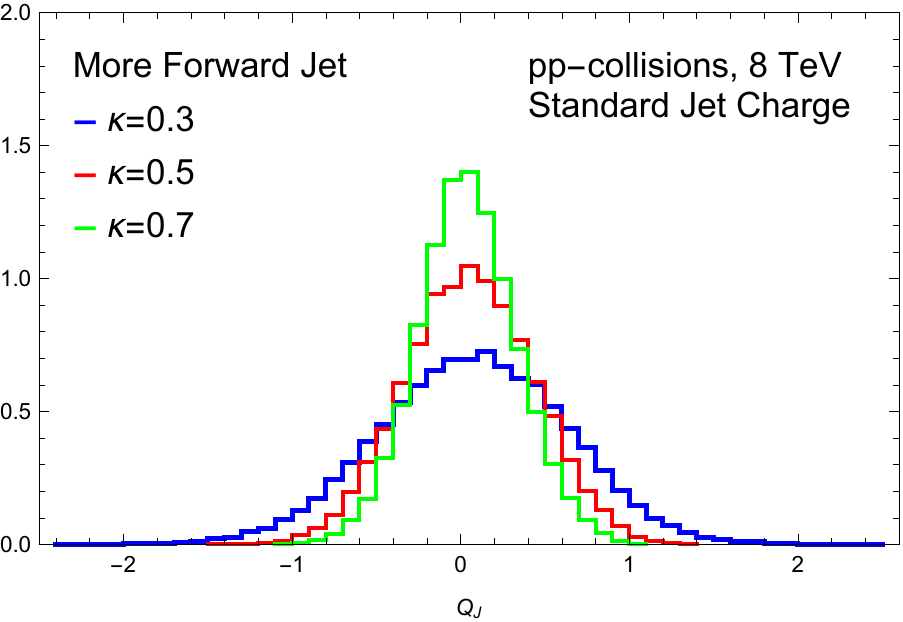}
        \includegraphics[scale=0.8]{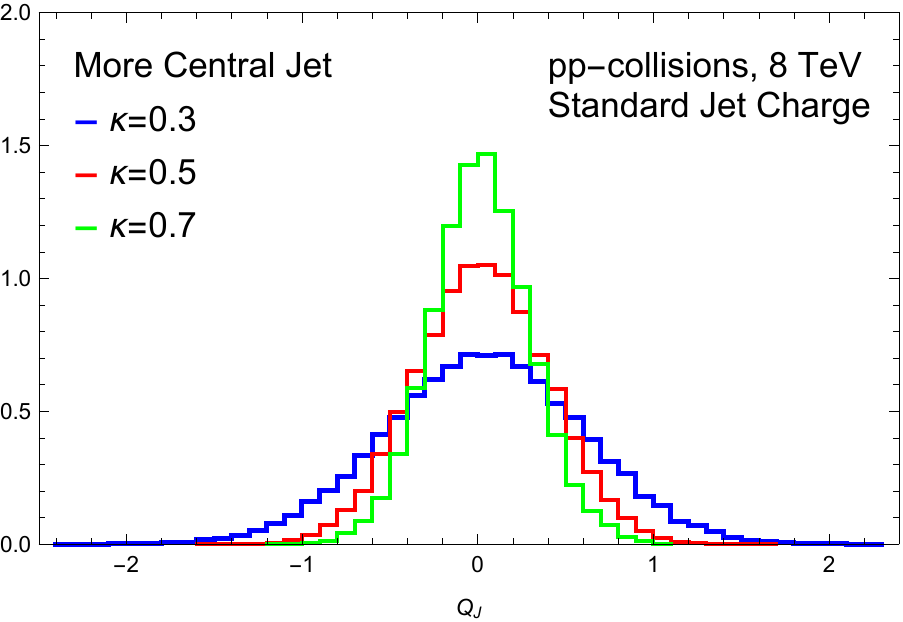}
         \caption{Normalized standard jet charge distributions for the more forward (top panel) and more central (bottom panel) of the leading jets in Pythia8 simulations of $pp\to j_1j_2 X$ at $\sqrt{s}=8 $ TeV with hadronization and MPI effects turned on.  Selection cuts of 200 GeV $< p_{T_{j_1,j_2}}<$ 300 GeV, $|\eta_{j1,j2}| <2.1$, and $p_{T_{j_1}}/p_{T_{j_2}} < 1.5$ on the leading ($j_1$) and subleading ($j_2$) anti-k$_T$ jets of jet radius $R=0.4$ are applied.
The blue, red, and green curves correspond to $\kappa=0.3,0.5,$ and 0.7, respectively.}
    \label{fig:StdJCkappa}
\end{figure}
%%%%%%%%%%%%%%%%%%%%%%%%%
\begin{figure}
    \centering
    \includegraphics[scale=0.8]{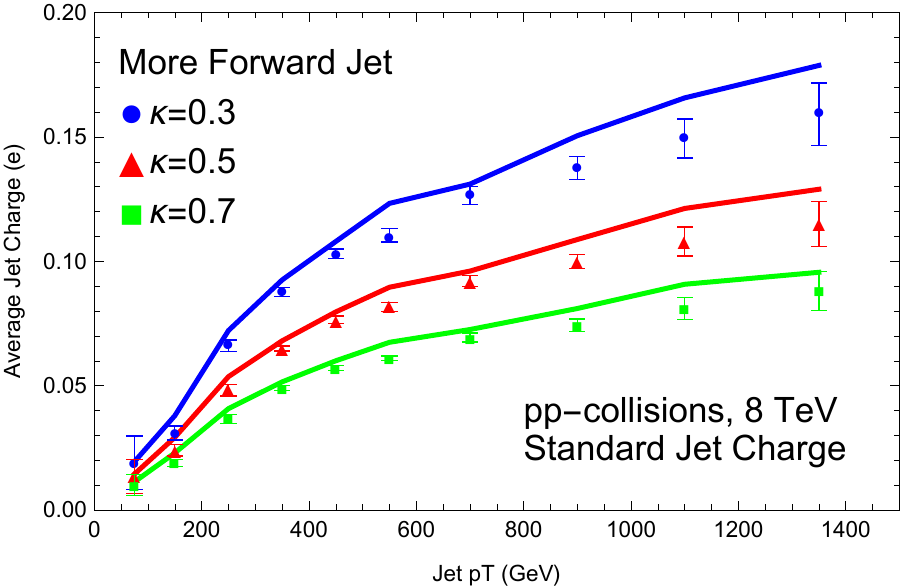}
    \includegraphics[scale=0.8]{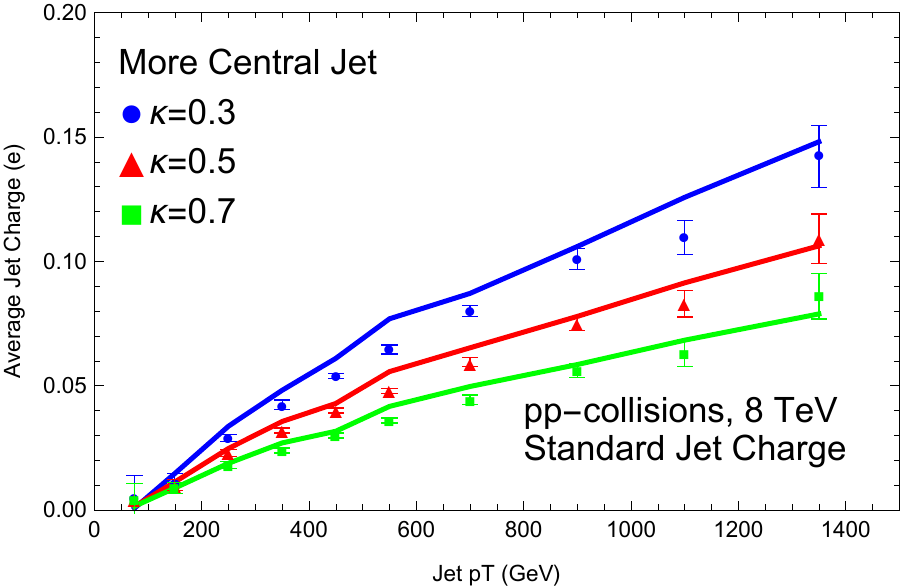}
    \caption{Comparison of the ATLAS data~\cite{Aad:2015cua} for the standard jet charge with Pythia 8.240 simulations (default tune). The average standard jet charge is plotted as a function of the jet $p_T$ in $pp\to j_1j_2 X$ at $\sqrt{s}=8 $ TeV. The average jet charge of the more forward (top panel) and more central (bottom panel) of the two leading jets are shown for  the values $\kappa=0.3,0.5,0.7$ corresponding to the blue, red, and green curves respectively. Selection cuts of 200 GeV $< p_{T_{j_1,j_2}}<$ 300 GeV, $|\eta_{j1,j2}| <2.1$, and $p_{T_{j_1}}/p_{T_{j_2}} < 1.5$ on the leading ($j_1$) and subleading ($j_2$) anti-k$_T$ jets of jet radius $R=0.4$ are applied. The jet charge values correspond to an average over the jet $p_T$ bins: [50 GeV,100 GeV], [100 GeV, 200 GeV], [200 GeV, 300 GeV], [300 GeV, 400 GeV], [400 GeV, 500 GeV], [500 GeV, 600 GeV], [600 GeV, 800 GeV], [800 GeV, 1000 GeV], [1000 GeV, 1200 GeV], and [1200 GeV, 1500 GeV].}
    \label{fig:avgJC}
\end{figure}
%%%%%%%%%%%%%%%%

The standard definition~\cite{Field:1977fa,Krohn:2012fg,Waalewijn:2012sv} of the jet charge is given by the weighted sum
\begin{align}
\label{eq:stdJC}
Q_\kappa^i=\sum_{h\in {i\text{-}\rm jet}} z_h^\kappa \>Q_h ,
\end{align}
for a jet initiated by the parton $i$ and the sum is over all hadrons $h$, with charge $Q_h$, in the jet. $\kappa >0$ is a constant parameter that is part of the jet charge definition, and $z_h$ is the jet transverse momentum or energy  fraction carried by the hadron $h$
\begin{align}
    z_h = \frac{p_{T_h}}{p_{T_J}} \qquad\text{or}\qquad  z_h = \frac{E_h}{E_J},
\end{align}
for $pp$ and $e^+e^-$ colliders, respectively. Here $p_{T_h}$ and $E_h$  denote the transverse momentum and energy of the hadron $h$ in the jet, respectively.
Similarly,   $p_{T_J}$ and $E_J$ denote the total jet transverse momentum and energy, respectively. In this work, we focus on the hadron collider environment and use the $p_T$-weighted jet charge definition.  

In this work, we propose the dynamic jet charge, denoted by $Q_{\rm dyn}$, defined as
\begin{align}
\label{eq:dynamicJC}
Q_{\rm dyn}^i=\sum_{h\in {i\text{-}\rm jet}} z_h^{\kappa({\rm P})} \>Q_h ,
\end{align}
for a jet initiated by parton $i$, where the constant parameter $\kappa$ in the standard definition is promoted to a function $\kappa({\rm P})$ of some property ``$P$" of the jet or each individual jet constituent. Here we focus exclusively on the scenario where the property $P$ is chosen to be the hadron momentum fraction  $z_h$ for each hadron in the jet. Thus,  each individual jet constituent contributes to the dynamic jet charge with the dynamically determined value $\kappa({z_h})$. We demonstrate that this dynamic jet charge definition can allow for enhanced jet discrimination and complement analyses based on the standard jet charge definition.

One might also consider scenarios where $\kappa$ is a function of other jet constituent properties such as the hadron transverse momentum with respect to the jet axis, $\kappa (k_h^\perp/p_{T_J})$, or more global dynamic jet properties such as the  jet mass, $\kappa(m_J/p_{T_J})$, or the groomed jet radius~\cite{Larkoski:2014wba,Kang:2019prh}, $\kappa(R_g)$, in defining the dynamic jet charge. Furthermore, one might generalize other jet observables, such as jet angularities~\cite{Berger:2003iw,Hornig:2009vb,Kang:2018qra,Kang:2018vgn,Bell:2018gce,Budhraja:2019mcz,Bauer:2020npd}, by transforming the constant parameters that appear in their definitions to dynamic parameters. If and how these dynamic jet charge definitions lead to increased discrimination power will be explored in future work. 

The theoretical prediction for the average standard jet charge is given by~\cite{Waalewijn:2012sv}
\begin{align}
\langle Q_\kappa^i \rangle = \int_0^1 dz\> z^\kappa \sum_{h\in {i\text{-}\rm jet}} Q_h \frac{1}{\sigma_{i\text{-}{\rm jet}}} \frac{d\sigma_{h\in {i\text{-}\rm jet}}}{dz},
\label{eq:avgjetcharge}
\end{align}
where $\sigma_{i\text{-}{\rm jet}}$ denotes the cross section for producing a jet initiated by the parton $i$ and $d\sigma_{h\in {i\text{-}\rm jet}}/dz$ is the differential cross section for producing the jet in which a hadron $h$ with momentum fraction $z$ is observed. This can be brought into the form~\cite{Waalewijn:2012sv}
\begin{align}
\langle Q_\kappa^i \rangle = \int_0^1 dz\> z^\kappa \sum_{h\in {i\text{-}\rm jet}}Q_h\sum_j  &\int _z^1 \frac{dz'}{z'} {\cal J}_{ij}(p_{T_J},R,z',\mu) \nn \\ 
&\times D_j^h\left (\frac{z}{z'},\mu\right ),
\label{eq:avgjetchargeFF}
\end{align}
where the ${\cal J}_{ij}(p_{T_J},R,z,\mu)$ are perturbatively calculable coefficients and the $D_j^h\left (z,\mu\right )$ are the nonperturbative fragmentation functions describing the fragmentation of the parton $j$ into the hadron $h$ which carries away momentum fraction $z$ from the parton $j$. The argument $R$ in the coefficients ${\cal J}_{ij}$ denotes the jet radius. The resummation of large logarithms in $\sim p_T R/\Lambda_{\rm QCD}$, due to the disparity in energy scales associated with jet dynamics and hadronization, is achieved by choosing the renormalization scale $\mu \sim p_T R$~\cite{Dasgupta:2016bnd,Kang:2016mcy} in the perturbative coefficients and using the standard DGLAP evolution to evaluate the fragmentation functions at this scale. Next-to-Leading Order (NLO) results for the ${\cal J}_{ij}$ coefficients can be found in Ref.~\cite{Waalewijn:2012sv}. At Leading Order (LO), ${\cal J}_{ij}^{(0)}(p_{T_J},R,z,\mu)=\delta_{ij}\delta(1-z)$ and the average jet charge in Eq.~(\ref{eq:avgjetchargeFF}) becomes

\begin{align}
\langle Q_\kappa^{i(0)} \rangle = \sum_{h\in {i\text{-}\rm jet}}Q_h\> \tilde{D}_i^h\left (\kappa,\mu\right ) + {\cal O}(\alpha_S),
\label{eq:avgjetchargeFFLO}
\end{align}
where $\tilde{D}_i^h\left (\kappa,\mu\right )$ is the Mellin moment of the fragmentation function $D_i^h\left (z,\mu\right )$
\begin{align}
\tilde{D}_i^h\left (\kappa,\mu\right ) = \int_0^1 dz\>z^\kappa \>D_i^h\left (z,\mu\right ),
\label{eq:MellinMomentFF}
\end{align}
with a multiplicative renormalization group evolution equation
\begin{align}
\mu \frac{d}{d\mu}\tilde{D}_i^h\left (\kappa,\mu\right ) = \frac{\alpha_s}{\pi} \tilde{P}_{ij}(\kappa) \>\tilde{D}_j^h\left (\kappa,\mu\right ),
\label{eq:MellinSpaceRGE}
\end{align}
where $\tilde{P}_{ij}(\kappa)$ are the standard splitting functions in Mellin space. 

For simulation results, we use Pythia8 (Pythia 8.240)~\cite{Sj_strand_2015} for event generation and FASTJET 3.3.2~\cite{Cacciari_2006,Cacciari_2012} for implementing jet algorithms and applying soft drop grooming~\cite{Larkoski_2014,frye2016precision,Frye_2016}. Jets are defined using the anti-k$_T$~\cite{Cacciari_2008} jet algorithm throughout the manuscript. All Pythia simulation results presented for $pp$-collisions and PbPb-collisions include underlying events corresponding to the Multi-Parton Interaction (MPI) switch being turned on, unless specified otherwise.

For calibration purposes, we perform Pythia simulations to compare with the ATLAS analysis~\cite{Aad:2015cua} for $pp$-collisions at $\sqrt{s} = 8$ TeV.  The two leading jets, $j_1$ and $j_2$, denoting the leading and subleading jets, respectively, are subject to pseudorapidity and transverse momentum selection cuts of $|\eta_{j1,j2}| <2.1$ and  $p_{T_{j_1}}/p_{T_{j_2}}<1.5$, respectively. The jets are defined with jet radius of $R=0.4$. These leading jets are classified as either ``more central" or ``more forward", corresponding to the jet with a smaller or larger magnitude of pseudorapidity, respectively. 

In Fig.~\ref{fig:StdJCkappa}, we show the simulation results for the standard jet charge distribution with the Pythia parton-level  setting `HardQCD:all = on' and hadronization and underlying event (MPI) turned on. The three curves in each panel correspond to the values $\kappa=0.3, 0.5,$ and 0.7, respectively. The top (bottom) panel corresponds to jet charge distributions for the more forward (central) of the two leading jets, each restricted to the jet $p_T$-bin:   $[200\>{\rm GeV},  300\>{\rm GeV}]$. Following the ATLAS analysis~\cite{Aad:2015cua}, Fig.~\ref{fig:StdJCkappa} uses jet charge bins of size 0.1 in units of the proton electric charge $e$.

In Fig.~\ref{fig:avgJC}, we show a comparison of the Pythia simulation results with ATLAS data~\cite{Aad:2015cua} for the average jet charge as a function of the jet  $p_T$-bin. The top and bottom panels correspond to the more forward and more central jets, respectively. We see that there is good agreement between the Pythia simulation results and the ATLAS data.

\section{Dynamic Jet Charge}

The properties of the standard jet charge definition in Eq.~(\ref{eq:stdJC}) and its  ability to discriminate between quark and gluon initiated jets and between quark jets of different flavors have been extensively studied~\cite{Krohn:2012fg,Waalewijn:2012sv,Fraser:2018ieu}. The dynamic jet charge is defined in Eq.~(\ref{eq:dynamicJC}) and we focus on the scenario where the constant parameter $\kappa$ in Eq.~(\ref{eq:stdJC}) is generalized to a function of the momentum fraction $z_h$, for each hadron $h$ in the jet
\begin{align}
\label{eq:dynJC}
Q^i_{\rm dyn}=\sum_{h\in {i\text{-}\rm jet}} z_h^{\kappa (z_h)} \>Q_h.
\end{align}
Thus, each hadron $h$ contributes to the jet charge with a \textit{dynamic} $\kappa$-parameter whose value is determined by its jet momentum fraction $z_h$.  Different functional forms of $\kappa(z_h)$ correspond to different definitions of the dynamic jet charge.
\begin{figure}
        \includegraphics[scale=0.8]{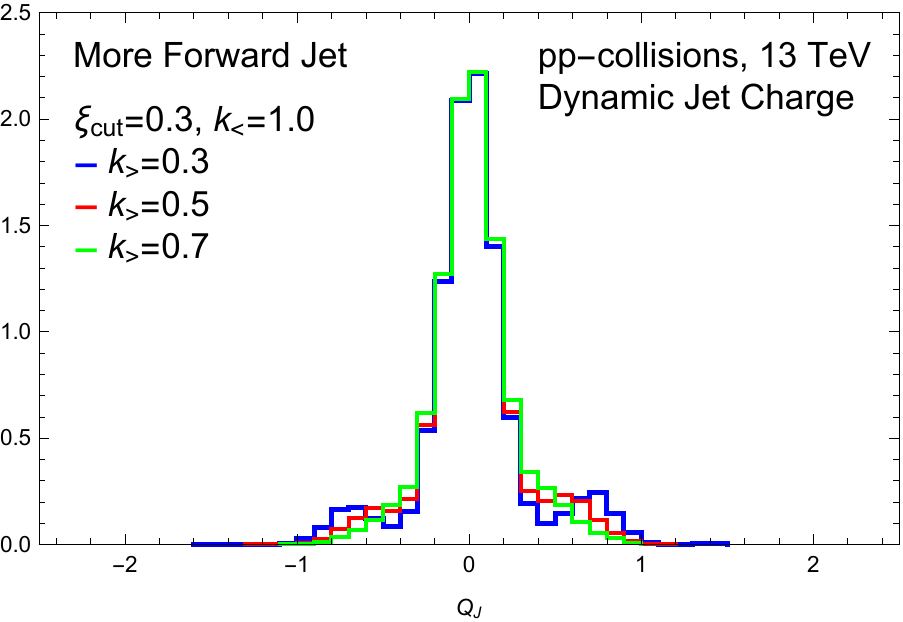}
        \includegraphics[scale=0.8]{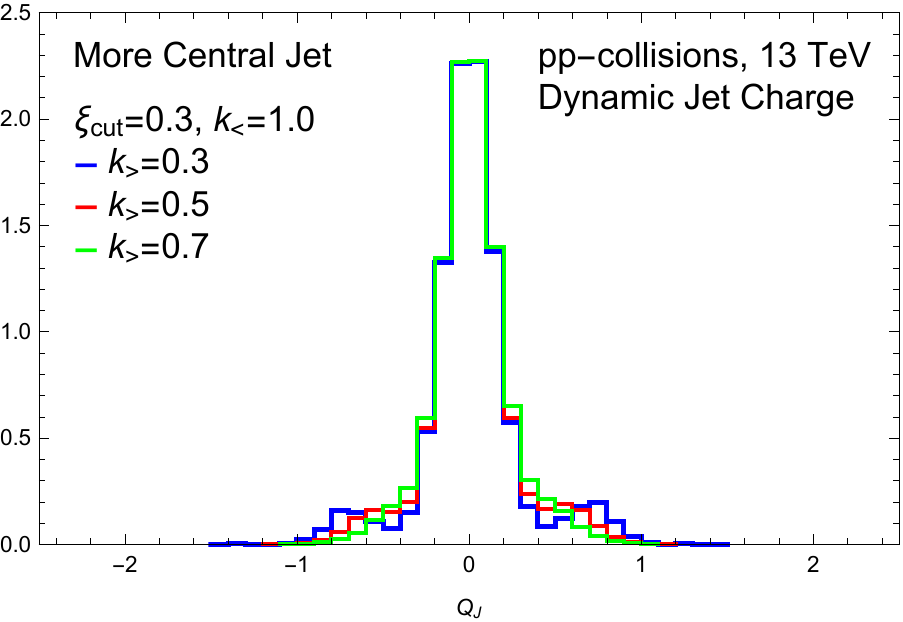}
         \caption{Normalized dynamic jet charge distributions for the more forward (top panel) and more central (bottom panel) jets from Pythia8 simulations of $pp\to j_1j_2 X$ at $\sqrt{s}=13 $ TeV with hadronization and MPI effects turned on. Selection cuts of 200 GeV $< p_{T_{j_1,j_2}}<$ 300 GeV, $|\eta_{j1,j2}| <2.1$, and $p_{T_{j_1}}/p_{T_{j_2}} < 1.5$ on the leading ($j_1$) and subleading ($j_2$) anti-k$_T$ jets of jet radius $R=0.4$ are applied. These distributions are for $\xi_{\rm cut}=0.3$, $k_<=1.0$ and the blue, red, and green curves correspond to $k_>=0.3, 0.5,$ and 0.7, respectively.}
    \label{fig:DynJCkgr}
\end{figure}

\begin{figure}
    \centering
    \includegraphics[scale=0.8]{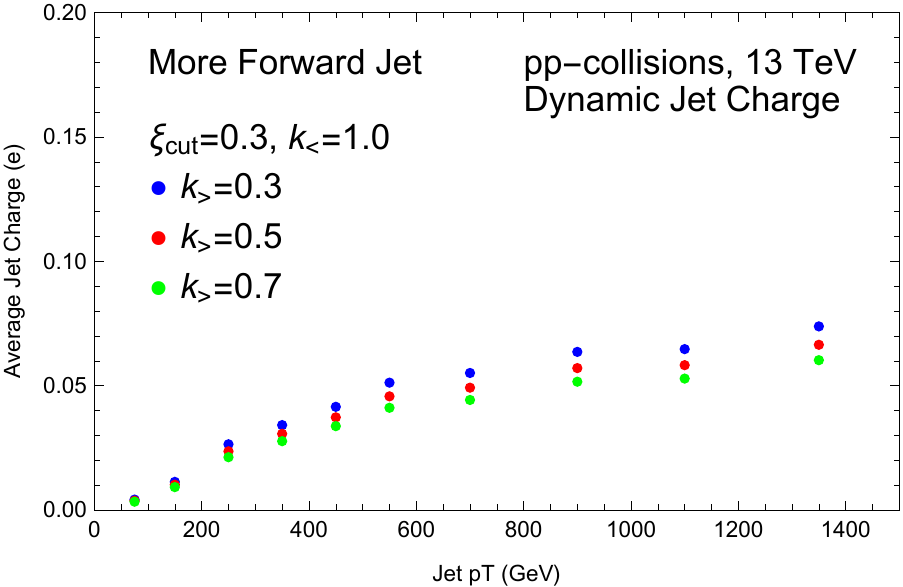}
    \includegraphics[scale=0.8]{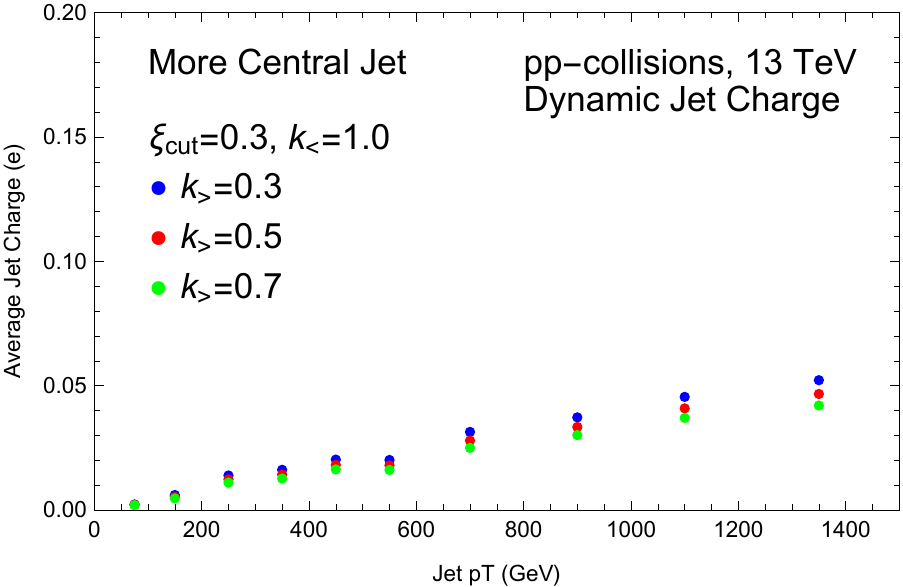}
\caption{The average dynamic jet charge from Pythia 8.240 simulations. The average dynamic jet charge is plotted as a function of the jet $p_{T_J}$ in $pp\to j_1j_2 X$ at $\sqrt{s}=13 $ TeV. The average dynamic jet charge of the more forward (top) and more central (bottom) of the two leading jets are shown for $\xi_{\rm cut}=0.3, k_<=1.0$ and $k_>=0.3,0.5,0.7$ corresponding to the blue, red, and green curves respectively.
Selection cuts of 200 GeV $< p_{T_{j_1,j_2}}<$ 300 GeV, $|\eta_{j1,j2}| <2.1$, and $p_{T_{j_1}}/p_{T_{j_2}} < 1.5$ on the leading ($j_1$) and subleading ($j_2$) anti-k$_T$ jets of jet radius $R=0.4$ are applied. The jet charge values correspond to an average over the jet $p_{T_J}$ bins: [50 GeV,100 GeV], [100 GeV, 200 GeV], [200 GeV, 300 GeV], [300 GeV, 400 GeV], [400 GeV, 500 GeV], [500 GeV, 600 GeV], [600 GeV, 800 GeV], [800 GeV, 1000 GeV], [1000 GeV, 1200 GeV], and [1200 GeV, 1500 GeV]}
    \label{fig:avgDynJC}
\end{figure}

This dynamic definition is motivated by the observed dependence of the shape of the standard jet charge distribution on the value of the constant parameter $\kappa$, as seen in Fig.~\ref{fig:StdJCkappa}. The standard jet charge distribution is characterized by a single peak structure that gets narrower  for increasing  values of the $\kappa$-parameter.  If the parameter $\kappa$ is generalized to a function of some dynamical property of individual hadrons in the jet, such as their momentum fraction $z_h$, then differences between quark and gluon jets in the distribution of this particle property will lead to differences in the average value $\langle \kappa\rangle_{\rm dyn}^i$\,,
\bea
\langle \kappa\rangle_{\rm dyn}^i =  \int_0^1 dz\> \kappa(z) \sum_{h\in {i\text{-}\rm jet}}  \frac{1}{\sigma_{i\text{-}{\rm jet}}} \frac{d\sigma_{h\in {i-\rm jet}}}{dz}\,,
\eea
for a jet initiated by a quark ($i=q$) or a gluon ($i=g$). Thus, such a dynamic parameter $\kappa(z_h)$ could give rise to enhanced differences in the jet charge distributions of quark and gluon jets. 
For example, gluon jets are typically characterized by  a higher multiplicity of hadrons compared to quark jets due to the larger color factor of the gluon. As a result, for a jet of given energy the average value of the momentum fraction $\langle z_h \rangle$  will be larger for quark jets compared to gluon jets. Correspondingly, the average value of the dynamic parameter $\langle \kappa  \rangle_{\rm dyn}^i$ will be different for quark and gluon jets, leading to enhanced differences in the dynamic jet charge distributions.

The theoretical prediction for the average dynamic jet charge $\langle Q^{i}_{\rm dyn} \rangle$ is given by
replacing the constant parameter $\kappa$ in Eq.~(\ref{eq:avgjetchargeFF}) by the dynamic function $\kappa(z)$
\begin{align}
\langle Q^i_{\rm dyn} \rangle = \int_0^1 dz\> z^{\kappa(z)} \sum_{h\in {i\text{-}\rm jet}}Q_h\sum_j  &\int _z^1 \frac{dz'}{z'} {\cal J}_{ij}(p_{T_J},R,z',\mu) \nn \\ 
&\times D_j^h\left (\frac{z}{z'},\mu\right ).
\label{eq:avgdynjetchargeFF}
\end{align}
Once again, using the LO result ${\cal J}_{ij}^{(0)}(p_{T_J},R,z,\mu)=\delta_{ij}\delta(1-z)$, the corresponding  LO  expression for the dynamic jet charge is given by
\begin{align}
\label{eq:LOdynjetcharge}
\langle Q^{i(0)}_{\rm dyn} \rangle &= \int_0^1 dz\> z^{\kappa(z)}\sum_{h\in {i\text{-}\rm jet}}Q_h\> D_i^h\left (z,\mu\right ) +{\cal O}(\alpha_s).
\end{align}
Note that due to the fact that $\kappa(z)$ is not a constant, the LO dynamic jet charge does not depend on simple  Mellin moments of the fragmentation function, as seen in Eqs.~(\ref{eq:avgjetchargeFFLO}) and (\ref{eq:MellinMomentFF}) for the standard jet charge. However, the standard
DGLAP evolution of the fragmentation function in Eq.~(\ref{eq:LOdynjetcharge}) can still be done directly in $z$-space.

The properties of the dynamic jet charge can be explored for different functional forms of $\kappa(z_h)$, corresponding to different definitions. For simplicity, in this work we restrict our analysis to the simple functional form $\kappa(z_h) = \kappa (z_h,\xi_{\rm cut}, k_<, k_>)$
\begin{equation}
  \kappa (z_h) =\begin{cases}  
  k_<, & z_h < \xi_{\rm cut} \\
  k_> , & z_h \geq \xi_{\rm cut} \\
  \end{cases}
\label{eq:dynfuncform}  
\end{equation}
where $\xi_{\rm cut},~ k_<$, and $k_>$ are three constant parameters that are part of the definition of the dynamic jet charge. We choose the default values to be $\xi_{\rm cut}=0.3, \>k_< =1.0, \>k_>=0.3$, except when we explicitly vary these parameters to study their impact on the dynamic jet charge distributions. Using the functional form in Eq.~(\ref{eq:dynfuncform}), the dynamic jet charge definition in Eq.~(\ref{eq:dynJC}) becomes
\begin{align}
\label{eq:dynJCfuncform}
Q^i_{\rm dyn}&=\sum_{h\in {i\text{-}\rm jet}} \Theta(\xi_{\rm cut} -z_h) \>z_h^{k_<} \>Q_h \nn \\
&+ \sum_{h\in {i\text{-}\rm jet}} \Theta(z_h-\xi_{\rm cut} ) \>z_h^{k_>} \>Q_h,
\end{align}
in terms of the $\xi_{\rm cut}, k_<,$ and $k_>$ parameters. Note that for $\xi_{\rm cut}=0$ or $\xi_{\rm cut}=1$, 
the dynamic jet charge reduces to the standard jet charge with $\kappa=k_>$ or  $\kappa=k_<$, respectively.

Comparing Eqs.~(\ref{eq:dynJCfuncform}) and (\ref{eq:stdJC}), we see that this definition is similar to the standard jet charge but with the modification that the low momentum hadrons ($z_h<\xi_{\rm cut}$) contribute with $\kappa=k_<$ and the high momentum hadrons ($z_h>\xi_{\rm cut}$) contribute with $\kappa=k_>$. For the default parameter choices $\xi_{\rm cut}=0.3, k_< =1.0,$ and $k_>=0.3$, the contribution of the low momentum hadrons relative to that of the high momentum hadrons is much more suppressed compared to that in the case of the standard jet charge. This leads to qualitatively distinct features for the dynamic jet charge. In Fig.~\ref{fig:DynJCkgr}, we show Pythia8 simulation results for  the more forward (top panel) and more central (bottom panel) leading jets in $pp$-collisions at $\sqrt{s}=13$ TeV. We see that the distribution  with default parameter choices (blue) has a central peak and two smaller speaks on either side.  The non-central peaks get smaller for increasing values of $k_>$.

The behavior of the peak structure in the dynamic jet charge distribution can be summarized as follows. The non-central peaks become more prominent when the high momentum hadrons ($z_h > \xi_{\rm cut}$) are given a higher weight (decreasing $k_>$) relative to the low momentum hadrons ($z_h < \xi_{\rm cut}$). The low momentum hadrons  tend to smear out the non-central peaks as seen in the distributions with the larger values of $k_>=0.5$ (red) and $k_>=0.7$ (green), corresponding to giving their contribution a larger weight. 

The theoretical prediction for the LO average dynamic jet charge in Eq.~(\ref{eq:LOdynjetcharge}), for the functional form in Eq.~(\ref{eq:dynfuncform}), is given by
\begin{align}
\langle Q^{i(0)}_{\rm dyn} \rangle &= \int_0^{\xi_{\rm cut}} dz\> z^{k_>}\sum_{h\in {i\text{-}\rm jet}}Q_h\> D_i^h\left (z,\mu\right ) \nn \\
&+ \int^1_{\xi_{\rm cut}} dz\> z^{k_<}\sum_{h\in {i\text{-}\rm jet}}Q_h\> D_i^h\left (z,\mu\right ) +{\cal O}(\alpha_S).
\label{eq:LOdynjetchargefuncform}
\end{align}
Analogous to Fig.~\ref{fig:avgJC},  we show Pythia simulation results for the average dynamic jet charge as a function of the jet $p_T$-bin in $pp$-collisions at $\sqrt{s}=13$ TeV in Fig.~\ref{fig:avgDynJC}.

\section{Quark-Gluon Discrimination}

In this section, we explore the use of the dynamic jet charge observable to discriminate between quark and gluon initiated jets.  We consider dijet events in $pp$-collisions at $\sqrt{s}=13$ TeV and heavy ion collisions (Pb-Pb) at $\sqrt{s}=2.76$ TeV within Pythia. Note that Pythia uses the so-called Angantyr model~\cite{Bierlich:2016smv,Bierlich:2018xfw} for heavy ion collisions, whose main idea is to stack parton level events, corresponding to individual nucleon-nucleon sub-collisions, on top of each other and hadronize them together. The model is able to give a good description of general final-state properties such as multiplicity and transverse momentum distributions in heavy ion collisions. However, it cannot describe any observables sensitive to collective behaviour, because the model generates events without assuming production of the thermalized medium after the collision, and thus the result is a quark-gluon-plasma-free simulation of heavy ion collisions. 
%\omit{We plan to study dynamics jet charge in the existence of the hot medium in the future utilizing %JEWEL~\cite{Zapp:2013vla} or JETSCAPE~\cite{Putschke:2019yrg} event generators.} 
Nevertheless, such Angantyr/Pythia simulations account for the increased underlying event activity in heavy ion collisions, and are still very useful to test the robustness of the dynamic jet charge as a first step. 

The jets are found using the anti-k$_T$~\cite{Cacciari_2008} jet algorithm with a jet radius, $R=0.4$. The two leading jets are subject to pseudorapidity cuts $|\eta_{j_1,j_2}| <2.1$ and $|\eta_{j_1,j_2}| <0.9$ for $pp$-collisions and PbPb-collisions,  
 respectively. They are also restricted to the $p_T$-bins $200\> {\rm GeV}< p_{T_{j_1,j_2}} < 300\>{\rm GeV}$ and $80\> \>{\rm GeV}< p_{T_{j_1,j_2}} < 150\>{\rm GeV}$ for $pp$-collisions and PbPb-collisions, respectively. Finally, the two leading jets are subject to the additional constraint $p_{T_{j_1}}/p_{T_{j_2}}<1.5$, following the ATLAS analysis~\cite{Aad:2015cua}.  All results are shown for the more central jet. Similar results are found for the more forward jet charge distributions and do not affect the discussion here.

\begin{figure}
    \centering
    \includegraphics[scale=0.8]{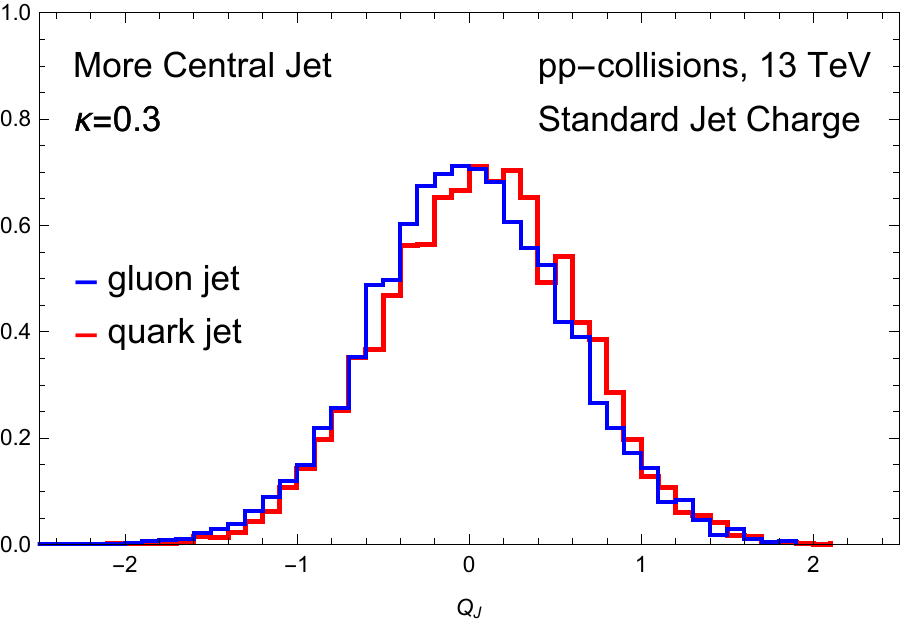}
    \includegraphics[scale=0.8]{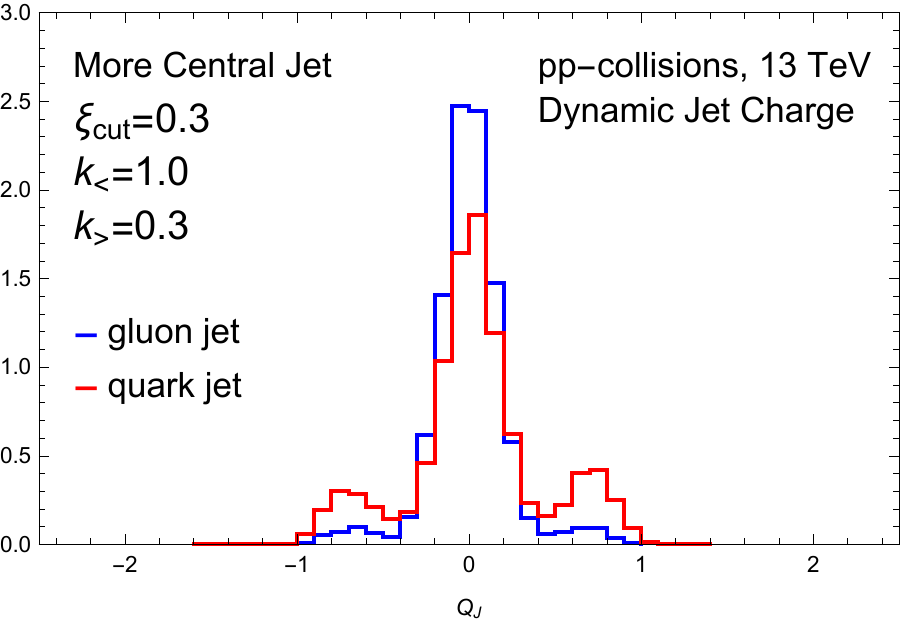}
    \includegraphics[scale=0.8]{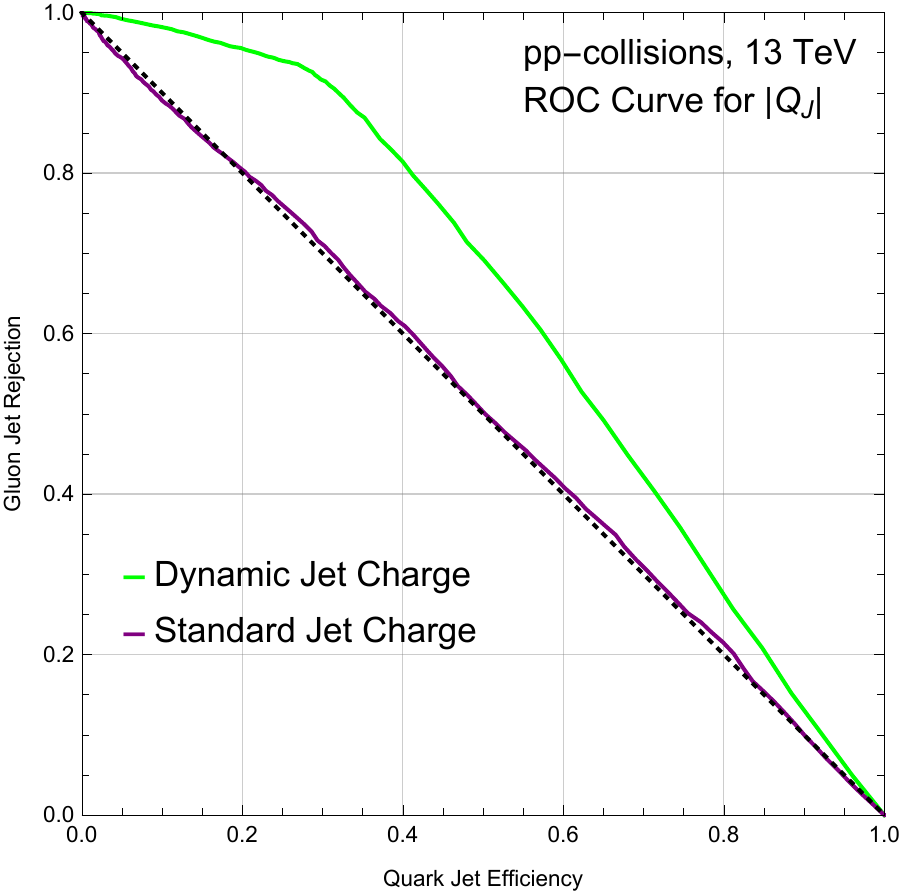}
    \caption{The standard (top) and dynamic (middle) jet charge distributions in $pp$-collisions at $\sqrt{s}=$13 TeV, for quark (red) and gluon (blue) initiated jets with selection cuts $p_{T_J} = [200,300]$ GeV, $|\eta_J| < 2.1$. A comparison of the corresponding  quark-gluon discrimination ROC curves for the standard (purple) and dynamic (green) jet charge distributions is given in the bottom panel.}
    \label{fig:StdvsDyn}
\end{figure}
\begin{figure}
    \centering
    \includegraphics[scale=0.8]{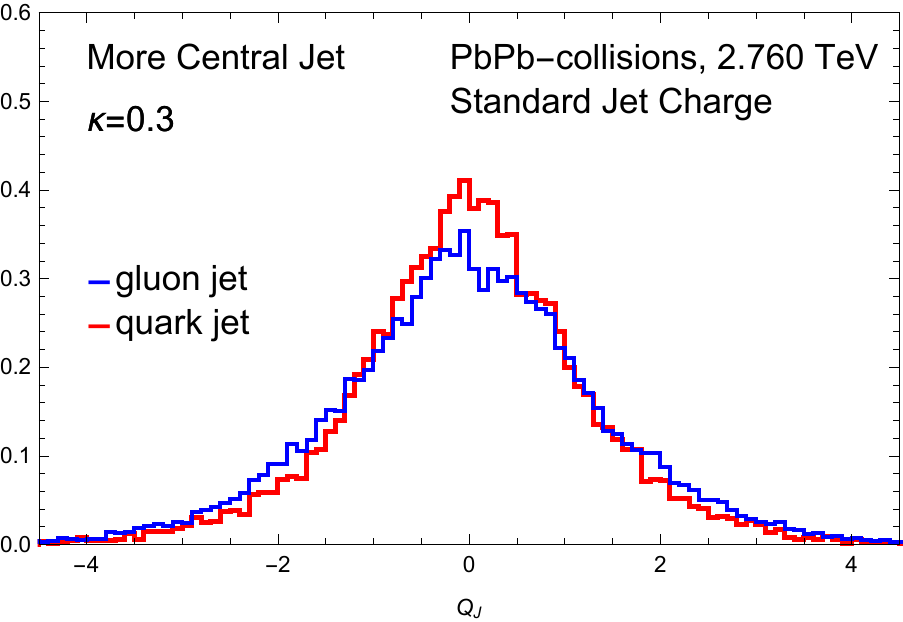}
    \includegraphics[scale=0.8]{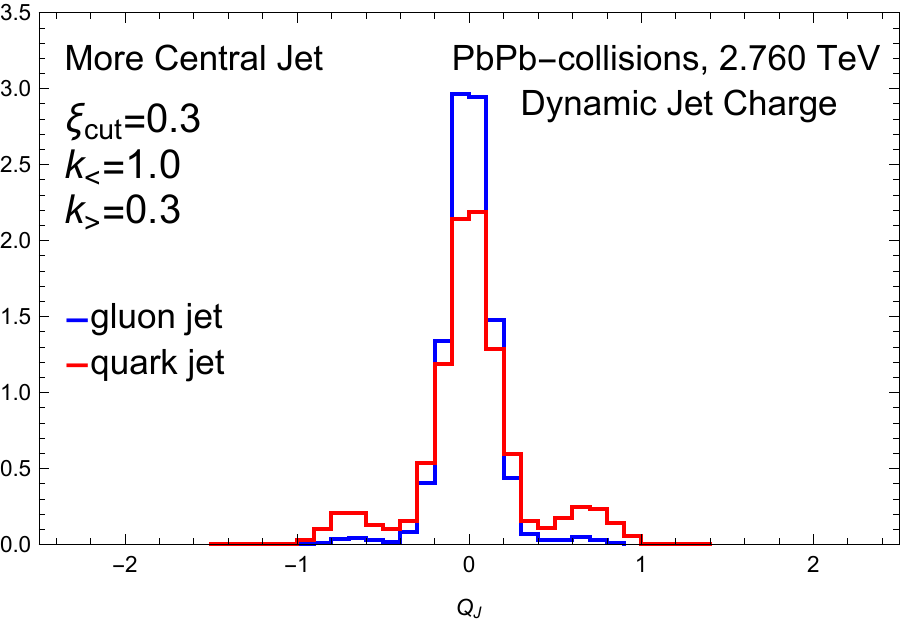}
    \includegraphics[scale=0.8]{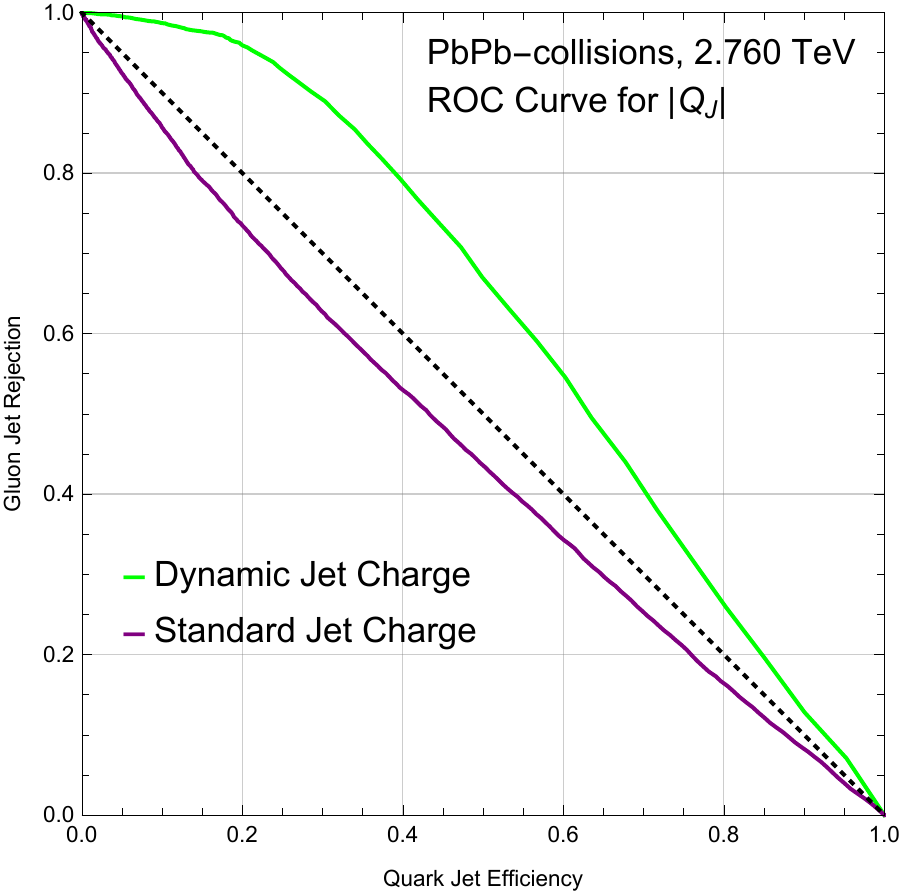}
    \caption{The standard (top) and dynamic (middle) jet charge distributions in Heavy Ion (Pb-Pb) collisions at $\sqrt{s}=$2.760 TeV, for quark (red) and gluon (blue) initiated jets with selection cuts $p_{T_J} = [80,150]$ GeV, $|\eta_J| < 0.9$. A comparison of the corresponding  quark-gluon discrimination ROC curves for the standard (purple) and dynamic (green) jet charge distributions is given in the bottom panel.}
    \label{fig:StdvsDynHeavyIon}
\end{figure}

 We compare jet charge distributions between quark-initiated  and gluon-initiated  jets, obtained through simulations of dijet events generated by the Pythia8 partonic channel settings `HardQCD:qq2qq=on' and `HardQCD:gg2gg=on', respectively.
In Fig.~\ref{fig:StdvsDyn}, we show simulation results of the normalized jet charge distributions for the standard (top panel) and dynamic (middle panel) jet charge definitions for quark (red) and gluon (blue) initiated jets in $pp$-collisions. We see that while the standard jet charge distributions are very similar for quark and gluon jets, there are distinct qualitative differences between their dynamic jet charge distributions.  Compared to the gluon jets, the quark jets have a relatively smaller peak height in the central region near zero jet charge and have relatively larger peak heights in the region of non-zero jet charge. This is found to be a consistent feature when evaluated over a wide range of kinematics and phase space selection cuts.

These qualitative differences between quark and gluon jets can be quantified by the Receiver Operating Characteristic (ROC) curve as shown in the bottom panel of Fig.~\ref{fig:StdvsDyn}. The ROC curves show the background (gluon jet) rejection  as a function of the signal (quark jet) efficiency, based on a cut on the absolute value of the jet charge $|Q_J|$. The background rejection is given by the fraction of gluon jets rejected and the signal efficiency corresponds to the fraction of quark jets kept, as a function of the cut on $|Q_J|$. The reference diagonal line (dashed black) corresponds to an ROC curve that shows no discrimination between the signal and background. ROC curves that lie above the diagonal correspond to non-zero discrimination power of signal over background. ROC curves that lie below the diagonal dashed line also indicate non-zero discrimination power and can be made manifest by interchanging what we define as the signal and background. We see that the ROC curve for the dynamic jet charge (green) shows  significantly improved discrimination between quark and gluon jets compared to the ROC curve (purple) for the standard jet charge.

In Fig.~\ref{fig:StdvsDynHeavyIon}, we show the same plots as in Fig.~\ref{fig:StdvsDyn} but for the heavy ion PbPb-collisions. We see that the standard jet charge distribution (top panel) is much broader compared to that in $pp$-collisions. On the other hand, the dynamic jet charge distribution (middle panel) is quite similar in PbPb-collisions and $pp$-collisions. The same distinguishing qualitative features between quark and gluon jets appear even for heavy ion collisions.  Correspondingly, the ROC curve (bottom panel) for the dynamic jet charge in heavy ion collisions looks similar to that in $pp$-collisions and gives better discrimination compared to the standard jet charge.  Thus, the quark-gluon discrimination power of the dynamic jet charge is largely unaffected by the significantly greater underlying event activity in heavy ion collisions. 
\begin{figure}
    \centering
    \includegraphics[scale=0.7]{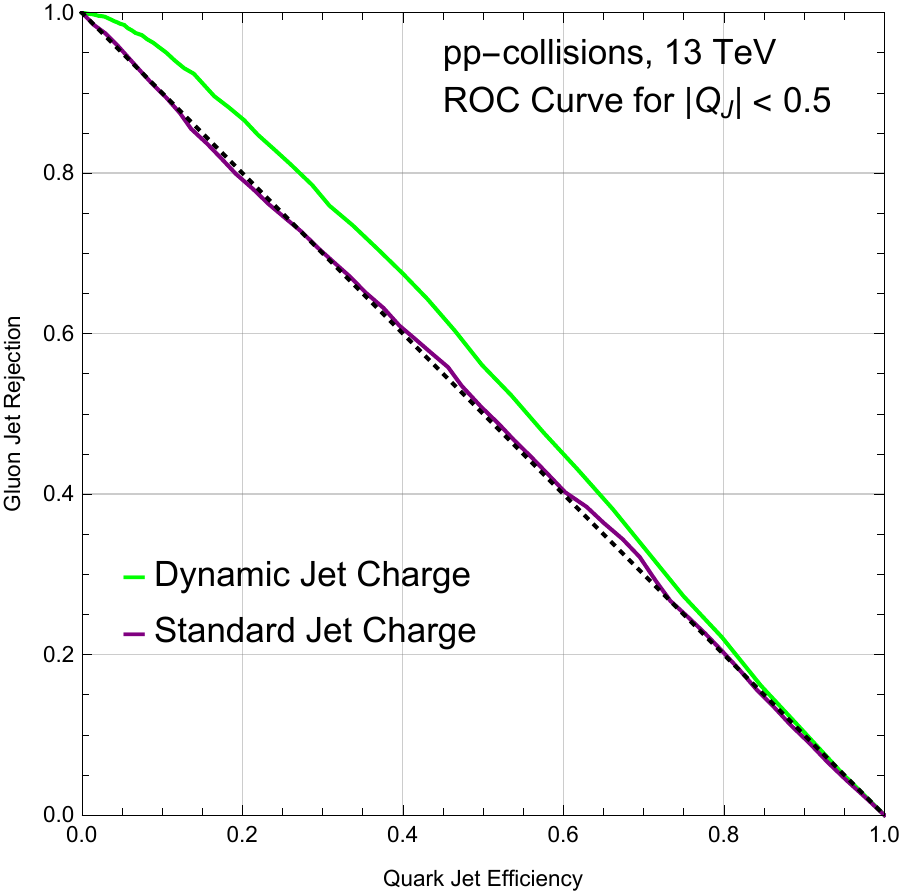}
    \includegraphics[scale=0.7]{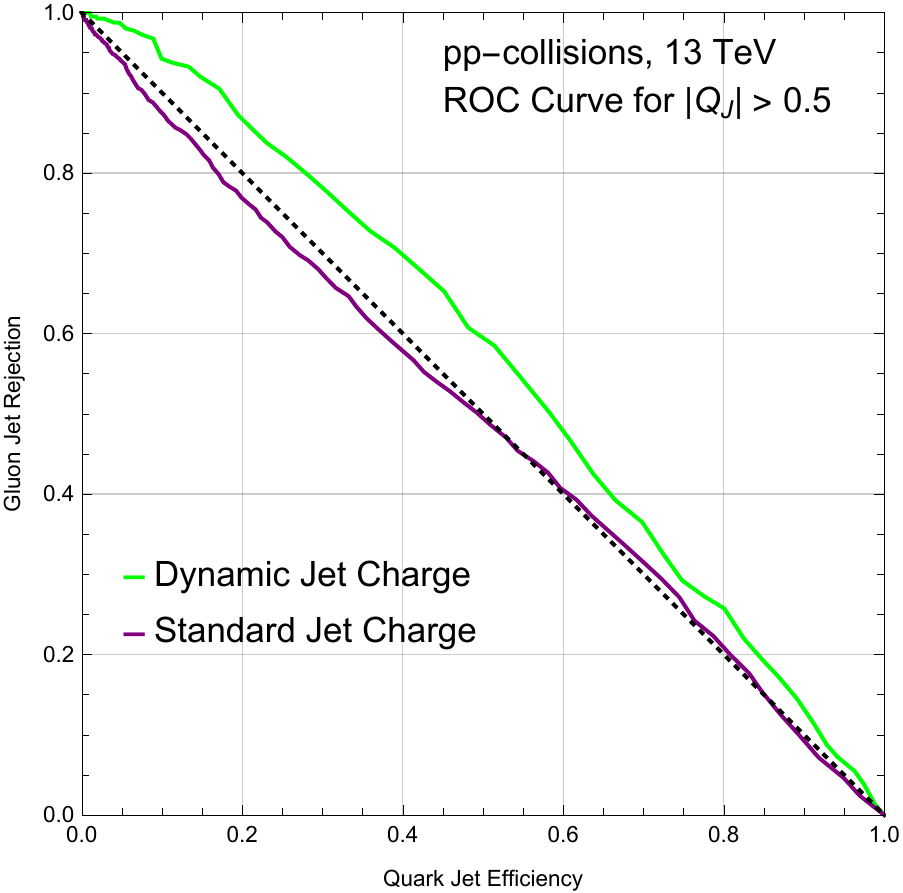}
    \caption{The local ROC curves for the standard (purple) and dynamic (green) jet charge distributions described in Fig.~\ref{fig:StdvsDyn} for for $pp$-collisions at $\sqrt{s}=$13 TeV, with selection cuts $p_{T_J} = [200,300]$ GeV, $|\eta_J| < 2.1$. The top and bottom panels correspond to the local ROC curves for the jet charge bins $|Q_J|<0.5$ and $|Q_J|>0.5$, respectively.  }
    \label{fig:binnedROCqvsg}
\end{figure}

We note that in PbPb-collisions the ROC curve for the standard jet charge (purple) in Fig.~\ref{fig:StdvsDynHeavyIon} is below the dashed diagonal line,  indicating slightly better discrimination (if one interchanges what is defined as signal and background) than in the case of $pp$-collisions. This can be understood as the result of the standard jet charge distribution for gluon jets being slightly broader than that for quark jets in PbPb-collisions, as seen in the top panel  of Fig.~\ref{fig:StdvsDynHeavyIon}.  

While the ROC curve provides one quantitative measure of the discrimination power of the dynamic jet charge, it may not fully capture it due to the multiple peak structure of the dynamic jet charge distributions.
The ROC curve best describes the discrimination power when comparing two distributions with a single peak structure and quantifies how well-separated  the peaks of the two distributions are. For distributions with multiple peaks, other ways of quantifying the discrimination power can be useful.

Due to the multiple peak structure of the dynamic jet charge distribution, additional quark-gluon discrimination is possible through an analysis of the jet data binned according to the dynamic jet charge value. In particular, the multiple peak structure naturally suggests binning the jet charge data such that the bins are centered around the peaks with bin-widths corresponding to the width of the peaks. A discrimination analysis can then be separately performed in each jet charge bin. For example, in Fig.~\ref{fig:StdvsDyn}, the dynamic jet charge distribution (middle panel) suggests the jet charge data be divided into two bins: $|Q_J| \leq 0.5$ and  $|Q_J| > 0.5$. In Fig.~\ref{fig:binnedROCqvsg}, we show the \textit{local} ROC curves for the $|Q_J| \leq 0.5$ bin (top panel) and the $|Q_J| > 0.5$ bin (bottom panel) for the standard (purple) and dynamic (green) jet charge distributions. We define the local ROC curve in each jet charge bin using only the events in that bin. This can be distinguished from the  \textit{global} ROC curve which is defined using the data in all bins, as was done Figs.~\ref{fig:StdvsDyn} and \ref{fig:StdvsDynHeavyIon}. The local ROC curve affords the standard interpretation for the dynamic jet charge since  each jet charge bin is selected to have at most a single peak.  We see from Fig.~\ref{fig:binnedROCqvsg} that some additional discrimination is possible using the dynamic jet charge in a binned analysis. While the ROC curve for the dynamic jet charge shows improved discrimination between quark and gluon jets, the improvement is only marginal. This can be understood as a result of the fact that in each bin, while there are significant differences between the peak heights for the quark and gluon dynamic jet charge distributions, the location of the peaks are about the same and ROC curves tend to probe differences in the peak positions. However, we will show later on that such a binned analysis with  local ROC curves can provide significant discrimination between $u$-quark and $d$-quark jets.

Noticing that the dynamic jet charge distribution characterizes the gluon (quark) jets with a higher (lower) central peak and  lower (higher) non-central peaks, one can quantify this difference through fractional counts in each bin. We introduce the definitions
\begin{align}
\label{eq:fracqvsg}
\epsilon_< = \frac{N( \>|Q_J|  \leq 0.5 \>)}{N_{\rm total}}, \qquad \epsilon_>= \frac{N( \>|Q_J|  > 0.5 \>)}{N_{\rm total}},
\end{align}
where $\epsilon_<$ and $\epsilon_>$ give the fraction of events in the $|Q_J|  \leq 0.5 $ and $|Q_J| > 0.5 $ jet charge bins, respectively. In Table~\ref{table:quarkvsgluon}, we show the values of $\epsilon_< $ and $\epsilon_>$ for quark and gluon jets  for  the standard and dynamic jet charge distributions shown in Figs.~\ref{fig:StdvsDyn} and \ref{fig:StdvsDynHeavyIon} for $pp$-collisions and heavy ion collisions, respectively.

%%%%%%%%%%%%%
\begin{table}[t!]
	\begin{center}
		\scalebox{1}{
			\begin{tabular}{|l c | c| c|}%| c}
				\hline \hline
				&Jet Charge $\>\>\>$ & & 
				\\
				&($pp$-collisions) $\>\>\>$& $\>\>\>\>\>\epsilon_<\>\>\>\>\>$ & $\>\>\>\>\>\epsilon_>\>\>\>\>\>$  
				\\[0.5ex] \hline\hline
				& Standard: quark  & 0.62   & 0.38   \\
				& $\>\>\>\>\>\>\>\>\>\>\>\>\>\>\>\>\>\>\>$ gluon & 0.63  & 0.37 
				\\[0.5ex] \hline
				& Dynamic: quark & 0.75 & 0.25  \\
				& $\>\>\>\>\>\>\>\>\>\>\>\>\>\>\>\>\>\>\>$ gluon   & 0.94 &0.06
				\\ \hline\hline
		\end{tabular}}
		%%%%%%%%%%%%%%%%%%%%%%%%%%%%%%%%
	\end{center}
		\begin{center}
		\scalebox{1}{
			\begin{tabular}{|l c | c| c| }%| c}
				\hline \hline
				&Jet Charge $\>\>\>$& &    
				\\
				&(PbPb-collisions) $\>\>\>$& $\>\>\>\>\>\epsilon_<\>\>\>\>\>$ & $\>\>\>\>\>\epsilon_>\>\>\>\>\>$ 
				\\[0.5ex] \hline\hline
				& Standard: quark  & 0.37   & 0.63  \\
				& $\>\>\>\>\>\>\>\>\>\>\>\>\>\>\>\>\>\>\>$ gluon & 0.32  & 0.68 
				\\[0.5ex] \hline
				& Dynamic: quark & 0.85 & 0.15 \\
				& $\>\>\>\>\>\>\>\>\>\>\>\>\>\>\>\>\>\>\>$ gluon   & 0.98 & 0.02
				\\ \hline\hline
		\end{tabular}}
	\end{center}

	\caption{Fraction of events for quark and gluon jets in the $|Q_J|  \leq 0.5 $  and $|Q_J| > 0.5 $ jet charge bins. The definitions of $\epsilon_< $ and $\epsilon_> $ are given in Eq.~(\ref{eq:fracqvsg}). The numbers correspond to the distributions in Fig.~\ref{fig:StdvsDyn} and \ref{fig:StdvsDynHeavyIon} for $pp$-collisions and PbPb-collisions, respectively. }
	\label{table:quarkvsgluon}
\end{table}
%%%%%%%%%%%%

We see that for the dynamic jet charge in $pp$-collisions only $\sim 6\%$ of gluon jets are found in the $|Q_J| > 0.5$ jet charge bin compared to $\sim 25\%$ for quark jets. On the other hand, $\sim 94\%$ of gluon jets and $\sim 75\%$ quark jets are found in the $|Q_J| \leq 0.5$ bin. This can be contrasted with the standard jet charge where almost the same fraction of quark and gluon jets are present in each jet charge bin. Similarly, as seen in Table~\ref{table:quarkvsgluon}, the same overall behavior is observed for PbPb-collisions where the fraction of quark and gluon jets is about the same in each bin for the standard jet charge but different for the dynamic jet charge. Thus, using the differences in the expected fraction of quark and gluon jets in each jet charge bin, the dynamic jet charge can allow for improved quark and gluon discrimination simply by sorting the jet data into jet charge bins.

\section{Quark Flavor Discrimination}

In this section, we explore the use of the dynamic jet charge to discriminate between $u$- and $d$-quark jets in Pythia8 simulations of $pp$-collisions at $\sqrt{s}=13$ TeV  and heavy ion collisions (Pb-Pb) at $\sqrt{s}=2.76$ TeV. The $u$-quark and $d$-quark jet samples were generated using the partonic channels $dg\to W^-u$ and $ug\to W^+ d$, respectively. Selection cuts of $|\eta_J| <2.1$ and $|\eta_J| <0.9$, and $200\>{\rm GeV} < p_{T_J} < 300 \>{\rm GeV}$ and  $80\>{\rm GeV} < p_{T_J} < 150 \>{\rm GeV}$   were used for $pp$-collisions and heavy ion collisions. The leading anti-k$_T$ jet radius was set to $R=0.4$. 

\begin{figure}
    \centering
    \includegraphics[scale=0.8]{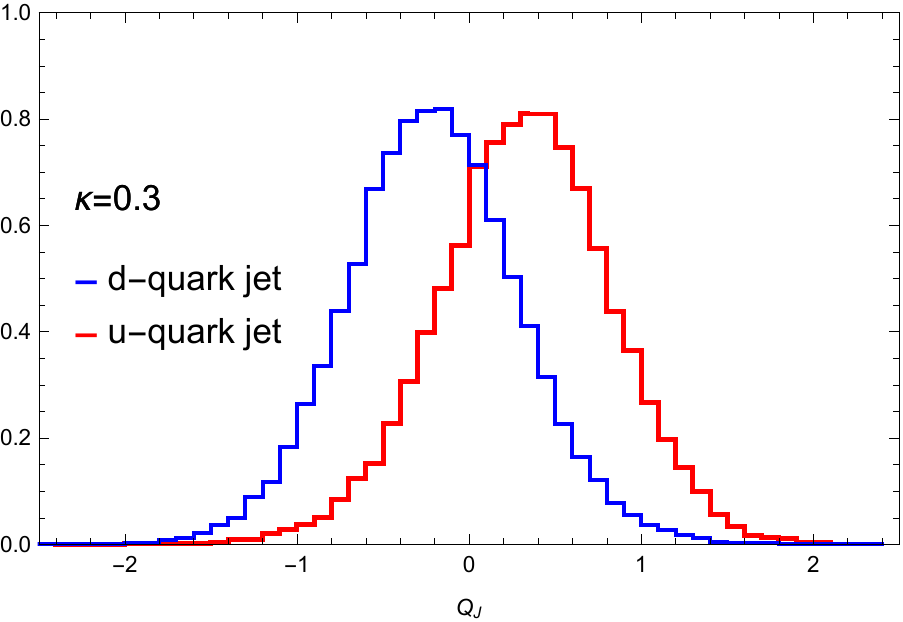}
    \includegraphics[scale=0.8]{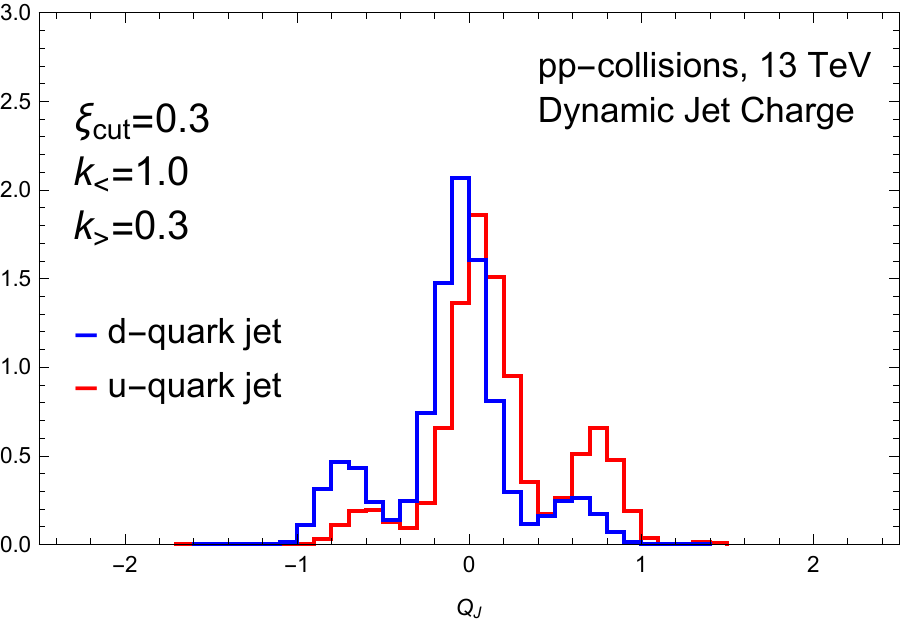}
    \includegraphics[scale=0.7]{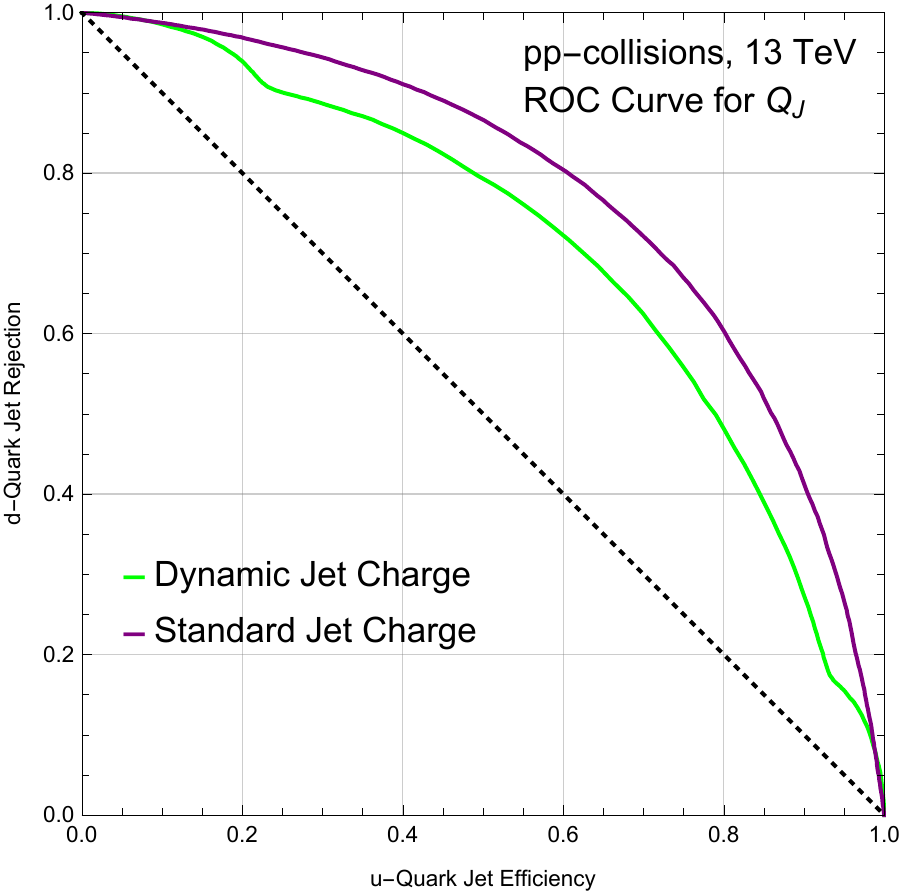}
    \caption{The standard (top) and dynamic (middle) jet charge distributions in $pp$-collisions at 13 TeV, for $u$-quark (red) and $d$-quark (blue) initiated jets with selection cuts $p_{T_J} = [200,300]$ GeV, $|\eta_J| < 2.1$. The channels $dg\to W^-u$ and $ug\to W^+ d$ partonic channels where used to select the $u$-quark and $d$-quark initiated jets respectively. A comparison of the corresponding  ``$u$" vs. ``$d$" discrimination ROC curves for the standard (purple) and dynamic (green) jet charge distributions is given in the bottom panel.}
    \label{fig:StdvsDynuvsd}
\end{figure}

In Fig.~\ref{fig:StdvsDynuvsd}, we show the standard (top panel) and dynamic (middle panel) jet charge distributions for $u$-quark (red) and $d$-quark (blue) jets. Once again we see that the dynamic jet charge distributions have qualitatively distinct features compared to the corresponding standard jet charge distributions. 

\begin{figure}
    \centering
    \includegraphics[scale=0.8]{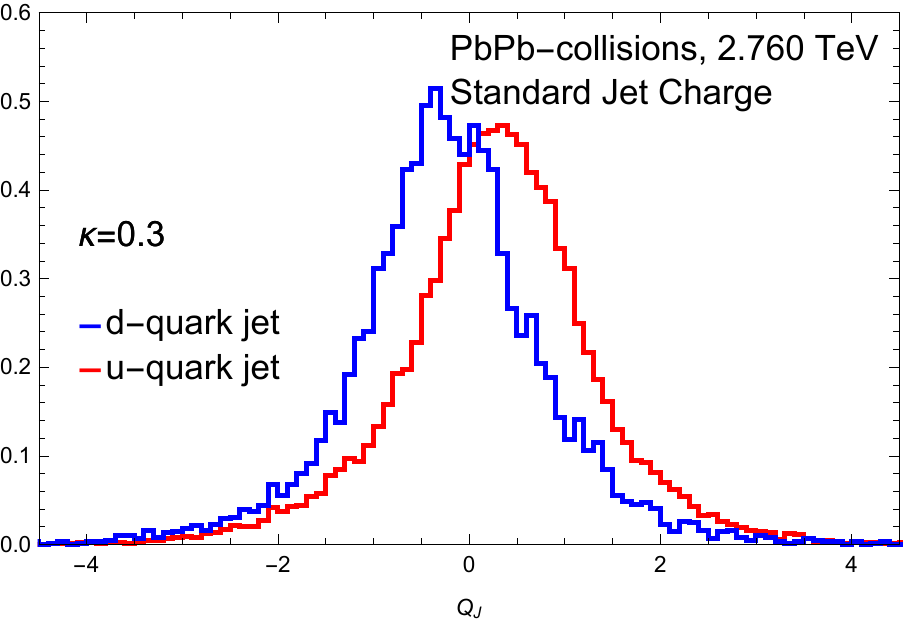}
    \includegraphics[scale=0.8]{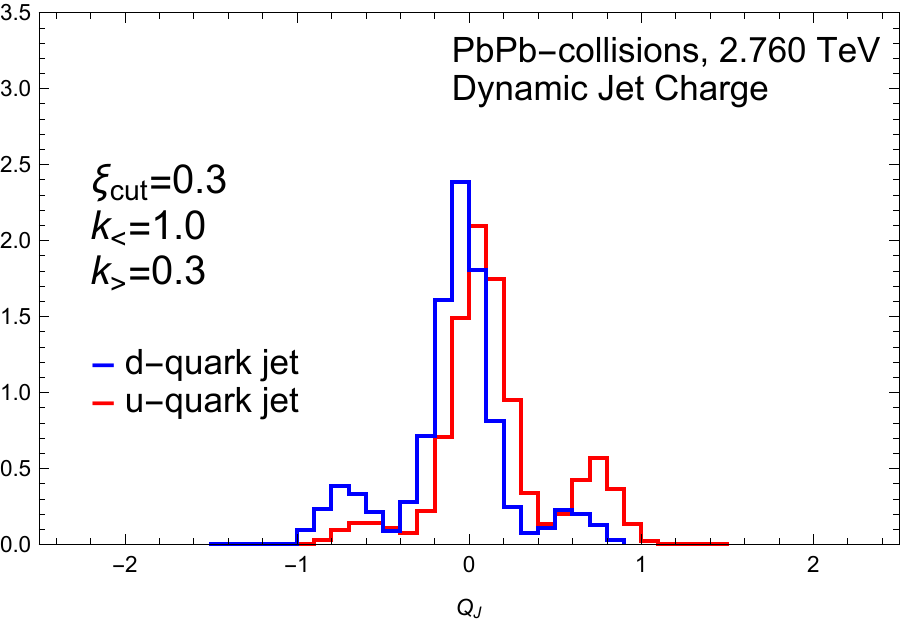}
    \includegraphics[scale=0.7]{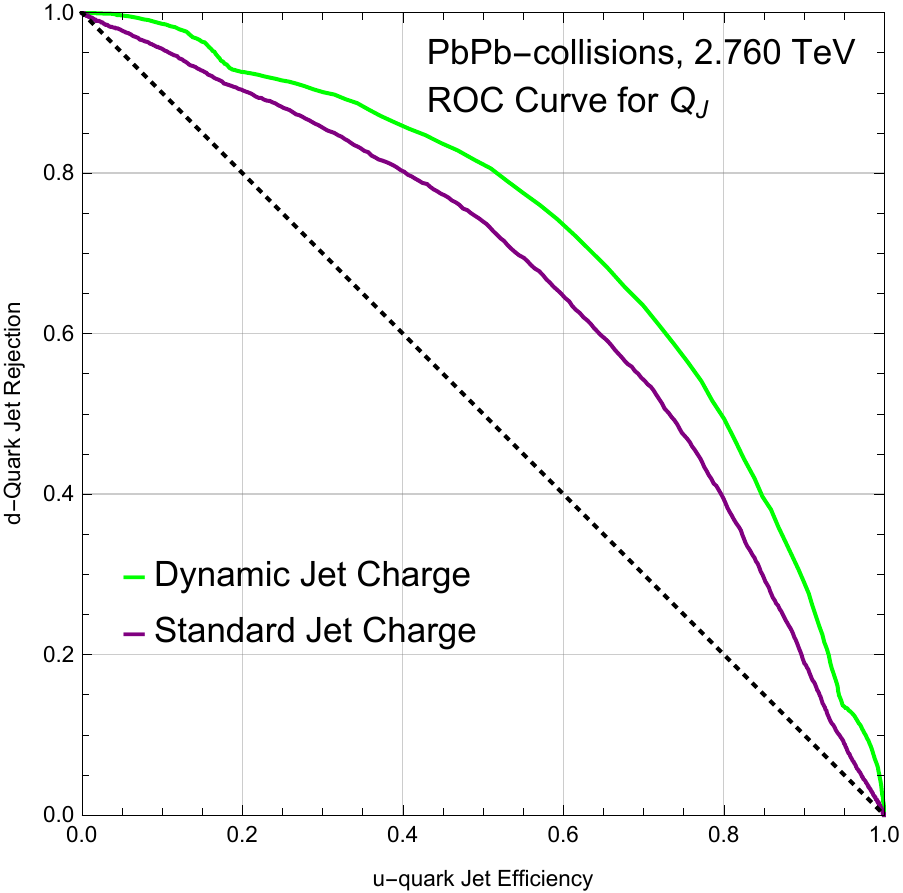}
    \caption{The standard (top) and dynamic (middle) jet charge distributions in PbPb-collisions at 2.760 TeV, for $u$-quark (red) and $d$-quark (blue) initiated jets with selection cuts $p_{T_J} = [80,150]$ GeV, $|\eta_J| < 0.9$. The channels $dg\to W^-u$ and $ug\to W^+ d$ partonic channels where used to select the $u$-quark and $d$-quark initiated jets respectively. A comparison of the corresponding  ``$u$" vs. ``$d$" discrimination ROC curves for the standard (purple) and dynamic (green) jet charge distributions is given in the bottom panel.}
    \label{fig:StdvsDynuvsdHI}
\end{figure}
 
The standard jet charge distribution for the $u$-quark ($d$-quark) jet has a single peak shifted to the right (left) of center, corresponding to the positive (negative) charge of the jet-initiating quark. The dynamic jet charge distributions are characterized by a multiple peak structure.  A shift of the central peak to the right (left) is also seen for the  $u$-quark ($d$-quark) jet but is less pronounced. However,  the secondary peak on the right (left) for the $u$-quark ($d$-quark) jet is much higher than that for the $d$-quark ($u$-quark) jet. In Fig.~\ref{fig:StdvsDynuvsd}, we also show the global ROC curves (bottom panel) for the standard and dynamic jet charge distributions. We see that in this case the global ROC curve for the standard jet charge shows better, although comparable, discrimination power compared to the dynamic jet charge. 
\begin{figure}
    \centering
    \includegraphics[scale=0.7]{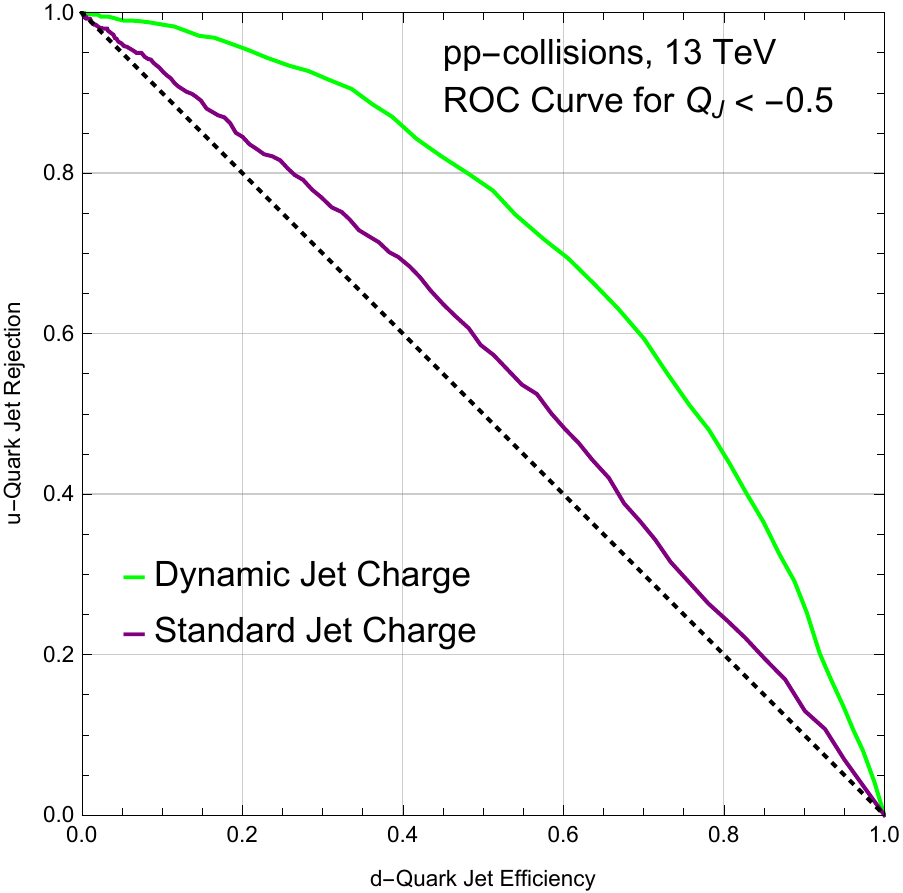}
    \includegraphics[scale=0.7]{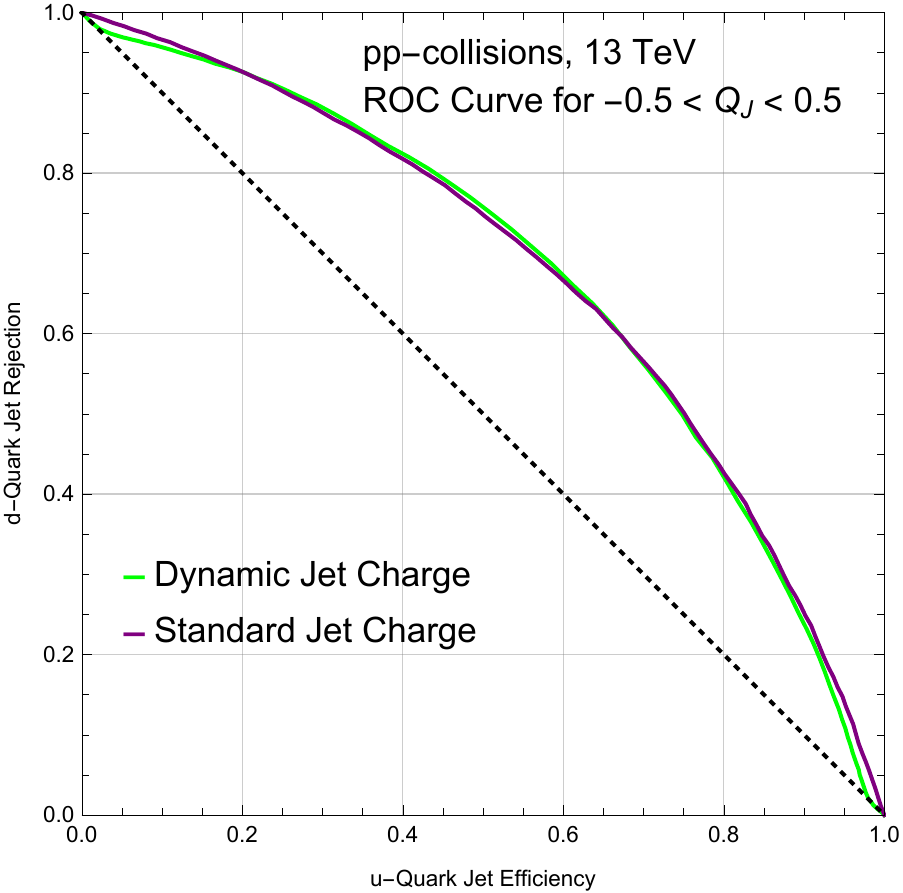}
    \includegraphics[scale=0.7]{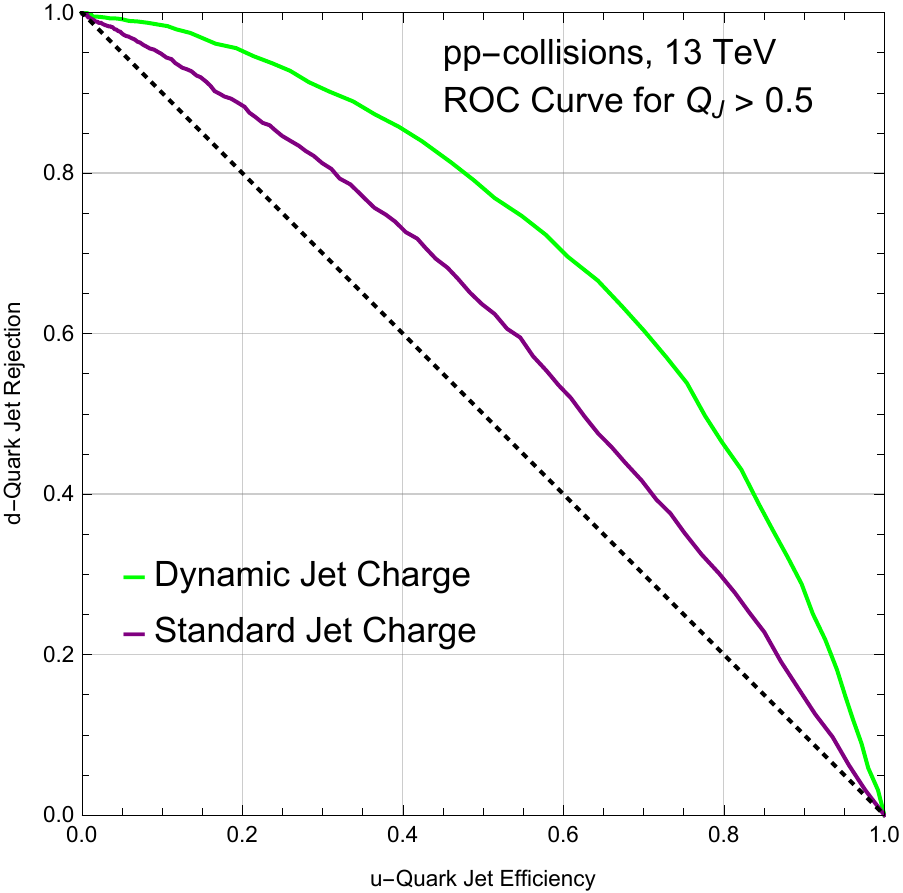}
    \caption{The local ROC curves for the standard (purple) and dynamic (green) jet charge distributions described in Fig.~\ref{fig:StdvsDynuvsd} for for $pp$-collisions at 13 TeV, with selection cuts $p_{T_J} = [200,300]$ GeV, $|\eta_J| < 2.1$. The top, middle, and bottom panels correspond to the local ROC curves for the jet charge bins $Q_J < -0.5$,  $|Q_J| <0.5$, and $Q_J >0.5$, respectively. }
    \label{fig:binnedROCuvsd}
\end{figure}
\begin{figure}
    \centering
    \includegraphics[scale=0.7]{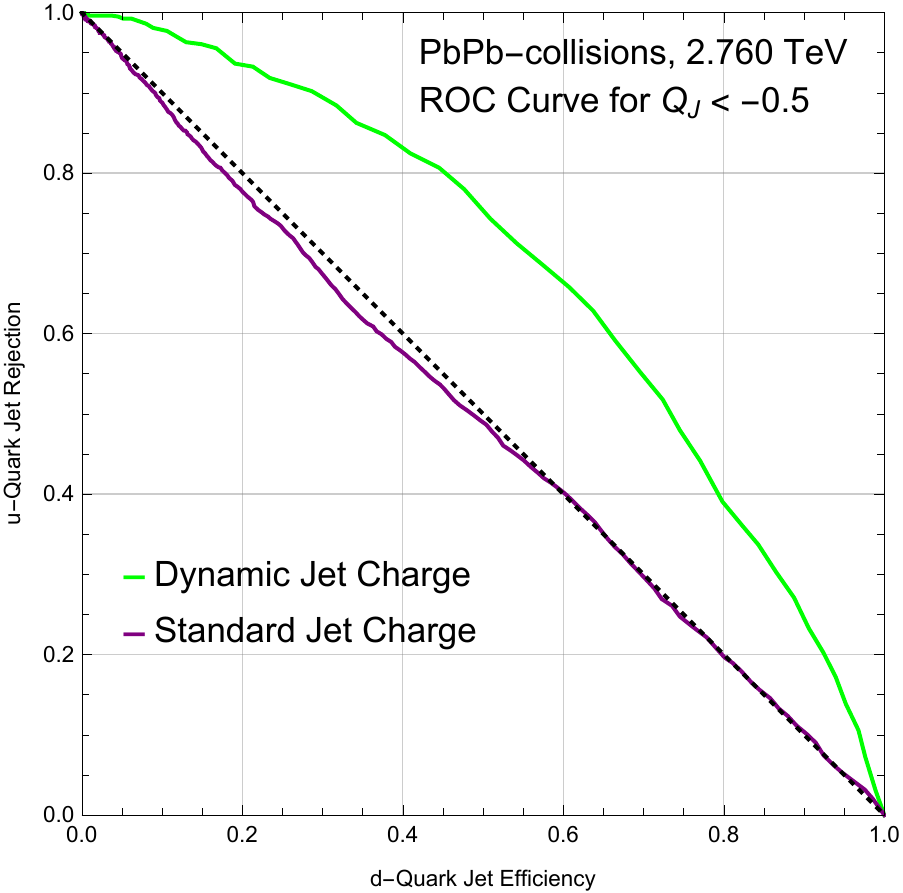}
    \includegraphics[scale=0.7]{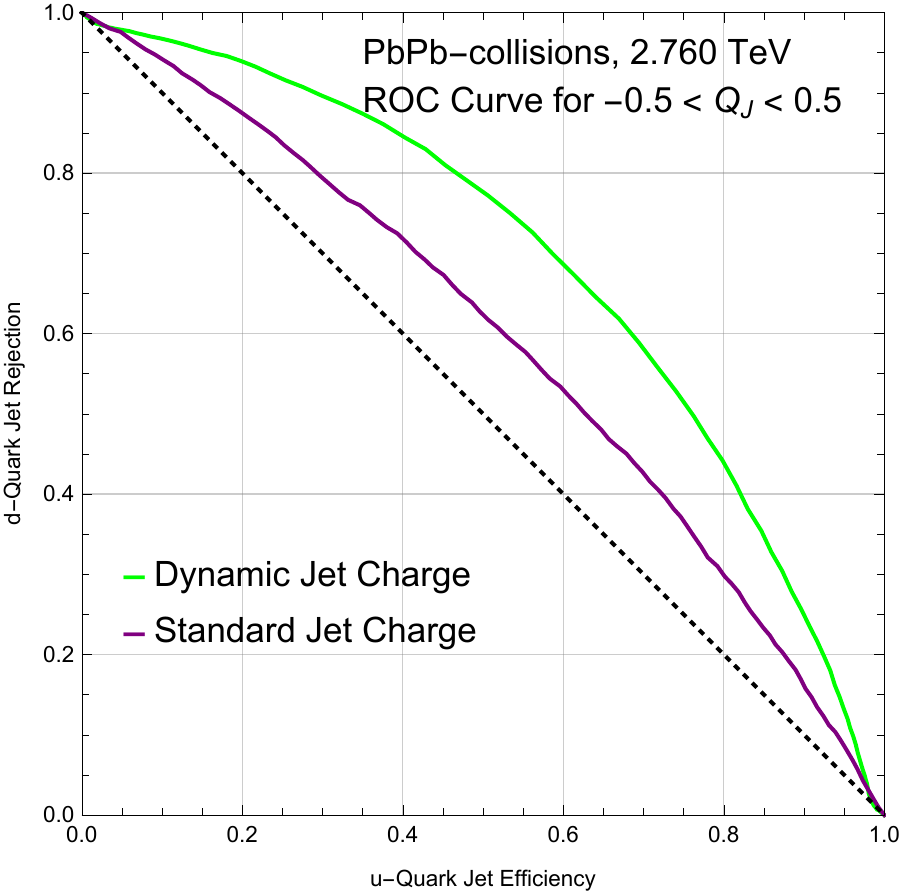}
    \includegraphics[scale=0.7]{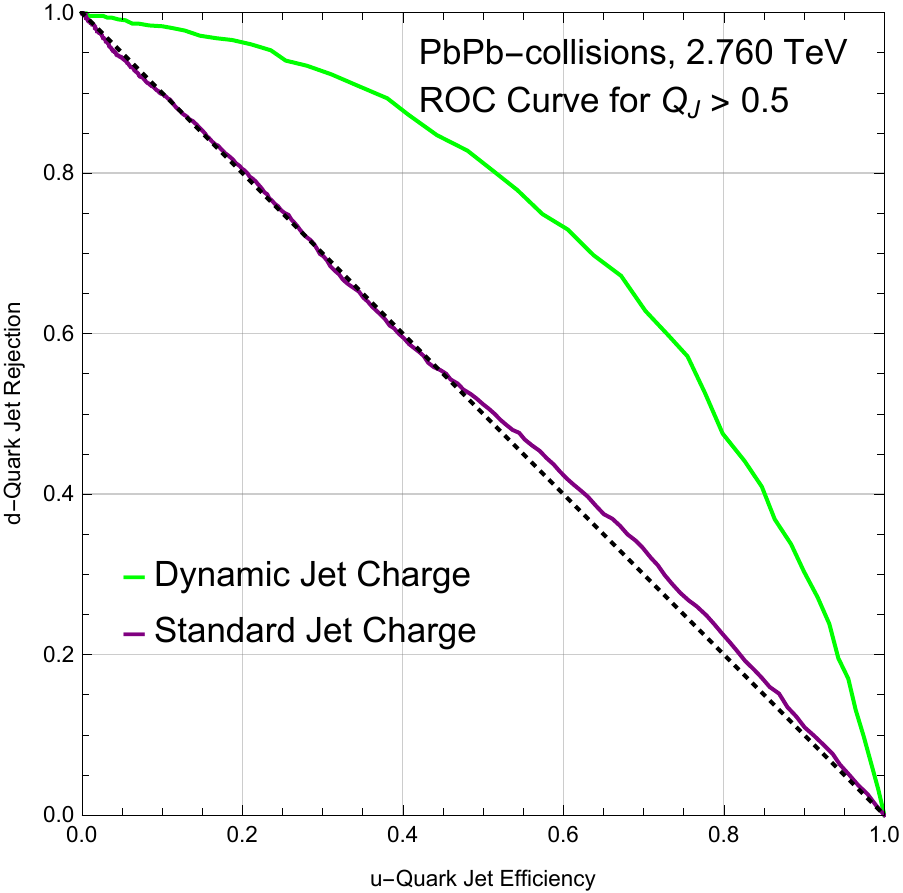}
    \caption{The local ROC curves for the standard (purple) and dynamic (green) jet charge distributions described in Fig.~\ref{fig:StdvsDynuvsdHI} for PbPb-collisions at 2.760 TeV, with selection cuts $p_{T_J} = [80,150]$ GeV, $|\eta_J| < 2.1$. The top, middle, and bottom panels correspond to the local ROC curves for the jet charge bins $Q_J < -0.5$,  $|Q_J| <0.5$, and $Q_J >0.5$, respectively.}
    \label{fig:binnedROCuvsdHI}
\end{figure}

In Fig.~\ref{fig:StdvsDynuvsdHI}, we show the same plots as in Fig.~\ref{fig:StdvsDynuvsd} but for heavy ion PbPb-collisions. We see that the standard jet charge distribution (top panel) is much broader compared to that in $pp$-collisions. Note that the jet charge distribution spans over $-4 < Q_J < 4$ in the plot for PbPb-collisions, while only over $-2 < Q_J < 2$ for $pp$-collisions. On the other hand, the dynamic jet charge distribution (middle panel) is quite similar in heavy ion and $pp$-collisions. Correspondingly, the global ROC curve (bottom panel) for the dynamic jet charge (green) in heavy ion collisions looks similar to that in $pp$-collisions. Thus, the $u$- vs. $d$-quark jet discrimination power of the dynamic jet charge is largely unaffected by the significantly greater underlying event activity in heavy ion collisions. By contrast, there is significant degradation in the discrimination power for the standard jet charge in PbPb-collisions compared to $pp$-collisions, as seen by comparing the global ROC curves in Figs.~\ref{fig:StdvsDynuvsd} and \ref{fig:StdvsDynuvsdHI}.

In fact, for PbPb collisions, the dynamic jet charge shows better, but comparable, discrimination power compared to the standard jet charge, as seen in the bottom panel of Fig.~\ref{fig:StdvsDynuvsdHI}.

Furthermore, once again, due to the multiple peak structure of the dynamic jet charge distribution, a binned analysis with local ROC curves can be performed. We revisit the dynamic jet charge distribution (middle panel) in Fig.~\ref{fig:StdvsDynuvsd} for $pp$-collisions. The peak structure suggests that the jet data be divided into three bins:  $(i)\> Q_J < -0.5, (ii)\> -0.5 < Q_J < 0.5,$ and $(iii)\> Q_J > 0.5$. We show the local ROC curves for the standard and dynamic jet charge distributions in each of these three jet charge bins in Figs.~\ref{fig:binnedROCuvsd} and ~\ref{fig:binnedROCuvsdHI} for $pp$- and PbPb-collisions, respectively. We see that the local ROC curves for the dynamic jet charge  shows better discrimination compared to standard jet charge in all bins for PbPb-collisions. For $pp$-collisions, the dynamic jet charge gives better discrimination in the leftmost ($Q_J < -0.5$) and rightmost ($Q_J > 0.5$) bins and about the same discrimination in the central bin ($-0.5 < Q_J < 0.5$), compared to the standard jet charge. 
%%%%%%%%%%%%%
\begin{table}[t!]
	\begin{center}
		\scalebox{1}{
			\begin{tabular}{|l c | c| c| c |}%| c}
				\hline \hline
				&Jet Charge $\>\>\>$& & &   
				\\
				&($pp$-collisions) $\>\>\>$& $\>\>\>\>\>\epsilon_L\>\>\>\>\>$ & $\>\>\>\>\>\epsilon_C\>\>\>\>\>$ & $\>\>\>\>\>\epsilon_R\>\>\>\>\>$  
				\\[0.5ex] \hline\hline
				& Standard: $u$-quark  & 0.05   & 0.59 & 0.36  \\
				& $\>\>\>\>\>\>\>\>\>\>\>\>\>\>\>\>\>\>\>$ $d$-quark & 0.28  & 0.65 &  0.08
				\\[0.5ex] \hline
				& Dynamic: $u$-quark & 0.05 & 0.73 & 0.21 \\
				& $\>\>\>\>\>\>\>\>\>\>\>\>\>\>\>\>\>\>\>$ $d$-quark   & 0.16 &0.77 & 0.08 
				\\ \hline\hline
		\end{tabular}}
		%%%%%%%%%%%%%%%%%%%%%%%%%%%%%%%%
	\end{center}
		\begin{center}
		\scalebox{1}{
			\begin{tabular}{|l c | c| c| c |}%| c}
				\hline \hline
				&Jet Charge $\>\>\>$& & &   
				\\
				&(PbPb-collisions) $\>\>\>$& $\>\>\>\>\>\epsilon_L\>\>\>\>\>$ & $\>\>\>\>\>\epsilon_C\>\>\>\>\>$ & $\>\>\>\>\>\epsilon_R\>\>\>\>\>$  
				\\[0.5ex] \hline\hline
				& Standard: $u$-quark  & 0.19   & 0.40& 0.41  \\
				& $\>\>\>\>\>\>\>\>\>\>\>\>\>\>\>\>\>\>\>$ $d$-quark & 0.38  &0.43 &  0.21
				\\[0.5ex] \hline
				& Dynamic: $u$-quark & 0.04 & 0.79 & 0.17 \\
				& $\>\>\>\>\>\>\>\>\>\>\>\>\>\>\>\>\>\>\>$ $d$-quark   & 0.12 &0.82 & 0.06
				\\ \hline\hline
		\end{tabular}}
	\end{center}

	\caption{Fraction of events for $u$-quark and $d$-quark jets in the $ Q_J < -0.5,  -0.5 < Q_J < 0.5,$ and $ Q_J > 0.5$  jet charge bins. The  $\epsilon_{L},\epsilon_{C},$ and $\epsilon_{R}$ fractions  are defined in Eq.~(\ref{eq:fracuvsd}). The numbers correspond to the jet charge distributions in Fig.~\ref{fig:StdvsDynuvsd} and \ref{fig:StdvsDynuvsdHI} for $pp$-collisions and PbPb-collisions, respectively. }
	\label{table:uvsd}
\end{table}

As done for the quark-gluon discrimination analysis, one can also quantify the $u$-quark vs. $d$-quark jet discrimination using the fraction of events in each jet charge bin. Based on the dynamic jet charge distributions in Figs.~\ref{fig:StdvsDynuvsd} and \ref{fig:StdvsDynuvsdHI}, we define the fraction of the total number of events in three jet charge bins as
\begin{align}
\label{eq:fracuvsd}
\epsilon_L &= \frac{N( \>Q_J < -0.5 \>)}{N_{\rm total}}, \nn \\
\epsilon_C &= \frac{N( -0.5 \leq Q_J  \leq 0.5 \>)}{N_{\rm total}}, \nn \\
\epsilon_R &= \frac{N( \>Q_J  > 0.5 \>)}{N_{\rm total}},
\end{align}
where $\epsilon_L,\epsilon_C,$ and $\epsilon_R$ give the fraction of the total number of events in the jet charge bins $Q_J < -0.5$, $-0.5 < Q_J < 0.5$, and $Q_J >0.5$, respectively.

In Table~\ref{table:uvsd}, we show the values of $\epsilon_L, \epsilon_C $, and $\epsilon_R$ for $u$-quark and $d$-quark jets  for  the standard and dynamic jet charge distributions shown in Figs.~\ref{fig:StdvsDynuvsd} and \ref{fig:StdvsDynuvsdHI} for $pp$-collisions and heavy ion collisions, respectively.  We see that for both $pp$-collisions and PbPb-collisions, there is a substantial difference in the event fractions of $u$-quark and $d$-quark jets in the left-most ($Q_J < -0.5$) and right-most ($Q_J > 0.5$) jet charge bins. On the other hand, the event fractions are about the same for both  $u$-quark and $d$-quark jets in the central jet charge bin ($-0.5 < Q_J < 0.5$). Thus, just as in the case of quark-gluon discrimination, sorting the data into jet charge bins can be used to discriminate between $u$-quark and $d$-quark jets in the left-most and right-most jet charge bins. We note that in this method, the standard jet charge seems to give better discrimination  than the dynamic jet charge for $pp$-collisions in the leftmost and rightmost bins. On the other hand, for PbPb-collisions the dynamic jet charge seems to give better discrimination in the leftmost and rightmost bins.

\section{Characteristics of the Dynamic Jet Charge}

\begin{figure}
    \centering
    \includegraphics[scale=0.7]{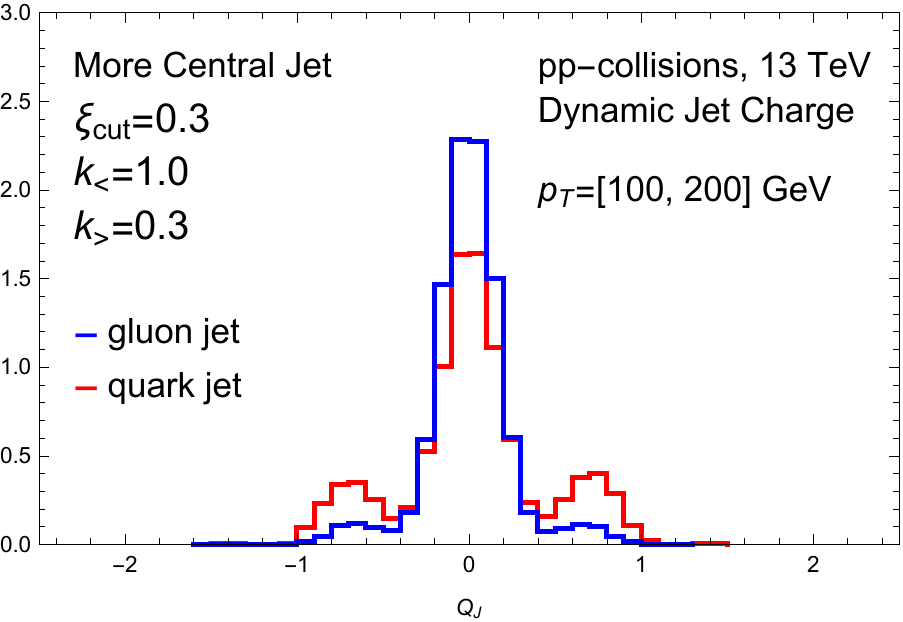}
    \includegraphics[scale=0.7]{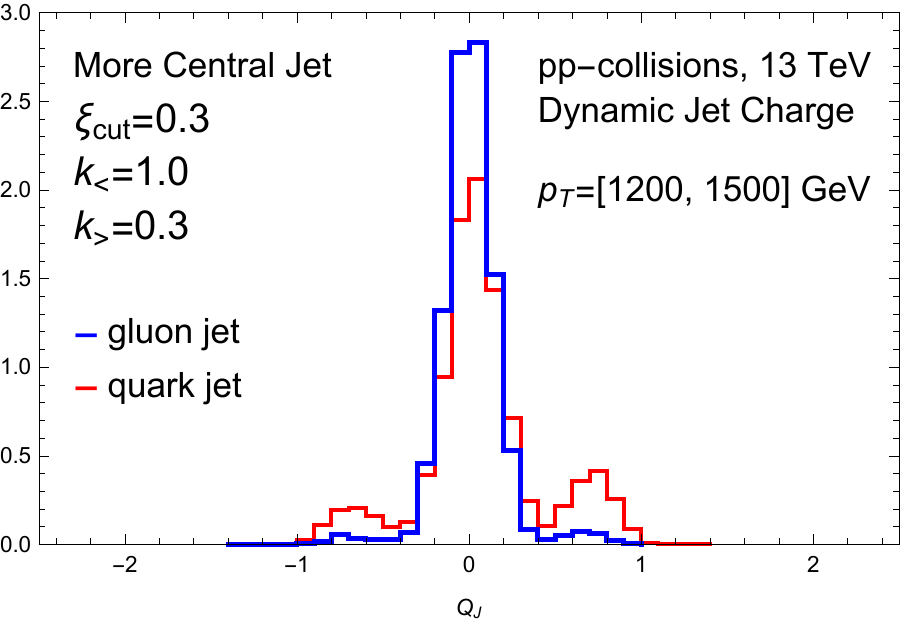}
    \caption{A comparison of the dynamic jet charge distributions for quark (red) and gluon (blue) jets in $pp\to j_1j_2 X$ at 13 TeV. Selection cuts of  $|\eta_{j1,j2}| <2.1$ and $p_{T_{j_1}}/p_{T_{j_2}} < 1.5$ on the leading ($j_1$) and subleading ($j_2$) anti-k$_T$ jets of jet radius $R=0.4$ are applied. The top and bottom panels correspond to jets in the $p_{T_J}$-bins [100 GeV, 200 GeV] and  [1200 GeV, 1500 GeV], respectively.}
    \label{fig:pTdependence}
\end{figure}

\begin{figure}
    \centering
    \includegraphics[scale=0.7]{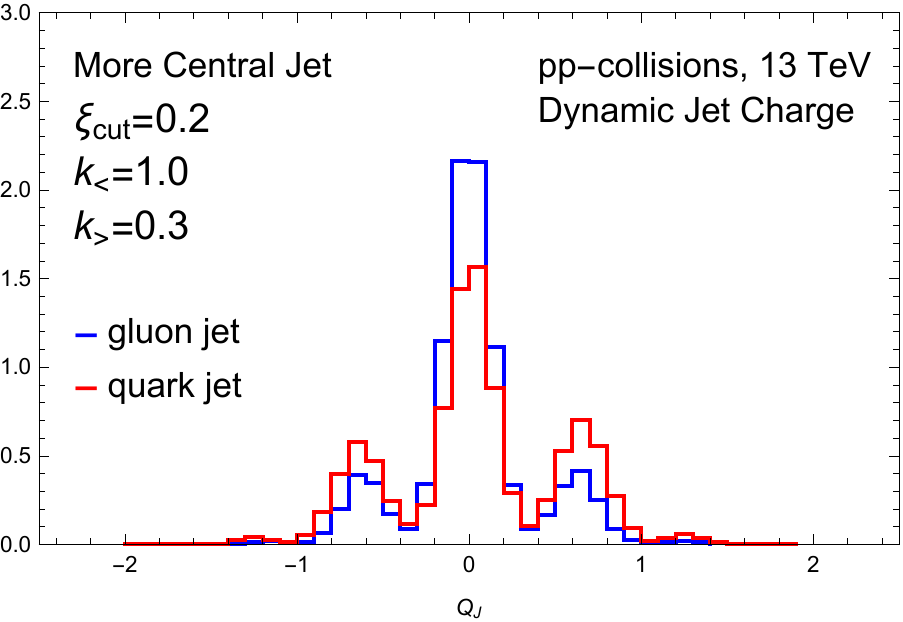}
    \includegraphics[scale=0.7]{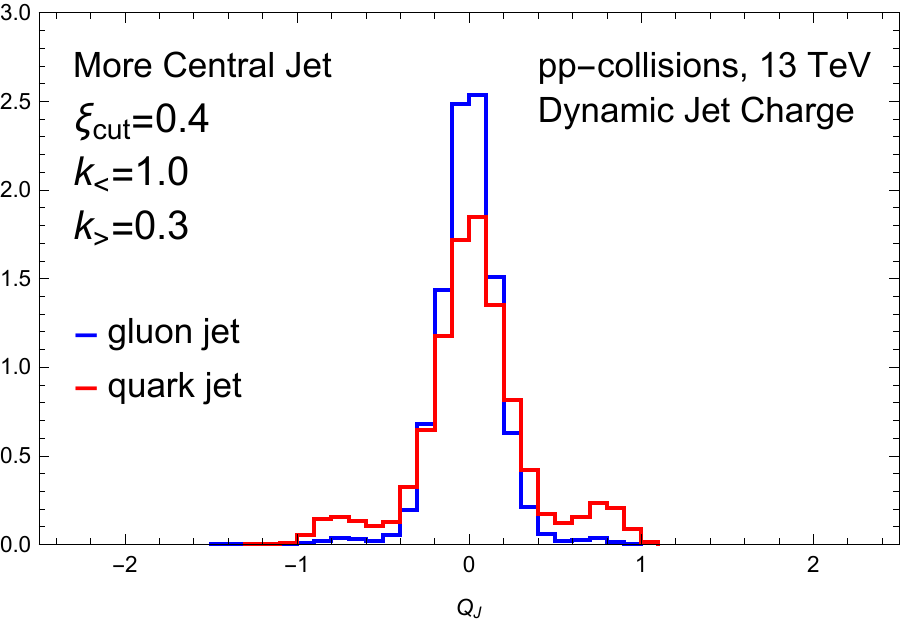}
     \includegraphics[scale=0.7]{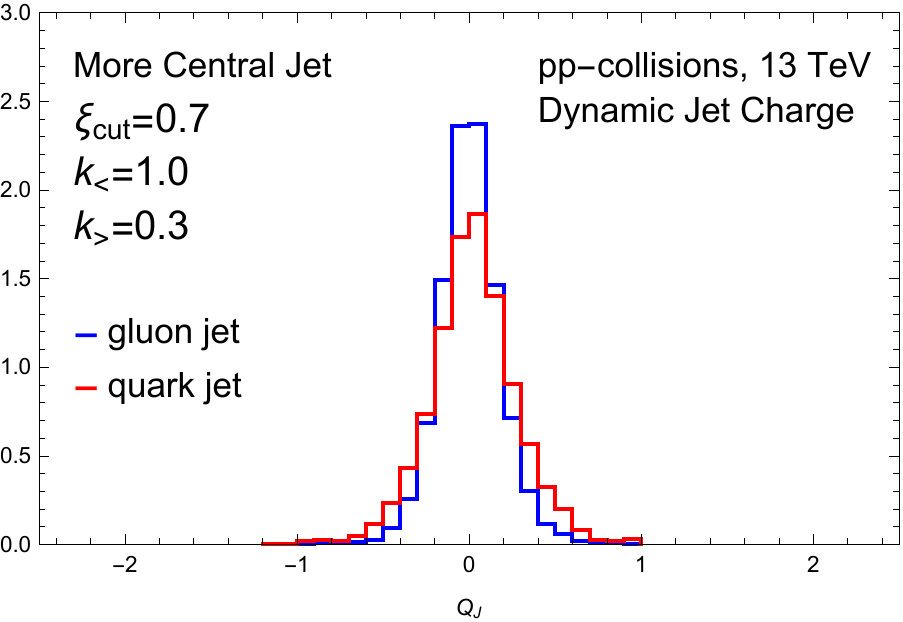}
    \caption{A comparison of the dynamic jet charge distributions for  different choices of the dynamic jet charge parameters, $\xi_{\rm cut}, k_< $, and  $k_>$, for quark (red) and gluon (blue) jets in $pp\to j_1j_2 X$  at 13 TeV in the jet $p_T$-bin: [200,300] GeV. Selection cuts of  $|\eta_{j1,j2}| <2.1$ and $p_{T_{j_1}}/p_{T_{j_2}} < 1.5$ on the leading ($j_1$) and subleading ($j_2$) anti-k$_T$ jets of jet radius $R=0.4$ are applied.}
    \label{fig:dynparameterdependence}
\end{figure}

In this section, we explore the typical behavior of the dynamic jet charge distribution for different choices of kinematic settings and jet charge parameters. We also present simulation studies of the impact of underlying event and jet grooming, on both the standard and dynamic jet charge distributions.

In Fig.~\ref{fig:pTdependence}, we show a comparison of the dynamic jet charge distribution for quark and gluon jets in the two different $p_{T_J}$-bins. The top and bottom panels correspond to $p_{T_J}=$[100 GeV, 200 GeV] and  $p_{T_J}=$[1200 GeV, 1500 GeV], respectively. We see that the distinctive multiple peak structure remains across a wide range of $p_{T_J}$. As seen in the bottom panel, at large $p_{T_J}$ an asymmetry between the peaks at negative and positive jet charge develops for the quark jets. This asymmetry can be understood as a consequence of the dominance of valence $u$-quark PDFs at large Bjorken-$x$, corresponding to large $p_{T_J}$, so that the quark jet sample has a larger fraction of $u$-quark jets compared to $d$-quark jets. As seen in Fig.~\ref{fig:StdvsDynuvsd}, the dynamic jet charge distribution for  $u$-quark and $d$-quark jets is tilted more towards positive and negative jet charge, respectively. The combined effect of the dominance of the $u$-quark PDFs at large $p_{T_J}$, and the bias of $u$-quark jets towards positive jet charge is reflected in the asymmetry of the overall quark jet distribution in the bottom panel of Fig.~\ref{fig:pTdependence}.

\begin{figure}
    \centering
    \includegraphics[scale=0.7]{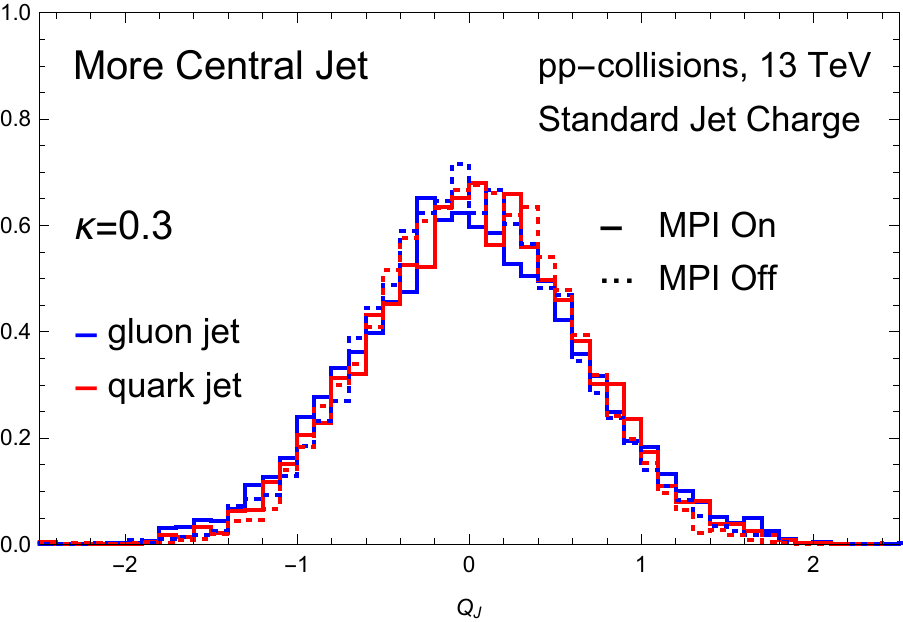}
    \includegraphics[scale=0.7]{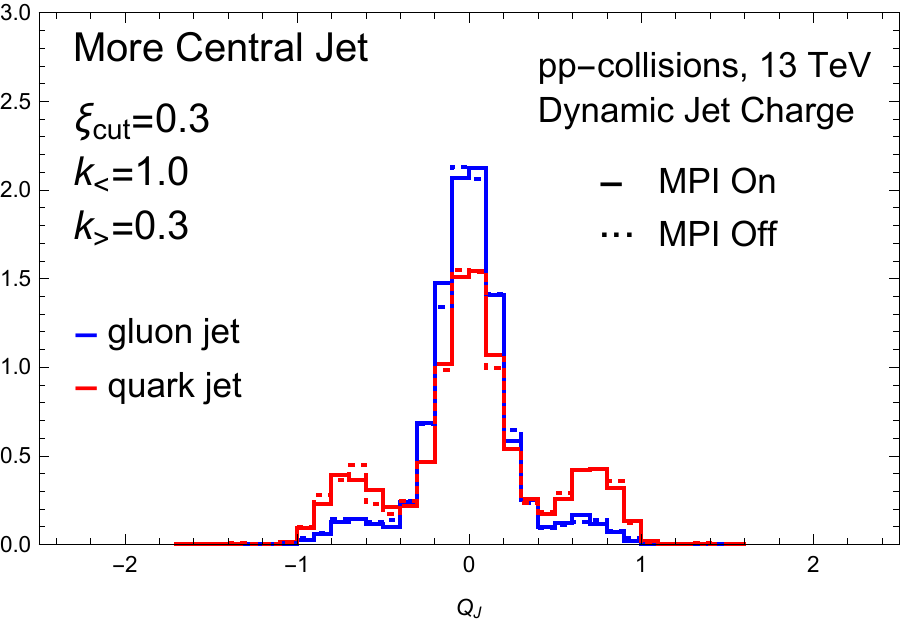}
    \caption{The standard (top) and dynamic (bottom) jet charge distributions  with MPI switch turned on (solid) and off (dashed) for quark (red) and gluon (blue) jets in $pp\to j_1j_2X$ at $\sqrt{s}=13$ TeV and in the jet $p_{T_J}$-bin: [50,100] GeV. Selection cuts of  $|\eta_{j1,j2}| <2.1$ and $p_{T_{j_1}}/p_{T_{j_2}} < 1.5$ on the leading ($j_1$) and subleading ($j_2$) anti-k$_T$ jets of jet radius $R=0.4$ are applied.}
    \label{fig:MPIOnvsOffStdandDyn}
\end{figure}

In Fig.~\ref{fig:dynparameterdependence}, we show the behavior of the dynamic jet charge distribution for different parameter choices. In particular, we set $k_< =1.0, k_> =0.3$ and vary $\xi_{\rm cut}$ over the three values $\xi_{\rm cut}=0.2$ (top panel), $\xi_{\rm cut}=0.4$ (middle panel), and $\xi_{\rm cut}=0.7$ (bottom panel). The plot for the default choice of $\xi_{\rm cut}=0.3$ is already shown in Fig.~\ref{fig:StdvsDyn} (middle panel). We see that for increasing values of $\xi_{\rm cut}$, the secondary non-central peaks in the distribution become less pronounced. For the largest value shown, $\xi_{\rm cut}=0.7$, the non-central peaks completely disappear. This can be understood as a consequence of the fact that there are typically very few hadrons in the jet with $z_h>  \xi_{\rm cut}=0.7$. As a result, as seen from Eq.~(\ref{eq:dynJCfuncform}), the jet charge distribution effectively reduces to the standard jet charge distribution with $\kappa=k_<=1.0$. Similarly, in the limit of $\xi_{\rm cut}\to 0$, the dynamic jet charge distribution will approach the standard jet charge distribution with $\kappa=k_>$.

\begin{figure}
    \centering
    \includegraphics[scale=0.7]{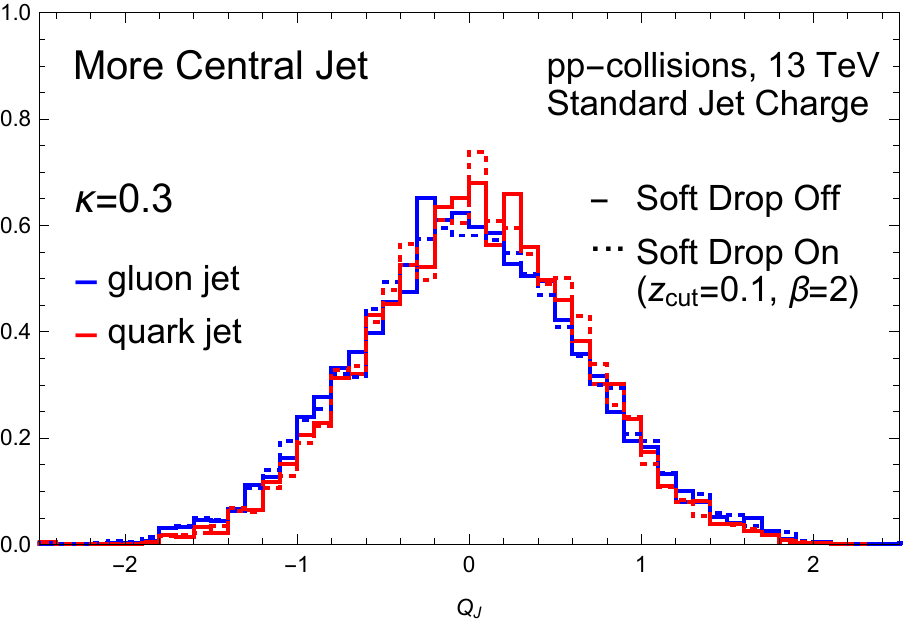}
    \includegraphics[scale=0.7]{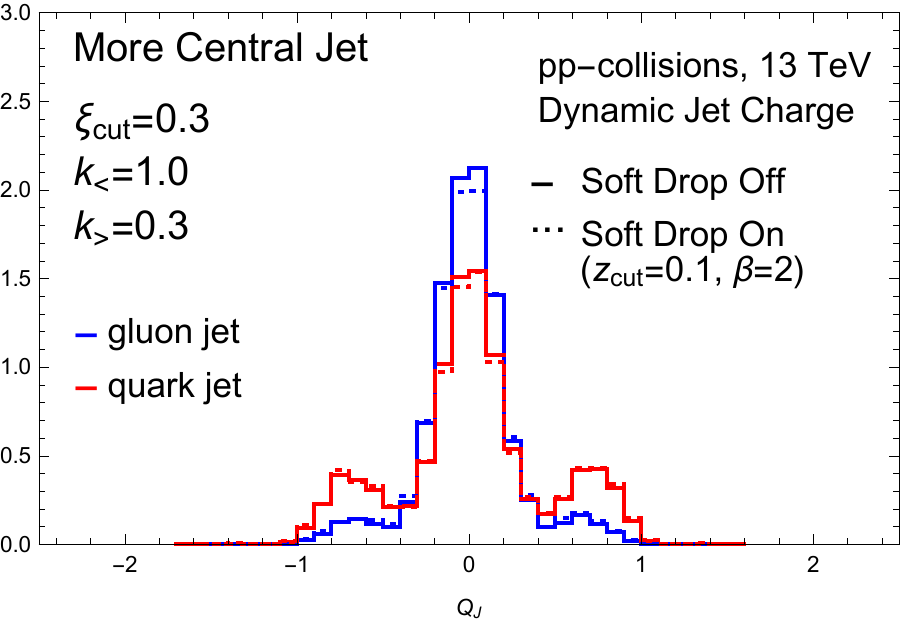}
    \caption{The standard (top) and dynamic (bottom) jet charge distributions  for ungroomed (solid) and groomed (dashed) quark and gluon jets in $pp\to j_1j_2X$ at $\sqrt{s}=13$ TeV and in the jet $p_{T_J}$-bin: [50,100] GeV. Selection cuts of  $|\eta_{j1,j2}| <2.1$ and $p_{T_{j_1}}/p_{T_{j_2}} < 1.5$ on the leading ($j_1$) and subleading ($j_2$) anti-k$_T$ jets of jet radius $R=0.4$ are applied.}
    \label{fig:SDOnvsOffStdandDyn}
\end{figure}

\begin{figure}
    \centering
    \includegraphics[scale=0.7]{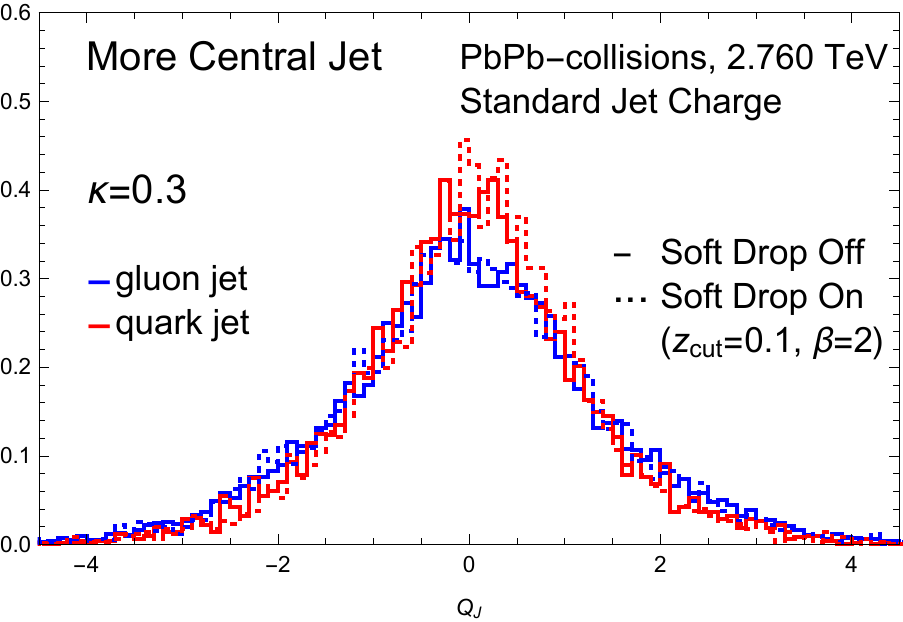}
    \includegraphics[scale=0.7]{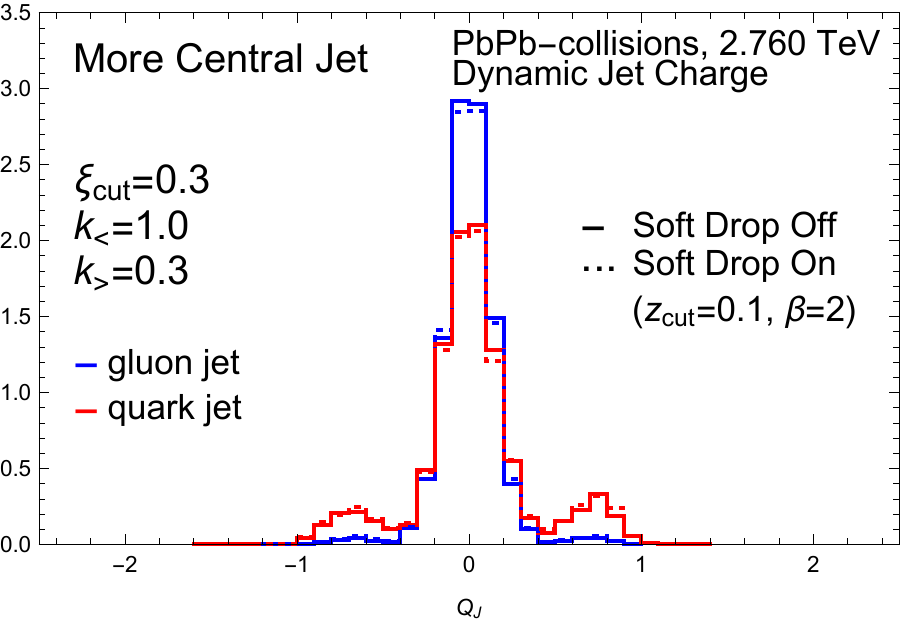}
    \caption{The standard (top) and dynamic (bottom) jet charge distributions  for ungroomed (solid) and groomed (dashed) quark and gluon jets in PbPb$\to j_1j_2X$ at $\sqrt{s}=2.760$ TeV and in the jet $p_{T_J}$-bin: [80,150] GeV. Selection cuts of  $|\eta_{j1,j2}| <0.9$ and $p_{T_{j_1}}/p_{T_{j_2}} < 1.5$ on the leading ($j_1$) and subleading ($j_2$) anti-k$_T$ jets of jet radius $R=0.4$ are applied.}
    \label{fig:SDOnvsOffStdandDynHI}
\end{figure}

\begin{figure}
    \centering
    \includegraphics[scale=0.7]{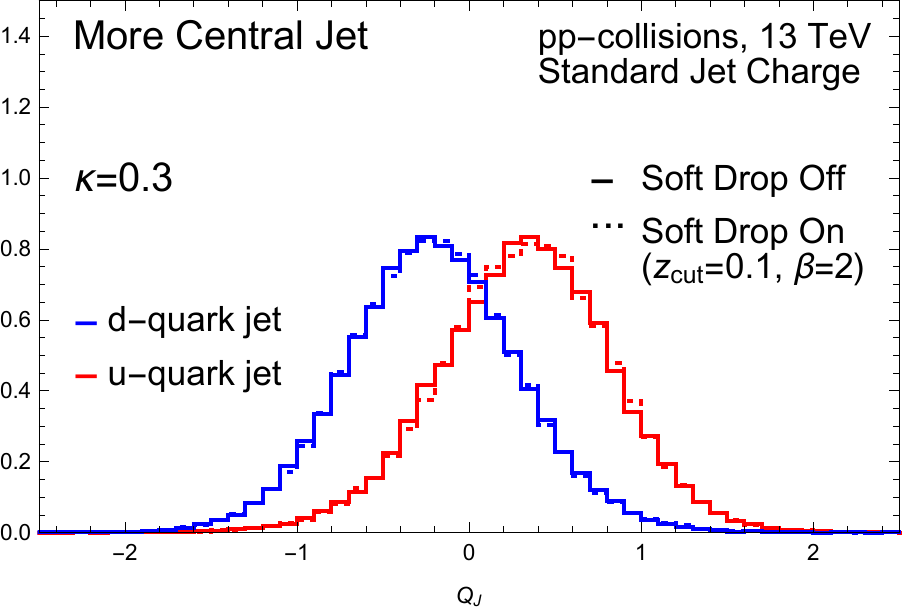}
        \includegraphics[scale=0.7]{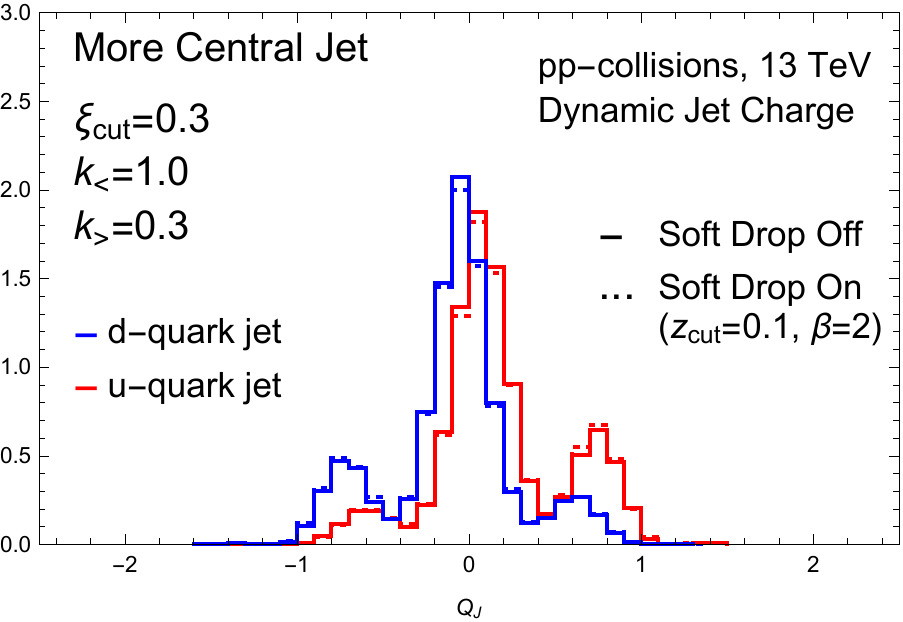}
    \caption{The standard (top) and dynamic (bottom) jet charge distributions for $u$-quark (red) and $d$-quark (blue) jets, with (dashed) and without (solid) grooming, in $pp$-collisions at  $\sqrt{s}=13$ TeV with $R=0.4$,  $p_{T_J} = [50,100]$ GeV, and $|\eta_J| < 2.1$. The partonic channels $dg\to W^-u$ and $ug\to W^+ d$ generated the $u$-quark and $d$-quark jets, respectively.}
    \label{fig:SDOnvsOffStdandDynUvsD}
\end{figure}

\begin{figure}
    \centering
    \includegraphics[scale=0.7]{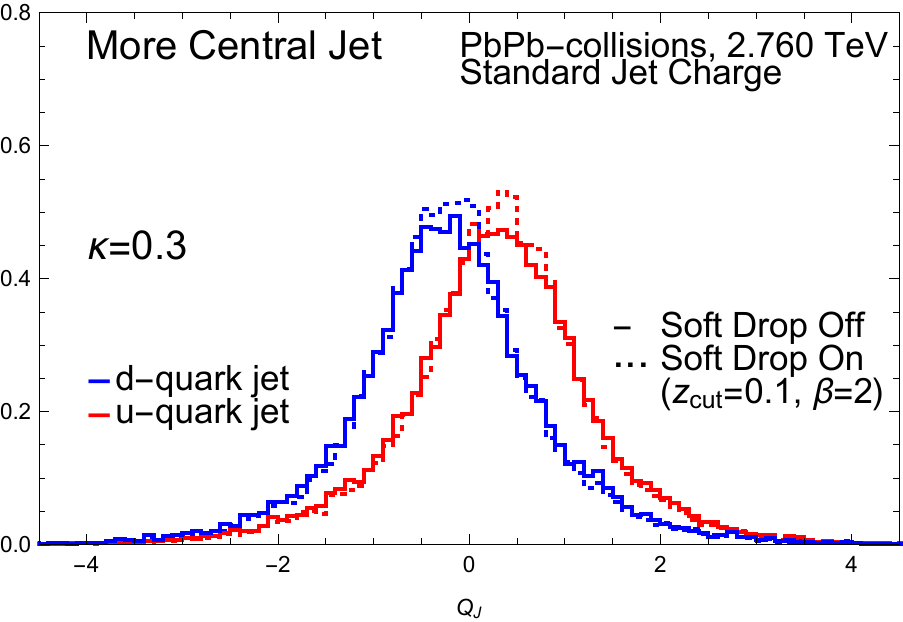}
        \includegraphics[scale=0.7]{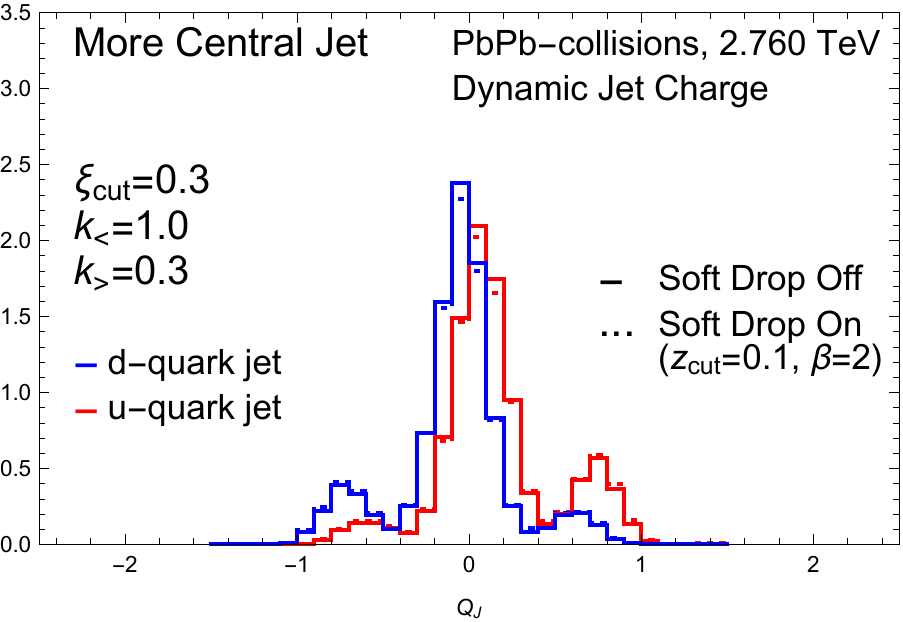}
    \caption{The standard (top) and dynamic (bottom) jet charge distributions for $u$-quark (red) and $d$-quark (blue) jets, with (dashed) and without (solid) grooming, in PbPb-collisions at  $\sqrt{s}=2.760$ TeV with $R=0.4$,  $p_{T_J} = [80,150]$ GeV, and $|\eta_J| < 0.9$. The partonic channels $dg\to W^-u$ and $ug\to W^+ d$ generated the $u$-quark and $d$-quark  jets.}
    \label{fig:SDOnvsOffStdandDynUvsDHI}
\end{figure}

As seen in Fig.~\ref{fig:MPIOnvsOffStdandDyn}, both the standard and dynamic jet charge distributions for quark and gluon initiated jets in $pp$-collisions are relatively unaffected by underlying event.  This can be understood as the result of the fact that, as seen in Eqs.~(\ref{eq:stdJC}) and (\ref{eq:dynJC}) the contributions of the individual jet constituents to the total jet charge are weighted by their jet transverse momentum or energy fractions. Thus, soft contamination effects on the jet charge are suppressed. As seen in Fig.~\ref{fig:MPIOnvsOffStdandDyn}, the soft contamination effects appear to be slightly more suppressed in the dynamic jet charge compared to the standard jet charge. This is simply understood as a result of the fact that for the choice of parameters $k_<=1.0$ and $k_>=0.3$, compared to the standard jet charge, the dynamic jet charge gives a much lower weighting to the low energy ($z_h < \xi_{\rm cut}$) jet constituents compared to those with higher energy ($z_h > \xi_{\rm cut}$). Thus, the dynamic jet charge allows the flexibility to choose parameters that make the jet charge even more robust against soft contamination. 

As a further demonstration of the relative insensitivity of the jet charge to soft contamination, in Figs.~\ref{fig:SDOnvsOffStdandDyn} and \ref{fig:SDOnvsOffStdandDynHI}, we show the effect of jet grooming on the standard and dynamic jet charge distributions in $pp$-collisions and PbPb-collisions, respectively.  We choose typical soft drop grooming parameters of  $z_{\rm cut}=0.1$ and $\beta=2$~\cite{Larkoski:2014wba}. Once again, we see that the standard and dynamic jet charge distributions for quark and gluon initiated jets are very similar for both groomed and ungroomed jets. As seen in Figs.~\ref{fig:SDOnvsOffStdandDynUvsD} and \ref{fig:SDOnvsOffStdandDynUvsDHI}, the same is true of $u$-quark and $d$-quark initiated jets. Thus, these simulation results indicate that the soft drop grooming of jets does not have much impact on the jet discrimination power of the standard and dynamic jet charge observables.
\begin{figure}
    \centering
    \includegraphics[scale=0.7]{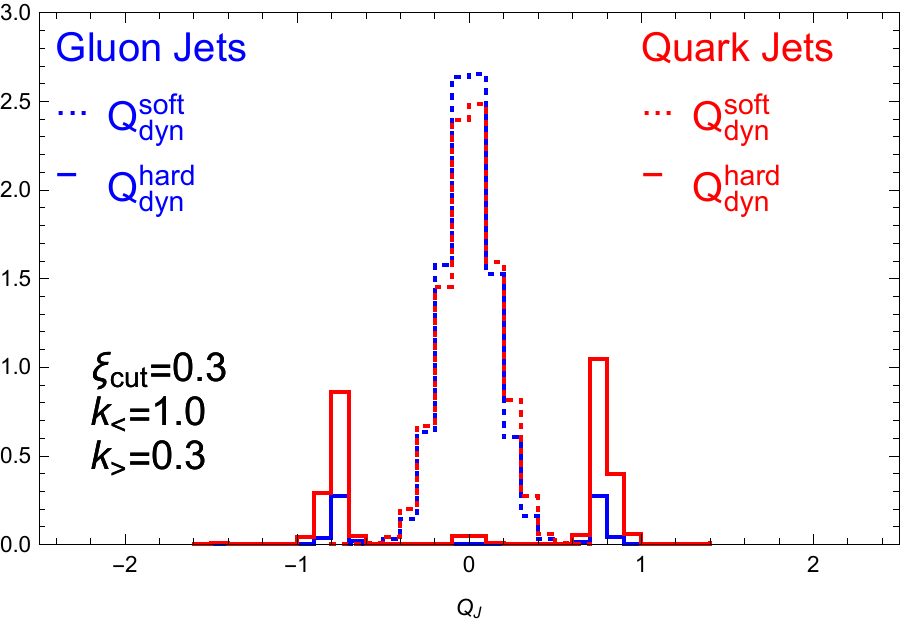}
     \includegraphics[scale=0.7]{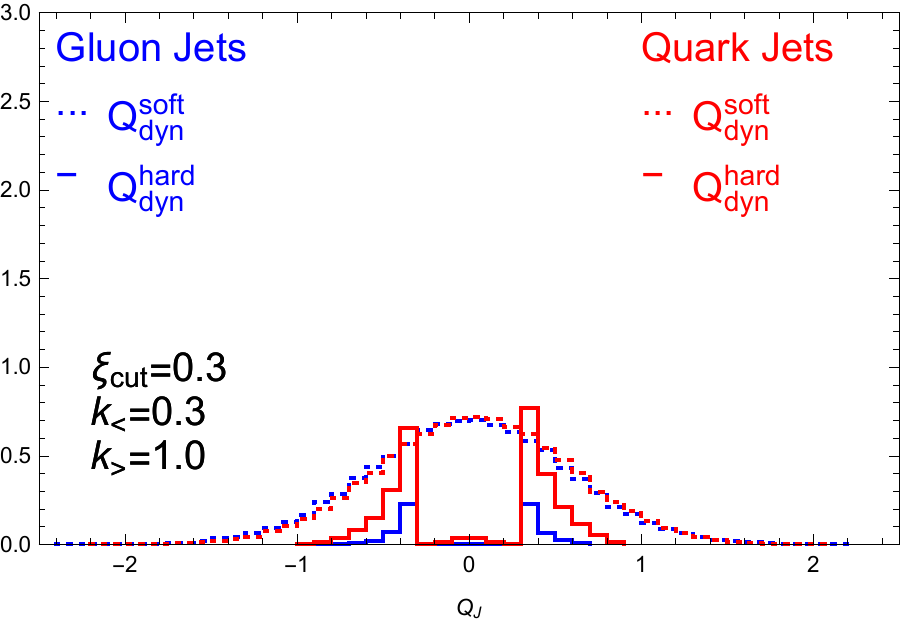}
      \includegraphics[scale=0.7]{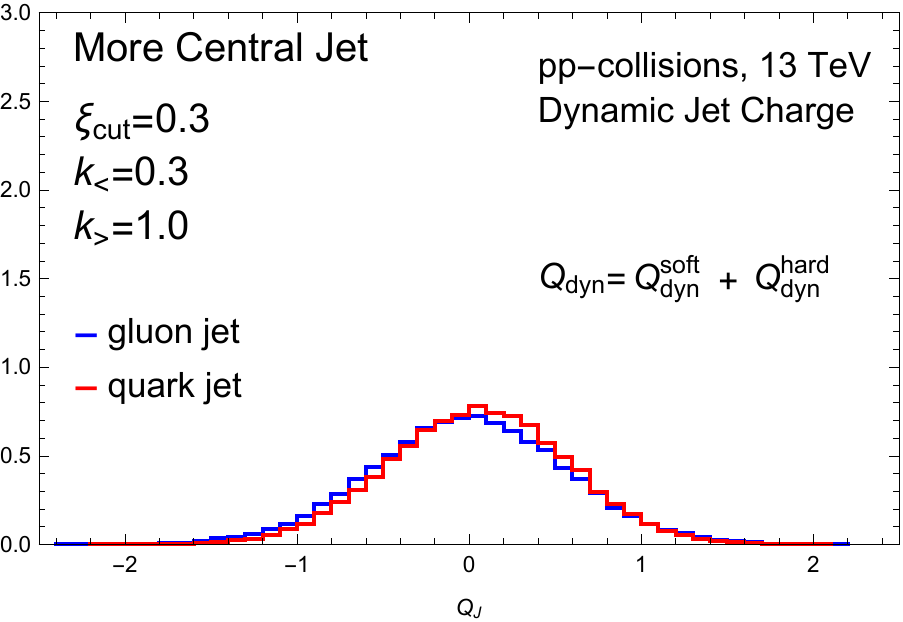}

    \caption{Top panel: The contribution to the dynamic jet charge from the soft (dashed) and hard (solid) particles in quark (red) and (gluon) jets for the choice of parameters: $\xi_{\rm cut}=0.3, k_<=1.0, k_>=0.3$.  Middle Panel: The contribution to the dynamic jet charge from the soft (dashed) and hard (solid) particles in quark (red) and (gluon) jets for the choice of parameters: $\xi_{\rm cut}=0.3, k_<=0.3, k_>=1.0$. Bottom Panel: The total dynamic jet charge distribution for $\xi_{\rm cut}=0.3, k_<=0.3, k_>=1.0$.  All the distributions above were generated for the more central jet in dijet production of $pp$-collisions at $\sqrt{s}=$13 TeV, with selection cuts $p_{T_J} = [200,300]$ GeV, $|\eta_J| < 2.1$, and $p_{T_{j_1}}/p_{T_{j_2}} < 1.5$ on the leading ($j_1$) and subleading ($j_2$) anti-k$_T$ jets of jet radius $R=0.4$. }
    \label{fig:HardvsSoft}
\end{figure}

Finally, we discuss in more detail the properties of the dynamic jet charge distribution that underly its discrimination power and characteristic shape with multiple peaks. The main idea is based on the observation (see Fig.~\ref{fig:StdJCkappa}) that the standard jet charge distribution gets broader (narrower) for smaller (larger) values of $\kappa$. This can be understood as the result of a large $\kappa$-value washing out the charge information of the individual particles, in the standard jet charge definition of Eq.~(\ref{eq:stdJC}), 
giving values closer to zero.  This feature is used in the dynamic jet charge to cleanly separate out the contributions of the hard ($z_h >\xi_{\rm cut}$) particles in the jet from the soft ($z_h <\xi_{\rm cut}$)  particles.   This is done by using $\kappa=k_>$ and $\kappa=k_<$ for the hard and soft particles, respectively, so that the dynamic jet charge can be defined as the sum of the contributions from the soft and hard particles in the jet:
\bea
Q_{\rm dyn}^i &=& Q_{\rm dyn}^{i,\text{soft}} +  Q_{\rm dyn}^{i,\text{hard}},  
\eea
where we have defined these contributions as:
\bea
Q_{\rm dyn}^{i,\text{soft}}&=&\sum_{h \in i\text{-jet}} \Theta (\xi_{\rm cut} - z_h) \>z_h^{k_<} \>Q_h ,\nn \\
Q_{\rm dyn}^{i,\text{hard}}&=&\sum_{h \in i\text{-jet}} \Theta (z_h-\xi_{\rm cut} ) \>z_h^{k_>} \>Q_h. 
\eea

In Fig.~\ref{fig:HardvsSoft}, we show separately the distributions of the $Q_{\rm dyn}^{i,\text{soft}}$ and $Q_{\rm dyn}^{i,\text{hard}}$ contributions to the total dynamic jet charge for the default choice, $\xi_{\rm cut}=0.3,~k_<=1.0,~k_>=0.3$ (top panel) and for a different choice, $\xi_{\rm cut}=0.3,~k_<=0.3,~k_>=1.0$ (middle panel).  The $Q_{\rm dyn}^{i,\text{soft}}$ distributions are normalized to unity. On the other hand, the $Q_{\rm dyn}^{i,\text{hard}}$ distributions are weighted by the relative fraction of events in which the jet contains at least one hard particle ($z_h > \xi_{\rm cut}$). Thus, the normalization of the $Q_{\rm dyn}^{i,\text{hard}}$ distribution relative to the unit normalization of $Q_{\rm dyn}^{i,\text{soft}}$, corresponds to the fraction of events in which the jet contains at least one hard particle ($z_h > \xi_{\rm cut}$). 

The default choice, $\xi_{\rm cut}=0.3, k_<=1.0, k_>=0.3$ (top panel), which gives a higher weight to the hard particles relative to the soft particles in the jet, causes the soft particles to be distributed in a narrow central peak. This corresponds to the narrow shape of the standard jet charge distribution with $\kappa=k_<=1.0$. On the other hand, the hard particles are accumulated near larger non-zero values of the jet charge, corresponding to the tails of the broader standard jet charge distribution with $\kappa=k_>=0.3$.  Note the clean separation of the effects of the hard and soft particles in the jet charge distribution. The choice of $ k_<=1.0, k_>=0.3$, localizes the effect of the soft particles to the central peak and the hard particles to the non-central peaks, giving rise to the characteristic multiple peak structure of the corresponding total dynamic jet charge, $Q_{\rm dyn}^i = Q_{\rm dyn}^{i,\text{soft}} +  Q_{\rm dyn}^{i,\text{hard}}$, as seen in Fig.~\ref{fig:StdvsDyn} (middle panel).

Since gluon jets have a higher multiplicity of particles than quark jets, it contains a higher fraction of soft particles compared to quark jets, for a jet of given energy. For this reason, the gluon (quark) jets are characterized by a higher (lower) central peak and lower (higher) non-central peaks, as seen in Fig.~\ref{fig:StdvsDyn} (middle panel). These differences give rise to the enhanced discrimination power of the dynamic jet charge between quark and gluon jets. On the other hand, for $u$-quark vs. $d$-quark jet discrimination, the higher weighting given to hard particles in the jet makes the dynamic jet charge more resilient at discrimination in heavy ion collisions where significant soft background activity can contaminate the jet.

The middle panel of Fig.~\ref{fig:HardvsSoft} shows the separate distributions of the $Q_{\rm dyn}^{i,\text{soft}}$ and $Q_{\rm dyn}^{i,\text{hard}}$ contributions to the dynamic jet charge for the different choice of parameters, $\xi_{\rm cut}=0.3, k_<=0.3, k_>=1.0$. For this choice, the contribution of the soft particles to the jet charge is given a higher weighting compared to the hard particles. Once again, $Q_{\rm dyn}^{i,\text{soft}}$ is characterized by a single peak structure, but this time corresponds to the much broader $\kappa =k_< =0.3$ standard jet charge distribution.  The distribution of $Q_{\rm dyn}^{i,\text{hard}}$ is still characterized by accumulations near non-zero jet charge values but which are now closer to the $Q_J \sim 0$ region, corresponding to the tails of the now much narrower standard jet charge distribution with $\kappa=k_>=1.0$. Thus, in this case, we no longer have a clean separation between the contributions of the soft and hard particles. The  $Q_{\rm dyn}^{i,\text{soft}}$  is much broader and the $Q_{\rm dyn}^{i,\text{hard}}$ distribution gives less pronounced non-central peaks that also appear closer to the  $Q_J\sim 0$ region. Thus, we no longer have a clean separation in the dynamic jet charge of the effects of the hard and soft particles, and correspondingly reduced discrimination power.  This is seen in Fig.~\ref{fig:HardvsSoft} (bottom panel), where the corresponding total dynamic jet charge distributions for the quark and gluon jets are nearly identical.

Thus, the default choices, $\xi_{\rm cut}=k_< =1.0$ and $k_>=0.3$, which enhance and cleanly separate the effect of the hard particles over the soft particles is better suited for jet discrimination. 
The most optimal choice could be found through a more detailed scan of dynamic jet charge parameters.

\section{Phenomenology}

In this section, we consider some phenomenological applications of the dynamic and standard jet charges.

\subsection{Flavor Separation of Nucleon TMDs using Dynamic Jet Charge}

Recently, a theoretical framework was introduced~\cite{Kang:2020fka} to use the standard jet charge observable as a unique tool to probe the flavor structure
in the nucleon spin program at the future Electron-Ion Collider (EIC). Similar directions are currently being explored~\cite{Aschenauer:2015eha,bnltalk} experimentally at the Relativistic Heavy Ion Collider (RHIC). 

The unpolarized electron-proton scattering process 
\bea
\label{eq:unep}
e+p\to e+{\rm jet}+X,
\eea
 in the back-to-back limit where the electron-jet transverse momentum imbalance $q_T$ is small, is sensitive to the unpolarized nucleon transverse momentum dependent parton distribution functions~(TMD PDFs)~\cite{Liu:2018trl,Arratia:2020nxw}. The TMD PDFs provide 3D imaging of the nucleon in momentum space~\cite{Accardi:2012qut}. The polarized scattering counterpart, $e+p(S_\perp)\to e+{\rm jet}+X$, in the small $q_T$ limit, where $S_\perp$ denotes the proton transverse spin vector, receives an additional contribution from a term that is sensitive to the Sivers function~\cite{Sivers:1989cc}, the polarized TMD PDF, that encodes additional quantum correlations between the motion of partons and the spin of the proton. The Sivers function can be directly accessed via the Sivers asymmetry $A_{UT} = [d\sigma(S_\perp^\uparrow) -d\sigma(S_\perp^\downarrow)]/[d\sigma(S_\perp^\uparrow) +d\sigma(S_\perp^\downarrow)]$. 

\begin{figure}
    \centering
    \includegraphics[scale=0.8]{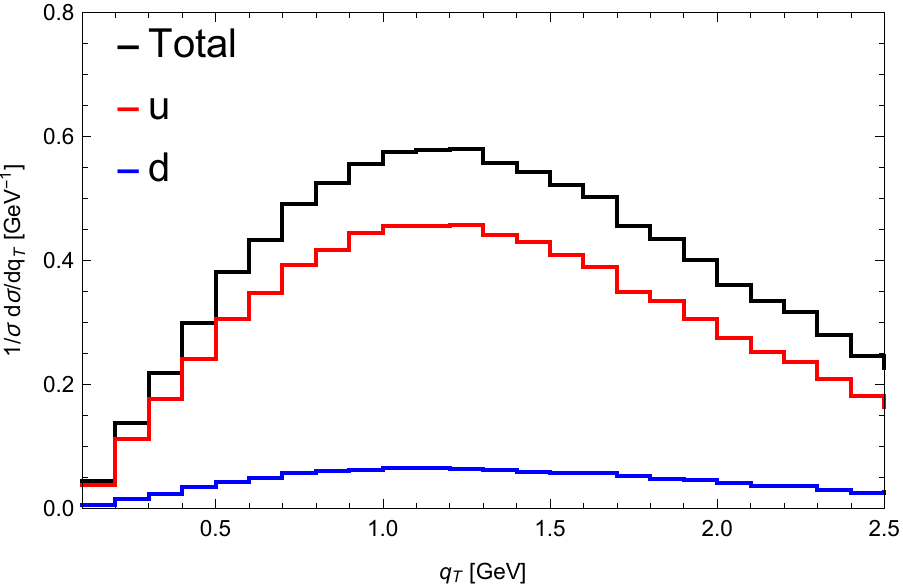}
    \caption{The relative size of contributions from the unpolarized $u$-quark (red), $d$-quark (blue), and sea quark (green) TMDs to the 
$q_T$-distribution integrated over all jet charge bins, corresponding to no jet charge measurement being made. }
    \label{fig:EICqTAllbins}
\end{figure}

While the electron-proton scattering cross section, in the small $q_T$ limit, probes polarized and unpolarized TMDs, it receives contributions from various partonic channels. Typically, the cross section is dominated by the $u$-quark channel so that it is sensitive primarily to the $u$-quark TMDs, as seen in Fig.~\ref{fig:EICqTAllbins} for the normalized $q_T$-distribution in unpolarized electron-proton scattering. Throughout this analysis we work in the center-of-mass (CM) frame with CM energy $\sqrt{s} = 105 \> {\rm GeV}$, and with event selection cuts of  $0.1\le y\le 0.85$, and  $15 \> {\rm GeV} \le p_{T}^e  \le 20 \>{\rm GeV}$,   $q_T < 2.5\>{\rm GeV}$, and $Q^2 > 10$ GeV$^2$, where $y$ denotes the inelasticity. Jets are constructed using the anti-$k_T$ jet algorithm~\cite{Cacciari_2008} with radius parameter $R = 1$.

One can enhance sensitivity to the $d$-quark TMD PDFs by making an additional measurement of the standard jet charge and restricting to a particular jet charge bin. For example, restricting to the standard jet charge bin $Q_J < -0.25$ increases the relative contribution of the $d$-quark TMD PDFs to the $q_T$-distribution, as seen in the bottom panel of Fig. 2 of Ref.~\cite{Kang:2020fka}. 

\begin{figure}
    \centering
    \includegraphics[scale=0.7]{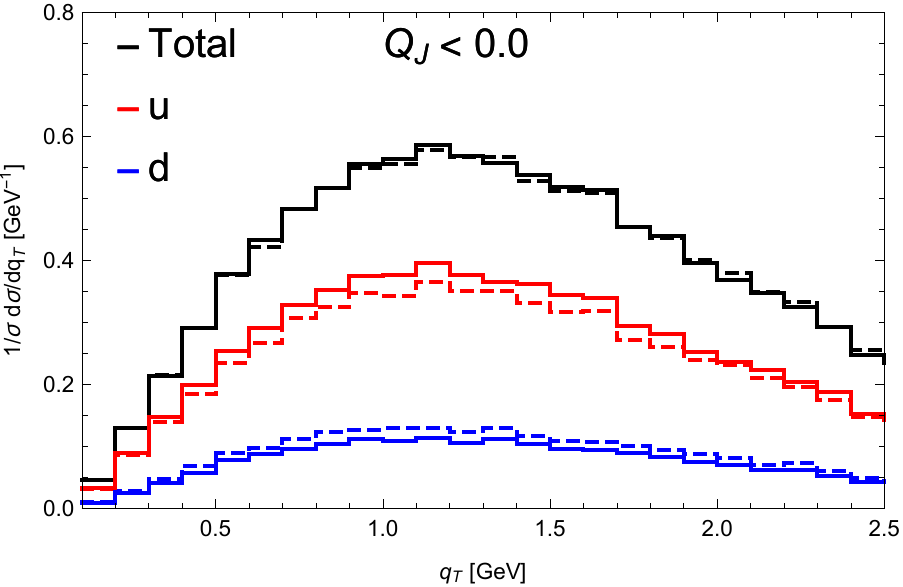}
    \includegraphics[scale=0.7]{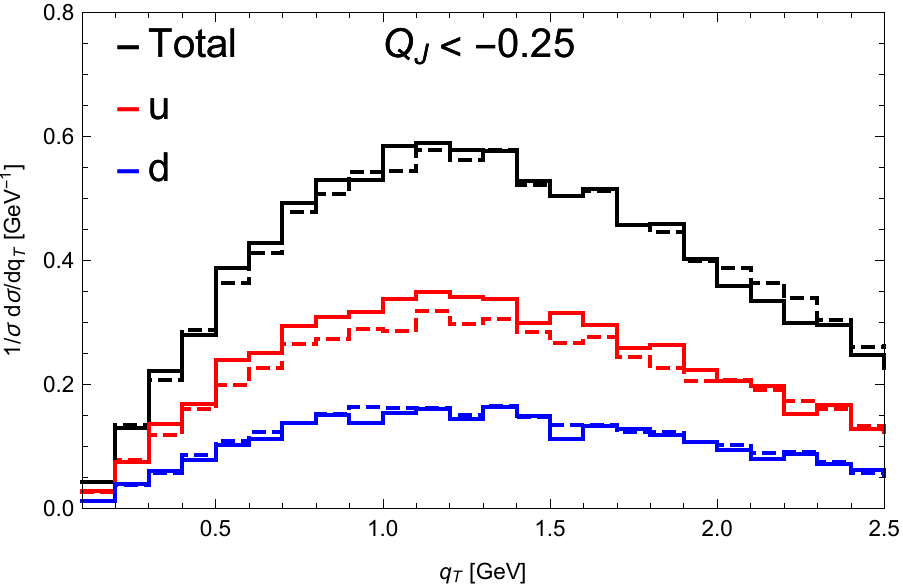}
    \caption{The relative size of contributions from the unpolarized $u$-quark (red) and $d$-quark (blue) TMDs to the 
$q_T$-distribution in different jet charge bins, for the dynamic (solid) and standard (dashed) jet charges. The top and bottom panels correspond to the jet charge bins $Q_J < 0.0$ and $Q_J < -0.25$, respectively.}
    \label{fig:EICqT}
\end{figure}

Here we show that the dynamic jet charge can also be used to probe the flavor structure of TMDs and can complement the standard jet charge analysis. In Fig.~\ref{fig:EICqT}, we show the normalized $q_T$ distribution for the unpolarized electron-proton scattering process in Eq.~(\ref{eq:unep}), restricted to specific jet charge bins. Here, the standard and dynamic jet charges are defined by restricting the sum over hadrons in the jet, in Eqs.~(\ref{eq:stdJC}) and (\ref{eq:dynJC}), to only include the charged pions $\pi^\pm$, so as to improve the sensitivity to the $u$- and $d$- quark TMDs. The top and bottom panels show the $q_T$-distribution restricted to the jet charge bins $Q_J < 0.0$ and $Q_J < -0.25$, respectively. The solid and dashed curves in Fig.~\ref{fig:EICqT} correspond to jet charge bin restrictions based on the dynamic and standard jet charge definitions, respectively. We see that the $q_T$-distribution receives a significantly higher relative contribution from the $d$-quark channel in these restricted jet charge bins, compared to Fig.~\ref{fig:EICqTAllbins} where no jet charge measurement is made. This enhanced sensitivity to $d$-quark TMD PDFs using the dynamic jet charge is similar to that found using the standard jet charge.

Thus, by sorting the data into jet charge bins, the dynamic jet charge can be used for flavor separation of nucleon TMDs and complement the standard jet charge analysis. The same strategy can be employed to use the dynamic jet charge for flavor separation of the Sivers function and we leave this for future work.

\subsection{Quark Jet Flavor Separation in Photon-Tagged Jet Production}

The RHIC spin program~\cite{Aschenauer:2015eha,Aschenauer:2016our,bnltalk} aims to further constrain the proton spin structure using longitudinally and transversely polarized proton beams to measure single and double spin asymmetries. The standard and dynamic jet charges can be used to disentangle the flavor structure of proton structure functions that arise in jet-production-based spin asymmetries. In addition, in heavy ion collisions, a jet charge analysis can probe the flavor dependence (quark vs.\ gluon or $u$-quark vs.\ $d$-quark) of jet modification~\cite{Li:2019dre,Chen:2019gqo,Sirunyan:2020qvi} due to parton propagation and showering in the nuclear medium.
\begin{figure}
    \centering
    \includegraphics[scale=0.7]{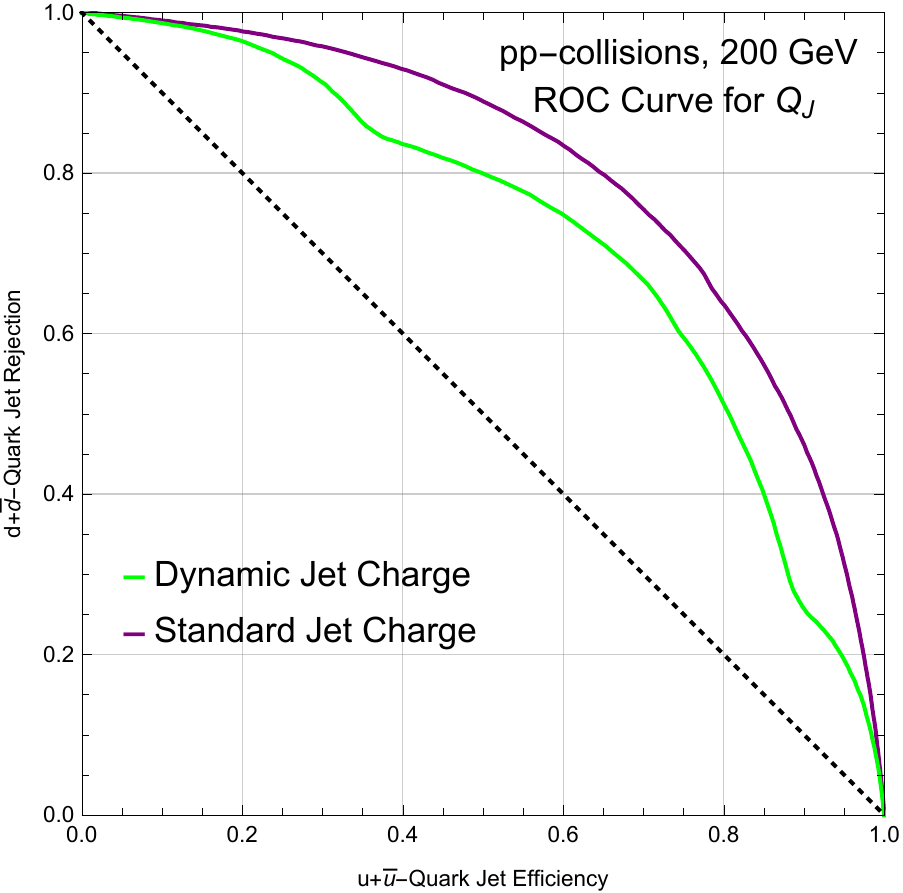}
    \includegraphics[scale=0.7]{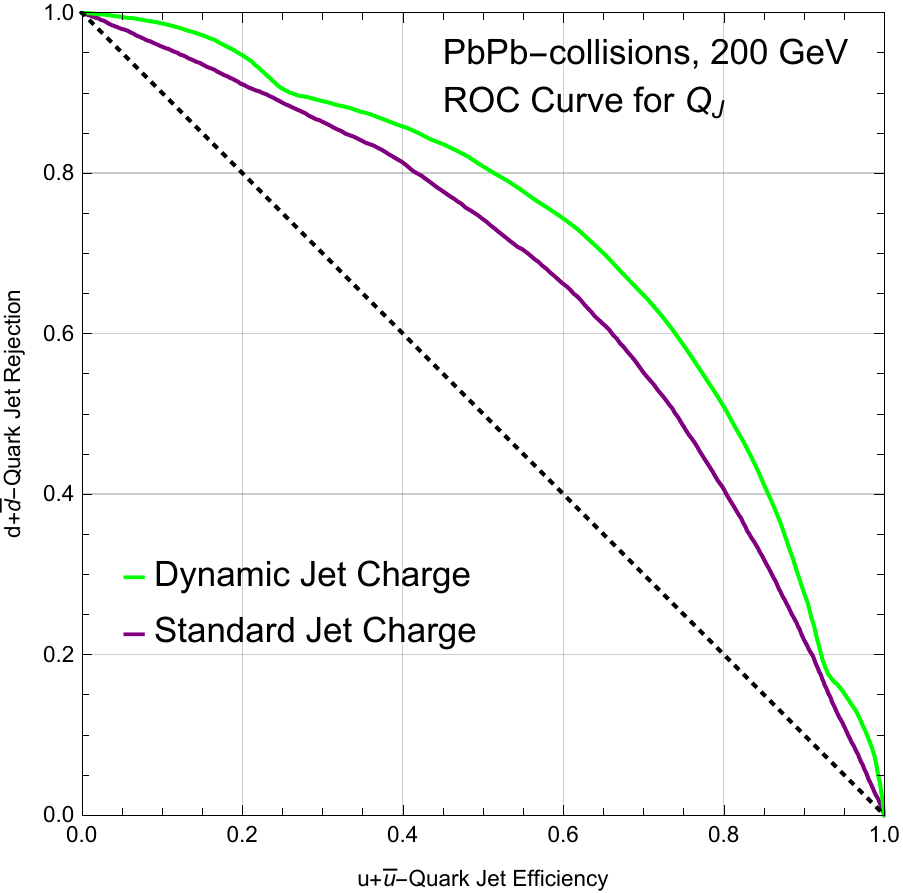}
    \caption{Global ROC curves for discrimination between $u$- or $\bar{u}$-jets and $d$- or $\bar{d}$-jets in $pp\to \gamma +  {\rm jet} + X$ (top panel) and PbPb$\to \gamma + {\rm jet} + X$ (bottom panel) based on Pythia simulations.  The collision center of mass energy and selection cuts used are: $\sqrt{s}=200$ GeV,  $p_{T} > 10.0$ GeV,  $p_{T}^\gamma > 30.0$ GeV, and $\Delta \phi > 7/8 \pi$. }
    \label{fig:ppvsPbPbgammajetuvsdNoqTcut}
\end{figure}

As an example, we consider the process of photon-tagged jet production in $pp$- and PbPb-collisions:
\bea
\label{eq:promptphoton}
p&+&p\to \gamma+{\rm jet}+X, \nn \\
 {\rm Pb} &+& {\rm Pb}\to \gamma+{\rm jet}+X,
\eea
A comparison of these processes provides valuable insights~\cite{Chatrchyan:2012vq,Dai:2012am,Kang:2017xnc,Aaboud:2018anc,Sirunyan:2018ncy} into jet modification since  the tagged photon leaves the strongly interacting medium relatively unaffected.  
\begin{figure}
    \centering
    \includegraphics[scale=0.7]{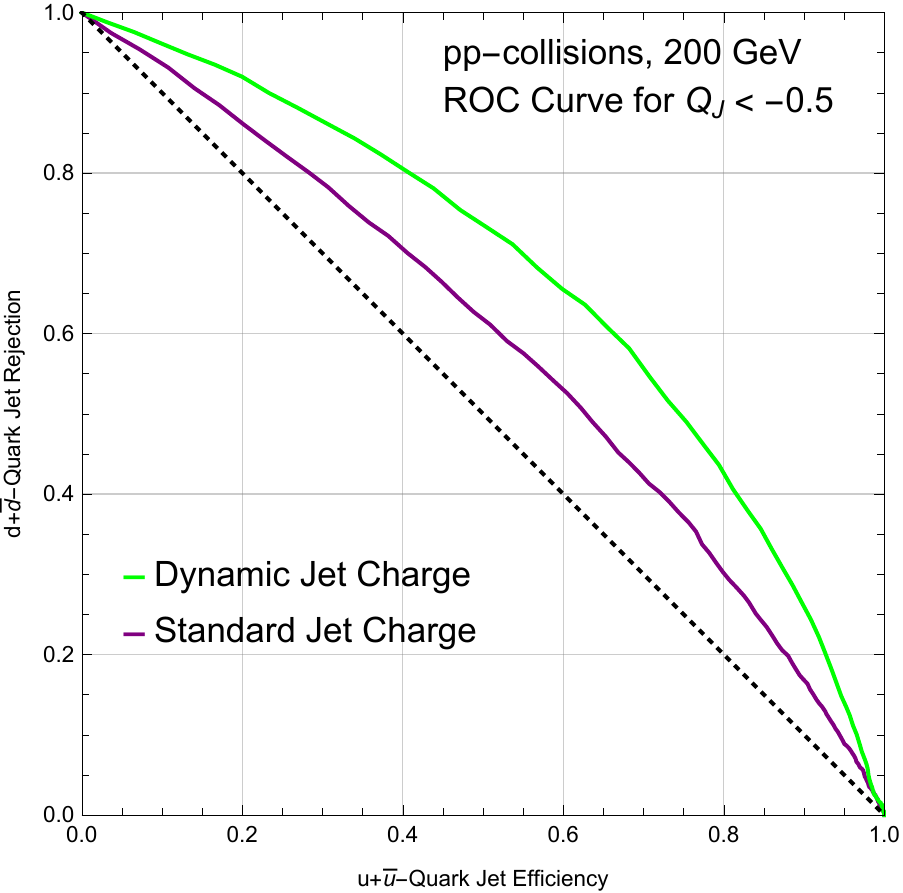}
    \includegraphics[scale=0.7]{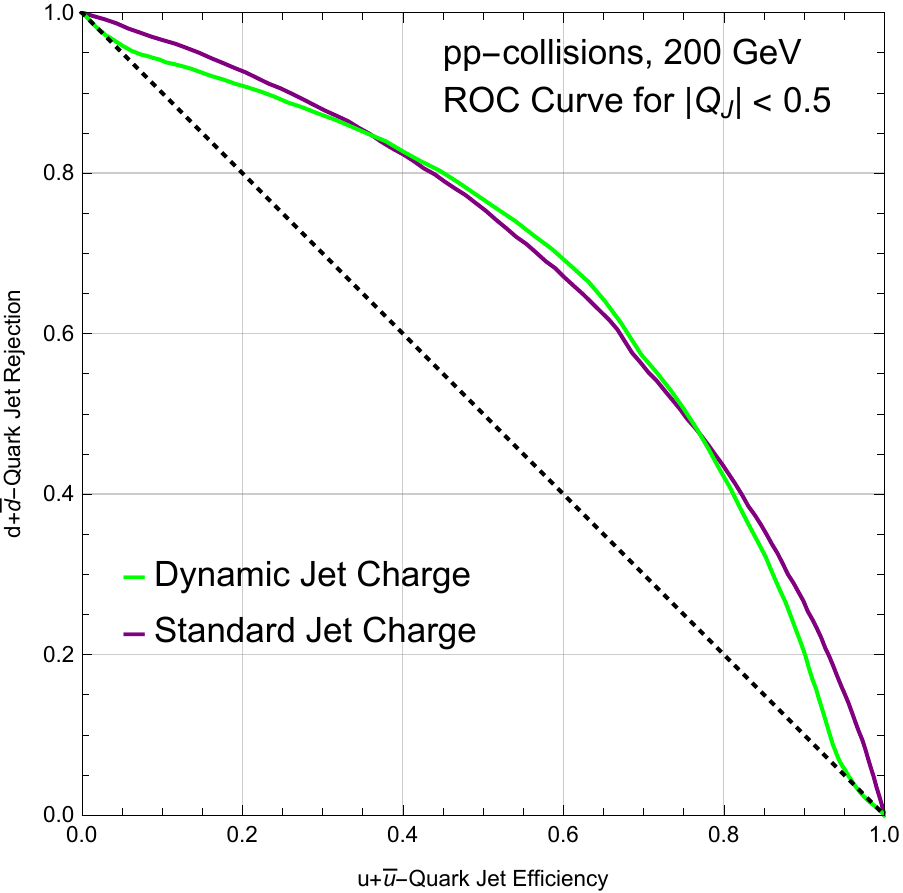}
    \includegraphics[scale=0.7]{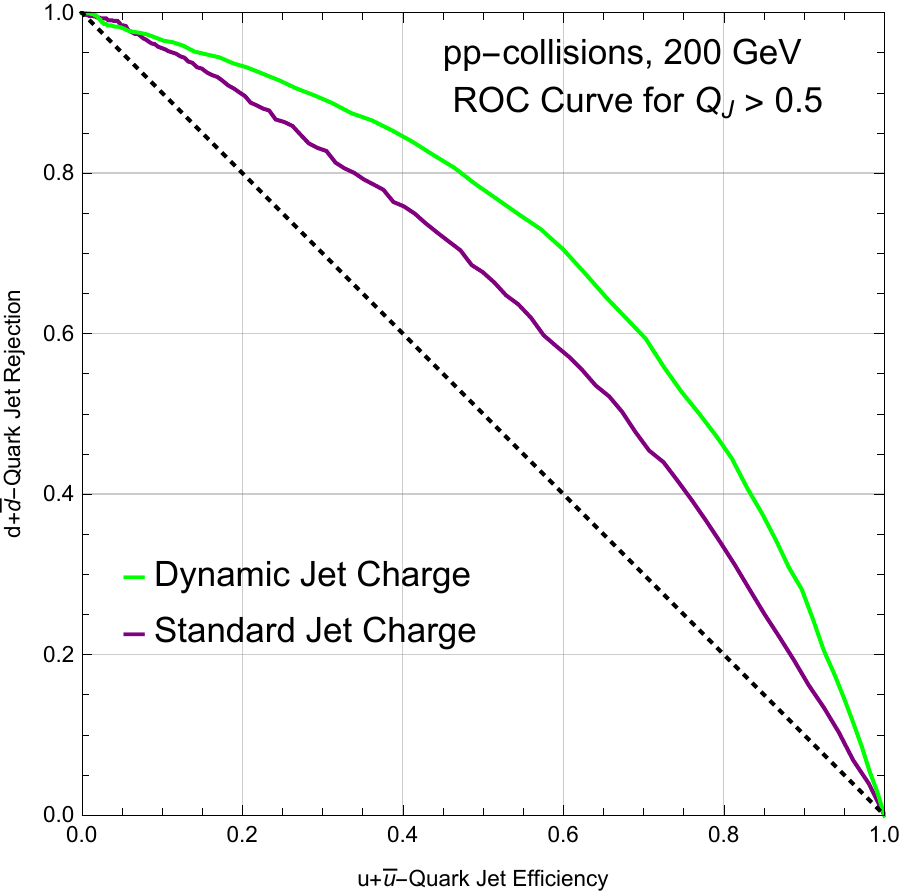}
\caption{Local ROC curves for discrimination between $u$- or $\bar{u}$-jets and $d$- or $\bar{d}$-jets in $pp\to \gamma +  {\rm jet} + X$ based on Pythia simulations.  The collision center of mass energy and selection cuts used are: $\sqrt{s}=200$ GeV,  $p_{T} > 10.0$ GeV,  $p_{T}^\gamma > 30.0$ GeV, and $\Delta \phi > 7/8 \pi$. The top, middle, and bottom panels correspond to the jet charge bins $Q_J < -0.5, |Q_J|  < 0.5$, and $Q_J >0.5$, respectively.}
    \label{fig:ppgammajetuvsdNoqTcutBinned}
\end{figure}
\begin{figure}
    \centering
    \includegraphics[scale=0.7]{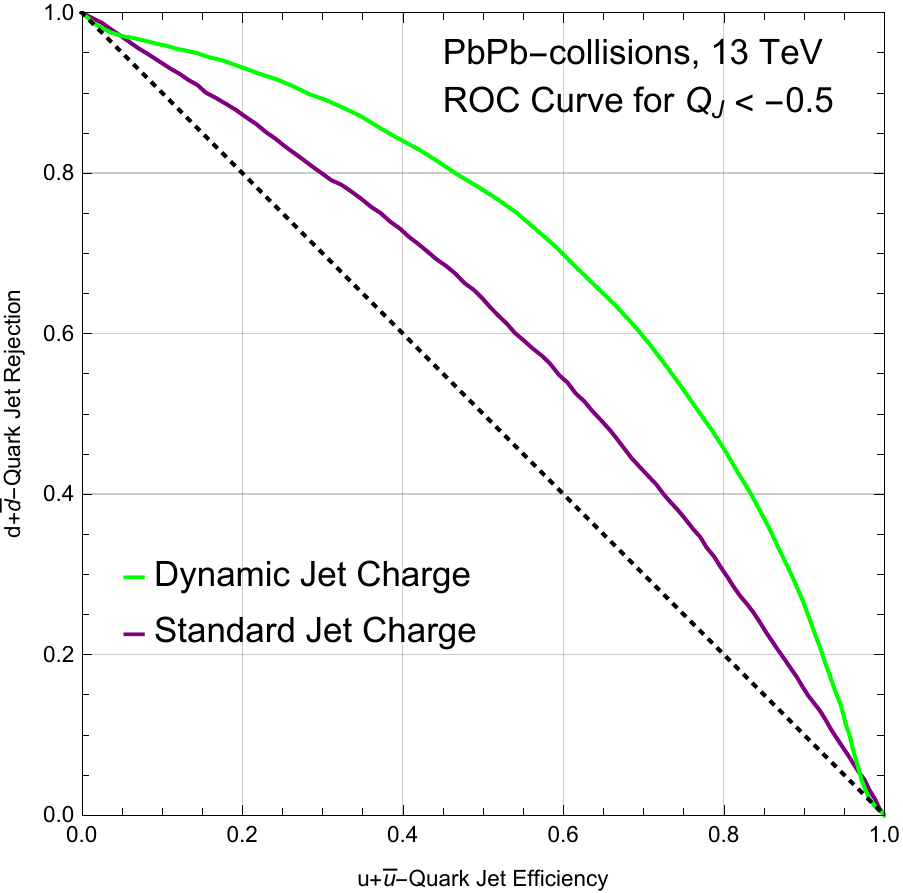}
    \includegraphics[scale=0.7]{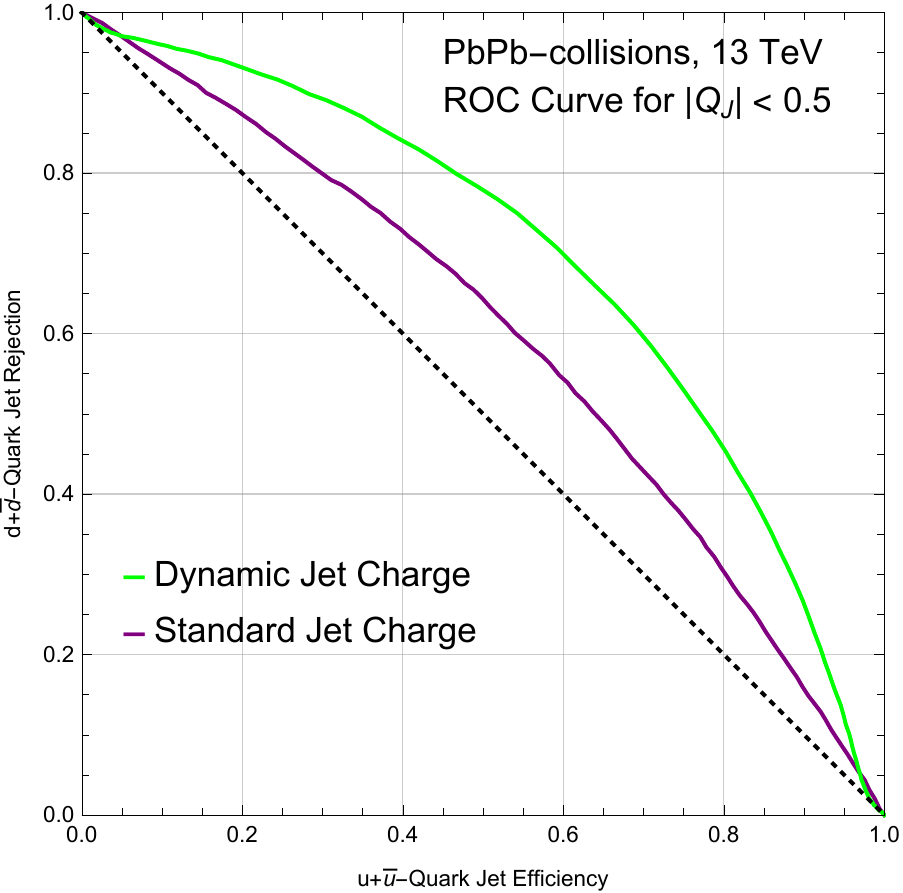}
    \includegraphics[scale=0.7]{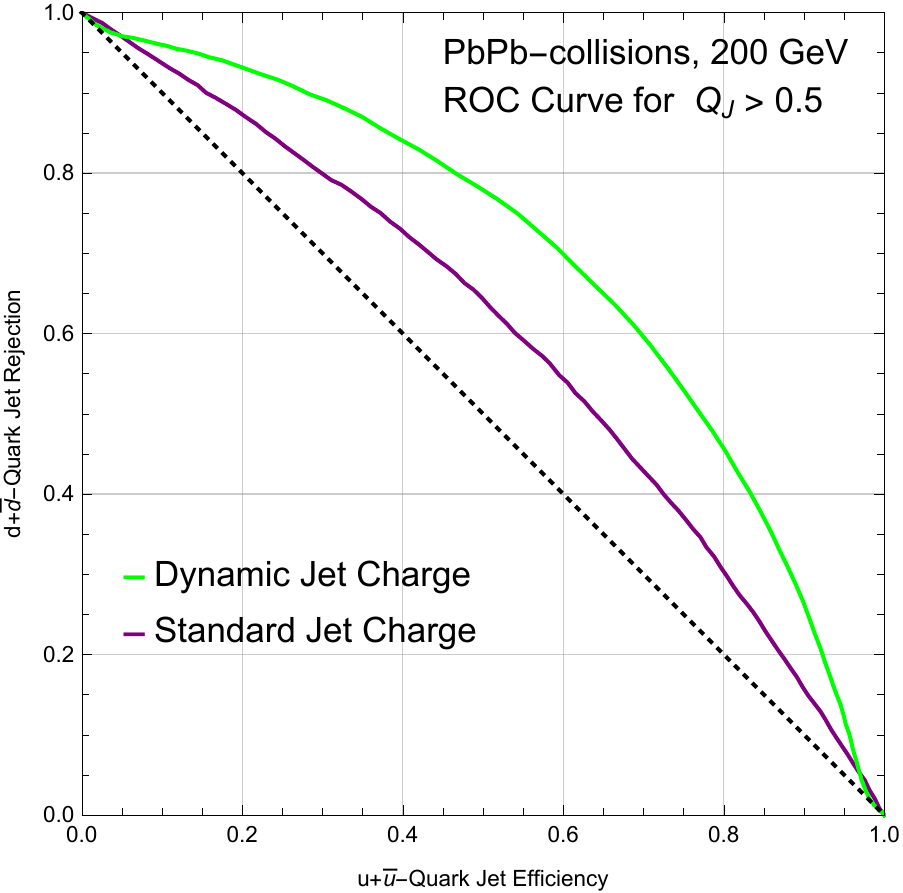}
\caption{Local ROC curves for discrimination between $u$- or $\bar{u}$-jets and $d$- or $\bar{d}$-jets in PbPb$\to \gamma +  {\rm jet} + X$ based on Pythia simulations.  The collision center of mass energy and selection cuts used are: $\sqrt{s}=200$ GeV,  $p_{T} > 10.0$ GeV,  $p_{T}^\gamma > 30.0$ GeV, and $\Delta \phi > 7/8 \pi$. The top, middle, and bottom panels correspond to the jet charge bins $Q_J < -0.5, |Q_J|  < 0.5$, and $Q_J >0.5$, respectively.}
    \label{fig:PbPbgammajetuvsdNoqTcutBinned}
\end{figure}

At the parton level, photon-tagged jet production is mediated by the channels:
\bea
\label{eq:promptphotonpartonic}
q &+& g\to \gamma+q , \nn \\
\bar{q} &+& g\to \gamma+\bar{q},  \nn \\
q&+&\bar{q}\to \gamma+g.
\eea
The $u,d$-quark jet production dominates due to the dominance of the quark and gluon PDFs  in the initial state. In the limit of small $q_T$, photon-tagged jet production can be a probe of the $u$-quark, $d$-quark, and gluon TMD PDFs. Similarly, asymmetries constructed from longitudinally and transversely polarized proton beams, can probe helicity PDFs~\cite{deFlorian:2014yva,Nocera:2014gqa} and the Sivers functions~\cite{Buffing:2018ggv,Kang:2020xez}, respectively.  The use of jet charge to discriminate between the production of $u$-quark and $d$-quark jets can thus provide complementary tools for flavor separation of proton structure functions.

We demonstrate the  $u$- vs. $d$-quark jet discrimination power of the dynamic jet charge in the photon-tagged jet production processes in Eq.~(\ref{eq:promptphoton}) through Pythia simulation results. We simulate $pp$- and PbPb-collisions at $\sqrt{s}= 200$ GeV with selection cuts of $p_T^\gamma > 30.0$ GeV, $p_{T}^{\rm jet} > 10.0$ GeV, and $\Delta \phi_{\gamma-\rm jet} > \frac{7}{8} \pi$, corresponding to the tagged photon transverse momentum, the leading jet momentum, and the azimuthal angular separation between the photon and the leading jet.

In Fig.~\ref{fig:ppvsPbPbgammajetuvsdNoqTcut}, we show the global ROC curves for discrimination between $u$- or $\bar{u}$-jets and $d$- or $\bar{d}$-jets in $pp\to \gamma + {\rm jet} + X$ (top panel) and PbPb$\to \gamma + {\rm jet}  + X$ (bottom panel), using the standard jet charge (purple) and the dynamic jet charge (green). The global ROC curves indicate that while the standard jet charge provides better discrimination in $pp$-collisions, the dynamic jet charge gives better discrimination in  PbPb-collisions.  Once again, this can be understood as a result of the robustness of the dynamic jet charge discrimination power in the presence of increased underlying activity in heavy-ion collisions.

However, even in $pp$-collisions, the dynamic jet charge can provide improved discrimination when the data is sorted into jet charge bins. As seen in Fig.~\ref{fig:ppgammajetuvsdNoqTcutBinned}, the dynamic jet charge gives better discrimination in the $Q_J < -0.5$ and $Q_J >0.5$ bins and comparable discrimination in the $|Q_J|< 0.5$ bin. Thus, the dynamic jet charge can be a complementary probe of the flavor structure of the proton spin. Fig.~\ref{fig:PbPbgammajetuvsdNoqTcutBinned} shows the same jet charge binned ROC curves for  PbPb-collisions. In this case, the dynamic jet charges provides improved discrimination in all jet charge bins.

\section{Likelihood Ratio Analysis of Jet Charge DIscrimination}

In our analysis so far, we have made use of ROC curves to quantify the jet charge discrimination power. This is standard practice in papers exploring new jet discrimination observables where the signal and background distributions are typically characterized by a single peak that may be translated relative to each other. In the presence of multiple peaks, as in the case of the dynamic jet charge, a sliding cut in the signal and background distributions may not be monotonically related to a cut in the likelihood ratio, as is often used in experimental analyses. Due to this multiple peak structure of the dynamic jet charge distribution, we also present here an analysis of signal versus background likelihood ratio distributions to quantify the absolute discrimination power. We find that the underlying conclusions derived from such an analysis of likelihood distributions are consistent with those based on the ROC analysis. 

We follow the procedure outlined in Refs.~\cite{Cousins:JHEP2005,Cousins:2018tiz}. For quark vs. gluon jet discrimination, we treat the quark jets as signal events and gluon jets as background events. For $u$-quark vs. $d$-quark jet discrimination, we treat the $u$-quark jets as signal events and $d$-quark jets as background events. We apply the procedure below:

\begin{itemize}
\item From the Pythia simulated standard and dynamic jet charge distribution data, we compute the binned jet charge probability values for quark, gluon, $u$-quark, and $d$-quark jets. This simply corresponds to the bin heights of the normalized jet charge distributions and provides us with the underlying signal and background standard and dynamic jet charge probability distributions.
\item From these underlying signal and background probability distributions, we generate 20,000 data signal and background samples, each of size $N=50$. 
\item For each of these samples, we calculate the likelihood ratio statistic:
\bea
-2\>\ln \lambda = \sum_{i=1}^N -2 \> \ln \left [\frac{p(Q_J|\text{signal})}{p(Q_J|\text{background})} \right ],
\eea
where $p(Q_J|\text{signal})$ and $p(Q_J|\text{background})$ give the probability for a given jet of charge $Q_J$ to be a signal jet or background jet, respectively. 
\item We generated the likelihood ratio distribution of $-2\>\ln \lambda$ for 20,000 samples taken from the signal and background data, and compared them. The likelihood ratio statistic approaches a Gaussian distribution as one increases the sample size, as expected from the Central Limit Theorem.
\item One criteria~\cite{Cousins:JHEP2005,Cousins:2018tiz} for determining whether a given data set should be classified as data corresponding to signal events or background events, is based on the value:
\bea
-2 \ln \lambda_{\rm cut} = \frac{M_{\rm sig.} S_{\rm bkg.}+M_{\rm bkg.} S_{\rm sig.}}{S_{\rm sig.}+S_{\rm bkg.}},
\eea
where $M_{\rm sig.}$ and $M_{\rm bkg.}$ are the mean values of the signal and background distributions, respectively. $S_{\rm sig.}$ and $S_{\rm bkg.}$ are the standard deviations of the signal and background distributions, respectively. The value of $-2 \ln \lambda_{\rm cut}$ corresponds to the value such that the number of standard deviations, $S_{\rm sig.}$, it is away from $M_{\rm sig.}$ is the same as the number of standard deviations, $S_{\rm bkg.}$, it is away from $M_{\rm bkg.}$. If the calculated value of $-2\ln \lambda$ for a given data sample, lies to the left (right) of $-2 \ln \lambda_{\rm cut}$, the data sample is classified as corresponding to signal (background) events.
\item Finally, we note that the values of $M_{\rm sig.}$ and $M_{\rm bkg.}$ scale with the sample size $N$ and the values of $S_{\rm sig.}$ and $S_{\rm bkg.}$ scale with $\sqrt{N}$~\cite{Cousins:JHEP2005,Cousins:2018tiz}. Thus, as one increases the sample size, the signal and background likelihood Gaussian distributions will move further apart from each other, on either side of $-2 \ln \lambda_{\rm cut}$, corresponding to a cleaner separation of these distributions. Thus, the smaller the required sample size to achieve a clean separation of signal and background likelihood distributions, the greater the discrimination power of the observable. 
\end{itemize}

\begin{figure}
    \centering
    \includegraphics[scale=0.7]{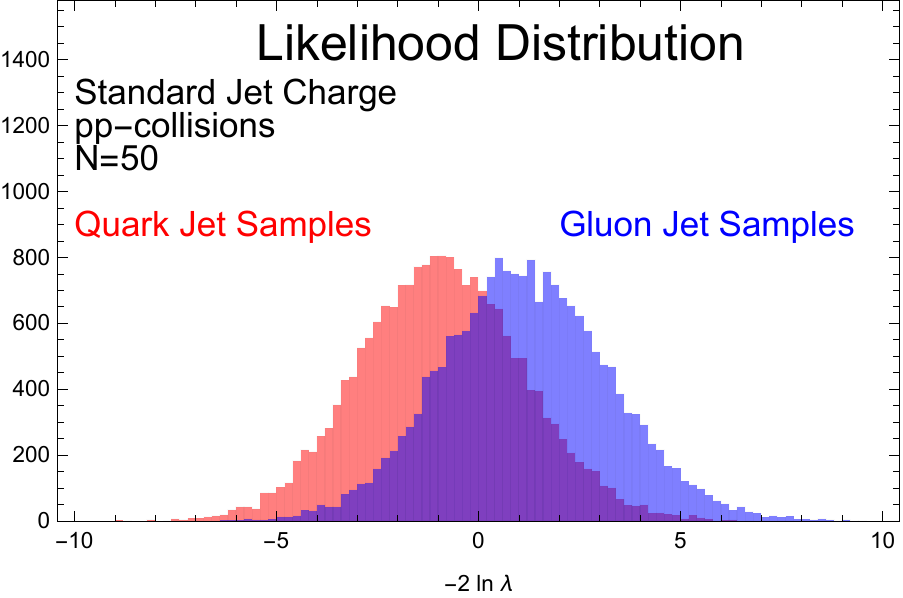}
    \includegraphics[scale=0.7]{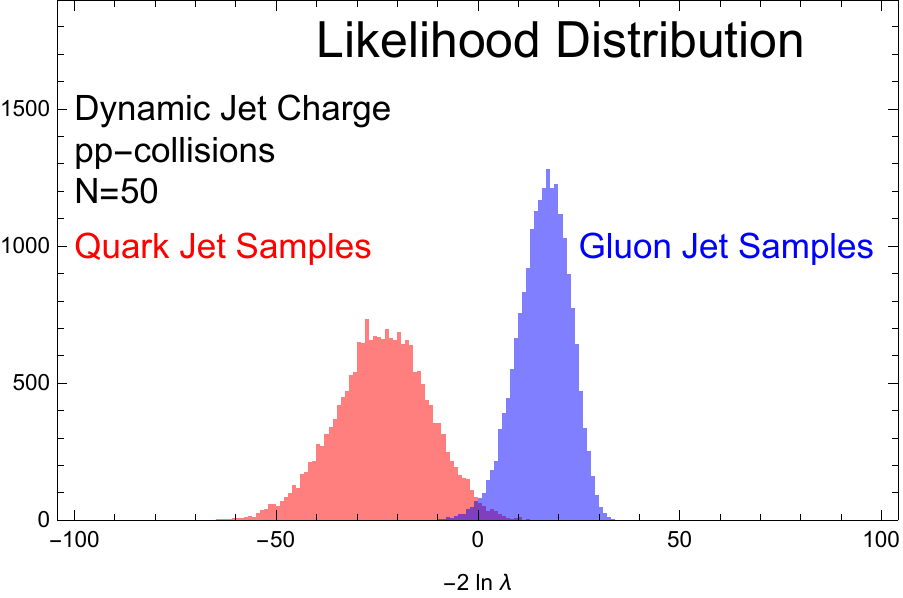}
\caption{The standard (top panel) and dynamic (bottom panel) jet charge likelihood distributions for quark (red) and gluon (blue) jet samples. The $-2 \ln \lambda_{\rm cut}$ values for the Standard and Dynamic likelihood distributions are 0.103  and 1.90, respectively. The distributions above were generated for the more central jet in dijet production of $pp$-collisions at $\sqrt{s}=$13 TeV, with selection cuts $p_{T_J} = [200,300]$ GeV, $|\eta_J| < 2.1$, and $p_{T_{j_1}}/p_{T_{j_2}} < 1.5$ on the leading ($j_1$) and subleading ($j_2$) anti-k$_T$ jets of jet radius $R=0.4$. }
    \label{fig:QvsGppLR}
\end{figure}

In Fig.~\ref{fig:QvsGppLR}, we show the standard (top panel) and dynamic (bottom panel) jet charge likelihood distributions for quark and gluon jet samples in $pp$-collisions. We see clearly that with a sample size of $N=50$, the dynamic jet charge already gives clean separation of the quark and gluon jet likelihood distributions. On the other hand, there is still significant overlap of the quark and gluon jet likelihood distributions for the standard jet charge. This indicates that the dynamic jet charge has higher discrimination power between quark and gluon jets, consistent with what was found using the ROC curve analysis, as seen in the bottom panel of Fig.~\ref{fig:StdvsDyn}.

\begin{figure}
    \centering
    \includegraphics[scale=0.7]{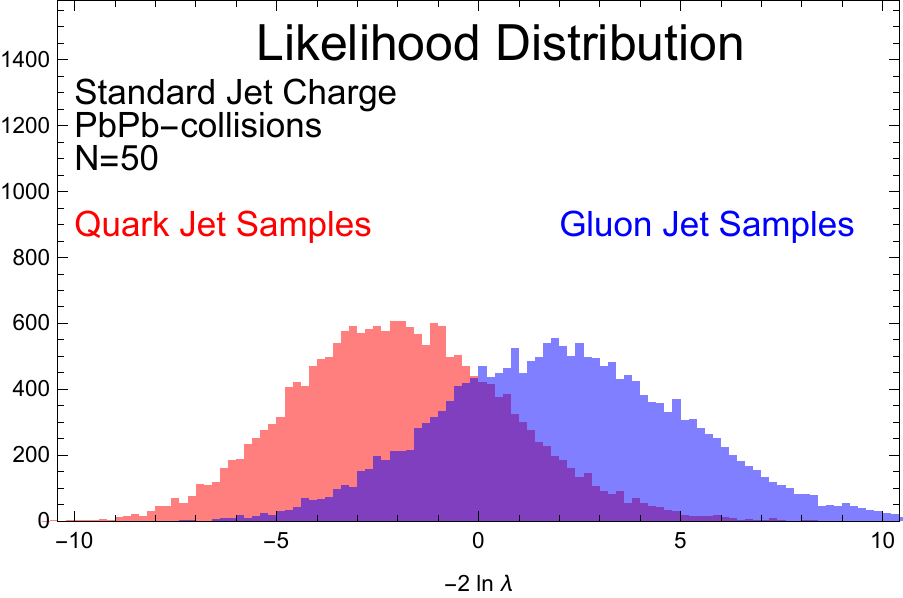}
    \includegraphics[scale=0.7]{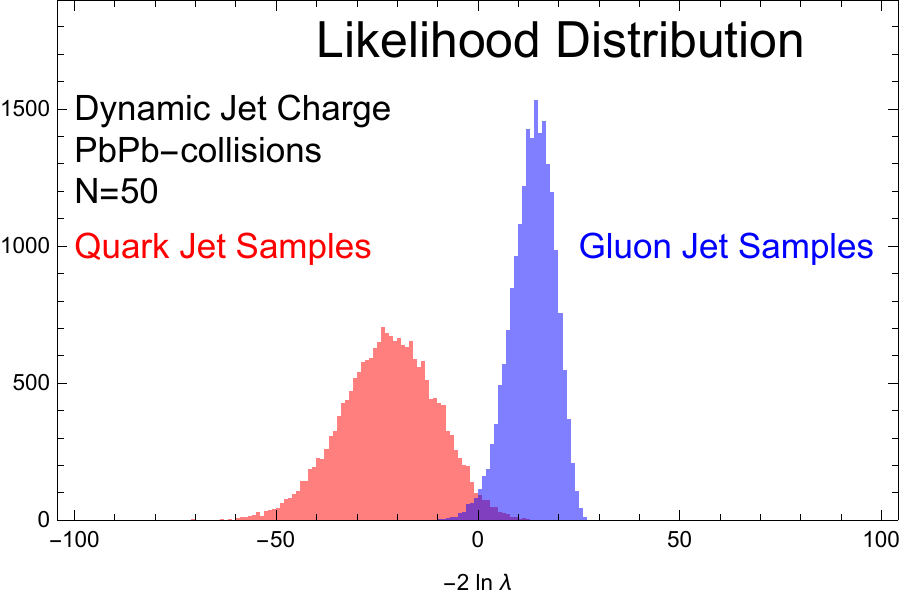}
\caption{The standard (top panel) and dynamic (bottom panel) jet charge likelihood distributions for quark (red) and gluon (blue) jet samples. The $-2 \ln \lambda_{\rm cut}$ values for the Standard and Dynamic likelihood distributions are -0.0308  and 2.17, respectively. The distributions above were generated for the more central jet in dijet production of PbPb-collisions at $\sqrt{s}=$2.760 TeV, with selection cuts $p_{T_J} = [80,150]$ GeV, $|\eta_J| < 0.9$, and $p_{T_{j_1}}/p_{T_{j_2}} < 1.5$ on the leading ($j_1$) and subleading ($j_2$) anti-k$_T$ jets of jet radius $R=0.4$. }
    \label{fig:QvsGPbPbLR}
\end{figure}

Similar results are observed for heavy-ion PbPb-collisions. In Fig.~\ref{fig:QvsGPbPbLR},  we show the standard (top panel) and dynamic (bottom panel) jet charge likelihood distributions for quark and gluon jet samples in PbPb-collisions.
Once again, we see that with $N=50$, the dynamic jet charge already gives clean separation of the quark and gluon jet likelihood distributions while there is still significant overlap for the standard jet charge. This conclusion is consistent with the ROC curves in Fig.~\ref{fig:StdvsDynHeavyIon}.

\begin{figure}
    \centering
    \includegraphics[scale=0.7]{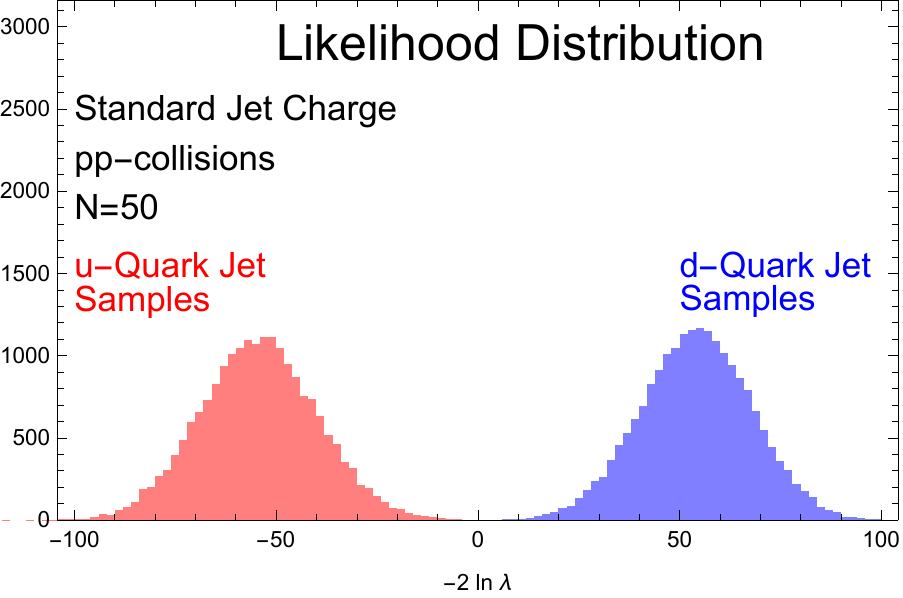}
    \includegraphics[scale=0.7]{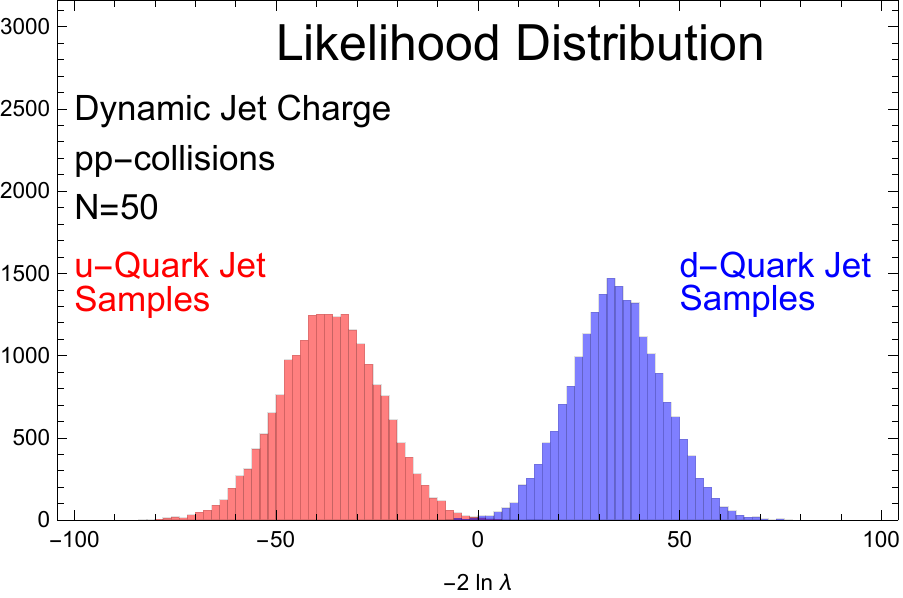}
\caption{The standard (top panel) and dynamic (bottom panel) jet charge likelihood distributions for $u$-quark (red) and $d$-(blue) jet samples. The $-2 \ln \lambda_{\rm cut}$ values for the Standard and Dynamic likelihood distributions are 1.34  and 0.340, respectively. The distributions above were generated for the more central jet in dijet production of $pp$-collisions at $\sqrt{s}=13$ TeV with $R=0.4$,  $p_{T_J} = [200,300]$ GeV, and $|\eta_J| < 2.1$. The partonic channels $dg\to W^-u$ and $ug\to W^+ d$ generated the $u$-quark and $d$-quark jets, respectively.}
    \label{fig:UvsDppLR}
\end{figure}

\begin{figure}
    \centering
    \includegraphics[scale=0.7]{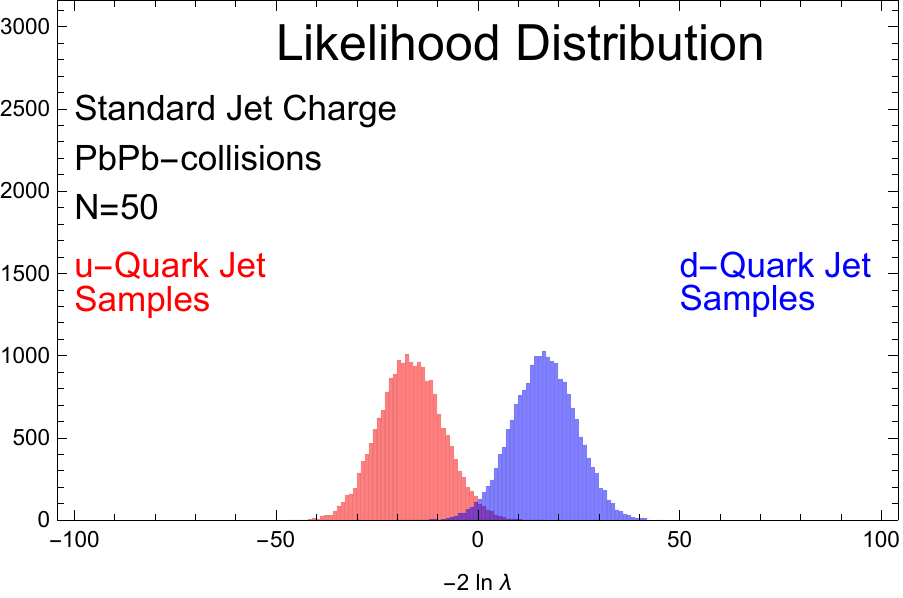}
    \includegraphics[scale=0.7]{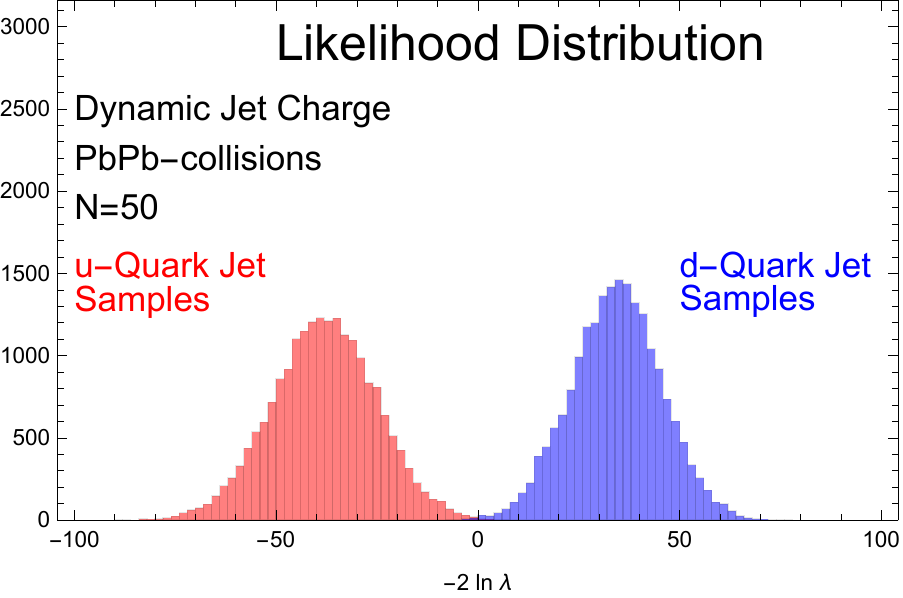}
\caption{The standard (top panel) and dynamic (bottom panel) jet charge likelihood distributions for $u$-quark (red) and $d$-(blue) jet samples. The $-2 \ln \lambda_{\rm cut}$ values for the Standard and Dynamic likelihood distributions are -0.0129  and 0.887, respectively. The distributions above were generated for the more central jet in dijet production of PbPb-collisions at $\sqrt{s}=2.760$ TeV with $R=0.4$,  $p_{T_J} = [80,150]$ GeV, and $|\eta_J| < 0.9$. The partonic channels $dg\to W^-u$ and $ug\to W^+ d$ generated the $u$-quark and $d$-quark jets, respectively.}
    \label{fig:UvsDPbPbLR}
\end{figure}

In Fig.~\ref{fig:UvsDppLR}, we show the standard (top panel) and dynamic (bottom panel) jet charge likelihood distributions for $u$-quark and $d$-quark jet samples in $pp$-collisions. In this case, we see that the standard jet charge has better (but comparable) discrimination between $u$-quark and $d$-quark jets, compared to the dynamic jet charge. This can be seen from the greater separation between the $u$-quark and $d$-quark likelihood distributions for the standard jet charge compared to the dynamic jet charge. Once again, this conclusion is consistent with the ROC curves in the bottom panel of Fig.~\ref{fig:StdvsDynuvsd}.

However, as seen in Fig.~\ref{fig:UvsDPbPbLR}, in heavy ion collisions the dynamic jet charge seems to give better (but comparable) discrimination between $u$-quark and $d$-quark jets, compared to the standard jet charge. We see that for $N=50$, the dynamic jet charge gives a cleaner (but comparable) separation between the likelihood distributions of the $u$-quark and $d$-quark jets in heavy ion collisions. This is consistent with the ROC curves in the bottom panel of Fig.~\ref{fig:StdvsDynuvsdHI} and is the result of the relative insensitivity of the dynamic jet charge to the significantly increased soft contamination in heavy ion collisions.

\section{Conclusions}

We  proposed a modified definition of the jet charge observable, the dynamic jet charge, which makes the parameter $\kappa$ that appears in the standard definition to be a function of some dynamic property of either the jet or the individual jet constituents.   We focused on the specific scenario where each hadron in the jet contributes to the dynamic jet charge with a dynamic parameter, $\kappa(z_h)$, determined by its jet transverse momentum or energy fraction. We have shown that the dynamic jet charge can complement analyses based on the standard jet charge and allow for improved jet discrimination.

The behavior of the dynamic jet charge was studied using Pythia8 simulations of $pp$-collisions and heavy ion PbPb-collisions. It was found that the discrimination power of the dynamic jet charge in PbPb-collisions is about the same as in $pp$-collisions, being largely unaffected by the significantly greater underlying event activity. The multiple peak structure observed in the dynamic jet charge distribution, also allows for a binned analysis where a discrimination analysis can be performed on jet data binned according to the jet charge. The jet charge bins are  centered around the peaks and with size corresponding to the width of the peaks. Such a binned analysis could provide additional discrimination compared to an unbinned analysis alone. We defined \textit{local} Receiver Operating Characteristic (ROC) curves, in the presence of multiple peak structures, to quantify the discrimination power within local jet charge bins that only involve a single peak structure, as opposed to global ROC curves that use data from all jet charge bins. Due to the multiple peak structure of the dynamic jet charge, we also presented a signal vs. background likelihood distribution analysis for quantifying the absolute discrimination power, even in the presence of multiple peak structures.
 
Pythia simulations indicated that the dynamic jet charge provides significantly improved discrimination between quark and gluon initiated jets, in both proton-proton and heavy ion collisions, compared to the standard jet charge. For discrimination between $u$-quark and $d$-quark initiated jets in proton-proton collisions, the global ROC curves and the likelihood distributions indicated that the standard jet charge has better, but comparable, discrimination power. However, the local ROC curves indicated that the dynamic jet charge can provide improved discrimination when the data is sorted into jet charge bins. Thus, the dynamic jet charge could complement jet flavor discrimination based on the standard jet charge analysis. For heavy ion collisions, it was found that the dynamic jet charge gives better, but comparable, discrimination as quantified by both the global and local ROC curves, and the likelihood analysis. This is a consequence of the improved resilience of the dynamic jet charge against contamination effects of the underlying event activity in heavy ion collisions. Future work includes studying the dynamic jet charge in heavy ion collisions using the JEWEL or JETSCAPE Monte Carlo simulation which, unlike the Pythia8 simulation, includes hot nuclear medium effects.  

We also presented phenomenological studies to show that the dynamic and standard jet charges can be a unique probe of the nuclear flavor structure through simulation studies of deep inelastic scattering in the back-to-back region between the final electron and the leading jet, and in photon-tagged jet production in proton-proton and heavy ion collisions. We envision many other phenomenological applications of both the standard and dynamic jet charges to probe various aspects of nuclear structure. We leave such explorations for future work. Finally, the underlying idea of using a dynamic parameter for the jet charge can, in principle, be applied to parameters in the definitions of other observables, such as jet angularities.

\section*{Acknowledgments} 
We thank Robert D. Cousins for helpful discussions on the likelihood ratio analysis. Z.K. is supported by the National Science Foundation under CAREER award~PHY-1945471. X.L. is supported by National Natural Science Foundation of China under Grant No.11775023. S.M., M.S., and T.W. thank UNG for support through the Faculty Undergraduate Summer Engagement (FUSE) grant.

\bibliographystyle{h-physrev3.bst}
\bibliography{jetcharge}

\end{document}